\documentclass[12pt,a4paper]{article}
 
\usepackage{jheppub}
\usepackage[usenames,dvipsnames]{xcolor}
\usepackage{amssymb} 
\usepackage{amsmath}
\usepackage{scalerel}
\usepackage{trimclip}
\usepackage{mathtools}
\usepackage{amsfonts}    
\usepackage{dsfont}
\usepackage{bm}
\usepackage{pdfpages}
\usepackage{verbatim}
\usepackage{stmaryrd}
\usepackage{graphicx}
\usepackage{relsize}
\usepackage{braket}
\usepackage[most]{tcolorbox}
\usepackage{tensor}
\usepackage{mathrsfs}
\usepackage{textgreek} 
\usepackage{enumitem}
\usepackage[mathscr]{euscript}
\usepackage[smalltableaux]{ytableau}
\ytableausetup{centertableaux}
\usepackage{tikz}
\usetikzlibrary{decorations.pathreplacing, decorations.markings,calc,shapes.misc,decorations.pathmorphing,patterns.meta, math}
\usepackage{slashed}
\usepackage{datetime}
\usepackage{hyperref}

\hypersetup{
    pdfencoding=unicode,
	colorlinks=true,
	urlcolor=Maroon,
	linkcolor=RoyalBlue,
	citecolor=Maroon,
	pdftitle={A universal sum over topologies in 3d gravity},
	pdfauthor={},
	pdfdisplaydoctitle=true,
	pdfstartview=FitH,
	linktocpage=true
}
\definecolor{wilsonred}{RGB}{255,4,34}
\definecolor{vert}{rgb}{0.1367 0.543 0.1367}

\usetikzlibrary{shapes}
\tikzset{
    partial ellipse/.style args={#1:#2:#3}{
        insert path={+ (#1:#3) arc (#1:#2:#3)}
    }
}

\newcommand{\ba}{\begin{align}}

\newcommand{\be}{\begin{equation}}
\newcommand{\ee}{\end{equation}}
\def\bd{\begin{tikzpicture}}
\def\ed{\end{tikzpicture}}

\allowdisplaybreaks

\newcommand\PSL{\text{PSL}}
\newcommand\SL{\text{SL}}
\newcommand\ZZ{\mathbb{Z}}

\renewcommand\d{\text{d}}
\newcommand{\e}{\mathrm{e}}

\newcommand{\U}{\mathrm{U}}

\renewcommand{\le}{\leqslant}
\renewcommand{\ge}{\geqslant}
\renewcommand{\leq}{\leqslant}
\renewcommand{\geq}{\geqslant}
\newcommand{\id}{\mathds{1}}

\makeatletter
\newcommand{\dbloverline}[1]{\overline{\dbl@overline{#1}}}
\newcommand{\dbl@overline}[1]{\mathpalette\dbl@@overline{#1}}
\newcommand{\dbl@@overline}[2]{%
  \begingroup
  \sbox\z@{$\m@th#1\overline{#2}$}%
  \ht\z@=\dimexpr\ht\z@-2\dbl@adjust{#1}\relax
  \box\z@
  \ifx#1\scriptstyle\kern-\scriptspace\else
  \ifx#1\scriptscriptstyle\kern-\scriptspace\fi\fi
  \endgroup
}
\newcommand{\dbl@adjust}[1]{%
  \fontdimen8
  \ifx#1\displaystyle\textfont\else
  \ifx#1\textstyle\textfont\else
  \ifx#1\scriptstyle\scriptfont\else
  \scriptscriptfont\fi\fi\fi 3
}
\makeatother

\DeclareMathOperator\Stab{Stab}

\newcommand{\fker}[6]{
\mathbb{F}_{#1 #2}{{\arraycolsep=0.2\arraycolsep\ensuremath{\begin{bmatrix}#4 & #3 \\ #5 & #6\end{bmatrix}}}}
}
\newcommand{\sker}[3]{
\mathbb{S}_{#1 #2}[#3]
}

\newcommand{\skerhat}[3]{
\widehat{\mathbb{S}}_{#1 #2}[#3]
}

\newcommand{\sixj}[6]{
\begin{Bmatrix} #1 & #2 & #3 \\ #4 & #5 & #6 \end{Bmatrix}
}

\title{A universal sum over topologies in 3d gravity}
\author[1,2]{Alexandre Belin}\emailAdd{alexandre.belin@unimib.it}
\author[3,4]{\!\!, Scott Collier}\emailAdd{scolli32@syr.edu}
\author[5]{\!\!, Lorenz Eberhardt}\emailAdd{l.eberhardt@uva.nl}
\author[6,7]{\!\!, Diego Liska}\emailAdd{diego.liska@epfl.ch}
\author[8]{\!\!, Boris Post} \emailAdd{boris.post@maths.ox.ac.uk}
\affiliation[1]{Dipartimento di Fisica, Universit\`a di Milano-Bicocca
I-20126 Milano, Italy}
\affiliation[2]{INFN, sezione di Milano-Bicocca, I-20126 Milano, Italy}
\affiliation[3]{Department of Physics, Syracuse University, Syracuse, NY, 13244, USA}
\affiliation[4]{Institute for Quantum \& Information Sciences, Syracuse University, Syracuse, NY, 13244, USA}
\affiliation[5]{Institute for Theoretical Physics,
University of Amsterdam, Amsterdam, 1098XH, NL}
\affiliation[6]{Department of Theoretical Physics, University of Geneva,
1211 Geneva, Switzerland}
\affiliation[7]{Laboratory for Theoretical Fundamental Physics, Institute of Physics,
Ecole Polytechnique F\'ed\'erale de Lausanne, 1015 Lausanne, Switzerland}
\affiliation[8]{Mathematical Institute, University of Oxford, Woodstock Road, Oxford, OX2 6GG, United Kingdom}
\abstract{
We explore the sum over topologies in $\mathrm{AdS}_3$ quantum gravity and its relationship with the statistical interpretation of the boundary theory. We formulate a statistical version of the conformal bootstrap that systematizes the universal statistical properties of high-energy CFT$_2$ data. We identify a series of surgery moves on bulk manifolds that precisely reflect the requirements of typicality and crossing symmetry of the boundary ensemble. These surgery moves generate a large number of bulk manifolds that have to be included in any reasonable definition of the gravitational path integral. We show that this procedure generates only on-shell (hyperbolic) manifolds, although it does not produce all of them. These proofs rely on structure theorems of 3-manifolds, which non-trivially interact with the requirements of the statistical boundary ensemble. We illustrate the application of this procedure with many examples, such as Euclidean wormholes, twisted $I$-bundles and handlebody-knots. Our findings reveal a large space of possible choices of which manifolds can be included in the gravitational path integral, reflecting a wide range of possible statistical ensembles consistent with crossing symmetry and typicality. 
}

\begin{document}

\setcounter{tocdepth}{3}
\maketitle
\setcounter{page}{3}
\pagebreak

%make math in all titles bold
\makeatletter
\g@addto@macro\bfseries{\boldmath}
\makeatother
%end code

%%%%%%%%%%%%%%%%%%%%%%%%%%%%%%%%%%%%
\section{Introduction}\label{sec:intro}
%%%%%%%%%%%%%%%%%%%%%%%%%%%%%%%%%%%%

The gravitational path integral continues to be an enigmatic yet essential tool in understanding aspects of holography. Low-dimensional models have served as valuable laboratories for uncovering some of its more intricate features. From these models, much evidence has been gathered in favour of the idea that the path integral for pure Einstein gravity can be understood as performing a \emph{statistical} average over the data that define the dual conformal field theory. Developing this idea in three-dimensional gravity has become an active direction of current research, but our understanding of the statistical holographic dictionary is still incomplete. 

Thanks to the absence of local bulk degrees of freedom \cite{Witten:1988hc}, pure AdS$_3$ quantum gravity is a particularly suitable playground to explore the statistical nature of the gravitational path integral, and significant progress has been made over the past few years \cite{Belin:2020hea,Chandra:2022bqq,Cotler:2020ugk,Mertens:2022ujr,DiUbaldo:2023qli,Belin:2023efa,Jafferis:2025vyp,Schlenker:2022dyo,deBoer:2023vsm,Collier:2023fwi, Collier:2024mgv,deBoer:2024mqg,deBoer:2025rct,Hartman:2025ula,Hartman:2025cyj,Chandra:2025fef,Yan:2023rjh,Chandra:2024vhm}. As a starting point, the main entry in the dictionary between pure 3-dimensional gravity and the statistical ensemble of CFT$_2$ data  posits that averaged boundary quantities are computed as a sum over topologies in the bulk:  
\begin{equation}\label{eq:sumovertopologies}
    \overline{Z[\Sigma]} = \sum_{\partial M = \Sigma} Z_\text{grav}[M]\,.
\end{equation}
On the right-hand side, $Z_\text{grav}$ denotes the gravitational path integral for pure Einstein gravity with a negative cosmological constant, on a fixed bulk topology $M$.\footnote{For notational clarity, we suppress the dependence on the moduli of the boundary Riemann surface and use $\Sigma$ as a placeholder for all the boundary data.} The sum runs over all distinct bulk 3-manifolds whose asymptotic boundary is $\Sigma$. If $\Sigma$ is a connected Riemann surface of genus $g$, the left-hand side computes the average genus $g$ partition function, in some yet to be specified statistical model of CFT$_2$ data. If $\Sigma$  consists of multiple disconnected components, then the right-hand side includes Euclidean wormhole topologies and the left-hand side computes a higher statistical moment of CFT$_2$ partition functions. In recent years, powerful new tools have been developed to compute $Z_\text{grav}[M]$ exactly, i.e.~to all orders in $G_N$, for a large class of 3-manifolds \cite{Collier:2023fwi,Collier:2024mgv}. Nevertheless, 3-manifold topology is a rich and complex subject, which means that understanding the full sum over topologies in \eqref{eq:sumovertopologies} is a difficult (and unresolved) problem.

In this paper, our principal aim is to systematize the correspondence between universal statistics of 2d conformal field theory and pure 3d quantum gravity. In particular, we want to understand what role the sum over topologies plays in the pure gravity/ensemble duality, and explain how the bulk sum over topologies arises from the \emph{boundary} perspective. We will explore how, and to what extent, the 3-manifolds contributing to \eqref{eq:sumovertopologies} are organized to furnish a consistent statistical ensemble of CFT$_2$ data, compatible with kinematic constraints, crossing symmetry, modular invariance and typicality at high energies.

\begin{figure}
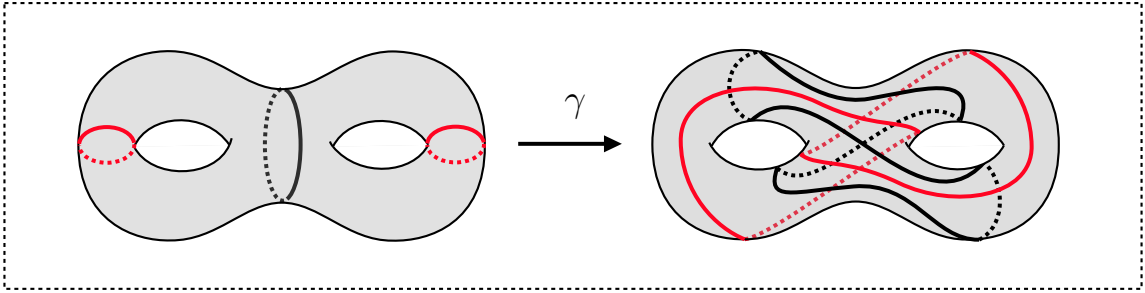

    \centering
    \tikzset{every picture/.style={line width=0.8pt}} %set default line width to 0.75pt 

    \caption{Example of a crossing transformation $\gamma$ on a genus-$2$ Riemann surface. The two sets of contractible cycles (the thick closed curves) lead to two distinct genus-$2$ handlebodies upon filling in the bulk. 
    }
    \label{fig:crossing}
\end{figure}

\paragraph{Handlebodies and crossing symmetry.} To illustrate the question that we want to ask, suppose that the asymptotic boundary $\Sigma$ is a Riemann surface of genus $g\geq 1$, and fix the boundary moduli $\Omega_i,\bar\Omega_i$. 
Then a natural class of 3-manifolds contributing to the sum over topologies consists of the genus-$g$ \emph{handlebodies}:\footnote{Here and in the rest of the paper, we use the symbol $\supset$ to denote that the right-hand side is a \emph{sub-sum} of the full sum over topologies.}
\begin{equation}\label{eq:handlebodysum}
        \overline{Z[\Sigma_{g}]} \quad \supset \sum_{\gamma \in \text{MCG}(\Sigma_{g})/\mathcal{H}_g} \!Z_\text{grav}[M^\gamma]\,.
    \end{equation}
Topologically, a genus-$g$ handlebody is homeomorphic to a `filled-in' genus-$g$ surface. Geometrically, it admits an infinite-volume hyperbolic metric after choosing a collection of $3g-3$ disjoint cycles on the boundary $\Sigma_g$ that become contractible in the bulk (a choice of identity channel, in CFT parlance). Different choices of cycles are related to each other by crossing transformations $\gamma$, which are elements of the mapping class group of $\Sigma_g$.\footnote{For an introduction to crossing transformations, see \cite{Eberhardt:2023mrq}.} An example of such a mapping class group transformation is shown in figure \ref{fig:crossing}.
The sum over handlebodies can hence be written as a sum over crossing transformations, modulo the bulk mapping class group of the handlebody $\mathcal{H}_g$ (the so-called handlebody group).

Note that in the $g=1$ case, the handlebody sum \eqref{eq:handlebodysum} was studied by Maloney and Witten \cite{Maloney:2007ud} and subsequently by Keller and Maloney  \cite{Keller:2014xba}. The resulting sum over solid tori takes the form of a Poincar\'e series, with $\text{MCG}(\Sigma_{1}) = \PSL(2,\mathbb{Z})$ and $\mathcal{H}_1 = \mathbb{Z}$. Each $Z_\text{grav}[M^\gamma]$ evaluates to a product of left- and right moving Virasoro vacuum characters, $|\chi_{\mathds{1}}(\gamma\cdot\tau)|^2$, and hence the role of the sum over topologies is quite clear: it renders the average partition function modular invariant. Restricting the sum to a smaller set of bulk saddles would violate modular invariance of the average.\footnote{A \emph{larger} set of bulk manifolds may be included, as long as it does not spoil modular invariance. In the genus-1 case, such additional contributions are conjectured to be necessary in order to cure the negativity in the spectral density in the near-extremal limit \cite{Benjamin:2019stq,Ghosh:2019rcj, Maxfield:2020ale,Benjamin:2020mfz,DiUbaldo:2023hkc}.}

In the higher-genus case, $g>1$, the gravitational path integral on a handlebody evaluates to the product of left- and right-moving vacuum Virasoro conformal blocks in the channel specified by $\gamma$ \cite{Collier:2023fwi}. Each vacuum block resums all contributions from descendant states, which are interpreted in 3d gravity as boundary gravitons \cite{Giombi:2008vd}. Similar to the genus-1 case, the sum over handlebodies renders the average genus-$g$ CFT partition function crossing symmetric (invariant under the action of the mapping class group). 

The sum over handlebodies has an interesting implication for the statistics of OPE coefficients in our model. In the original statistical model of \cite{Belin:2020hea, Chandra:2022bqq}, a Gaussian ansatz was used to describe the statistics of heavy OPE coefficients. This simplified description captures many important features of gravity, but it generates only a finite subset of handlebodies (for the genus-$2$ computation, see section 8.3 of \cite{Chandra:2022bqq}). Hence, the average is only invariant under a finite subset of all crossing transformations. Instead, insisting on a fully crossing symmetric average $\overline{Z[\Sigma_g]}$ implies the existence of \emph{non-Gaussianities} in the statistical description of the CFT$_2$ data \cite{Belin:2023efa, Belin:2021ryy,Anous:2021caj,Collier:2023fwi,Collier:2024mgv,deBoer:2024mqg, deBoer:2025rct, Hartman:2025ula,Hartman:2025cyj,deBoer:2023vsm,Chandra:2025fef}.
 In section \ref{sec:3}, we will show how to systematically include non-Gaussian OPE statistics in order to reproduce the full (infinite) sum over handlebodies, for all genera $g\geq 1$. One useful consequence of this systematic approach is that the spins $J_i = h_i-\bar h_i$ of all propagating heavy operators are appropriately quantized.

    \paragraph{Why include non-handlebodies?} Now we come to the central question of this paper. In three dimensions, there are many single-boundary topologies that are not handlebodies: the classic example of such a non-handlebody (in the 3d gravity literature) is the twisted $I$-bundle of \cite{Yin:2007at}. However, since the handlebody sum in equation \eqref{eq:handlebodysum} already furnishes a crossing symmetric average (for any choice of connected genus-$g$ boundary), one could ask why non-handlebodies are needed at all, from the perspective of the boundary CFT. Said differently, is there a redundancy in the possible consistent statistical ensembles dual to pure 3d gravity? 
    
    This question is especially compelling for a connected boundary of genus $g\geq 2$. In the $g=1$ case, all non-handlebodies are off-shell \cite{Maloney:2007ud}, so there is no semiclassical saddlepoint approximation for $Z_\text{grav}$ on such topologies. These include single-boundary Seifert manifolds \cite{Maxfield:2020ale,deBoer:2025rct}, for which it is currently not known how to  compute the gravitational path integral from first principles. By contrast, for $g\geq 2$ there are many non-handlebodies that support well-defined \emph{on-shell} solutions of pure Einstein gravity with the correct asymptotic boundary behaviour, so there is no a priori reason to discard their contributions. Moreover, they can be added in a manifestly crossing symmetric way, by summing over mapping class group images of a given `seed' non-handlebody, as will be shown in section \ref{sec:minimal completion}. The question is: does the statistical theory of CFT$_2$ data require these saddles for internal consistency, or are they optional? We view this question as a statistical version of the conformal bootstrap, applied to 3d gravity.

    Let us emphasize that the above question is distinct from previous work on Euclidean wormholes and the factorization puzzle \cite{Belin:2020hea,Chandra:2022bqq,Maldacena:2004rf,Marolf:2021kjc,Saad:2019lba,Marolf:2020xie,Stanford:2020wkf}. When the boundary $\Sigma$ has multiple disconnected components, there are well-known wormhole topologies connecting them (the simplest example being the Maldacena-Maoz solution \cite{Maldacena:2004rf}), and these are always non-handlebodies. Moreover, it is by now well understood that these Euclidean wormholes probe higher statistical moments (such as the variance) in the ensemble of OPE coefficients \cite{Belin:2020hea,Chandra:2022bqq, Collier:2024mgv, deBoer:2024mqg}. So, in the multi-boundary case, non-handlebodies are clearly needed for the ensemble to have non-zero fluctuations around the average. 
    On the other hand, the \emph{single}-boundary non-handlebodies that we address in this paper have so far remained elusive.

\begin{figure}
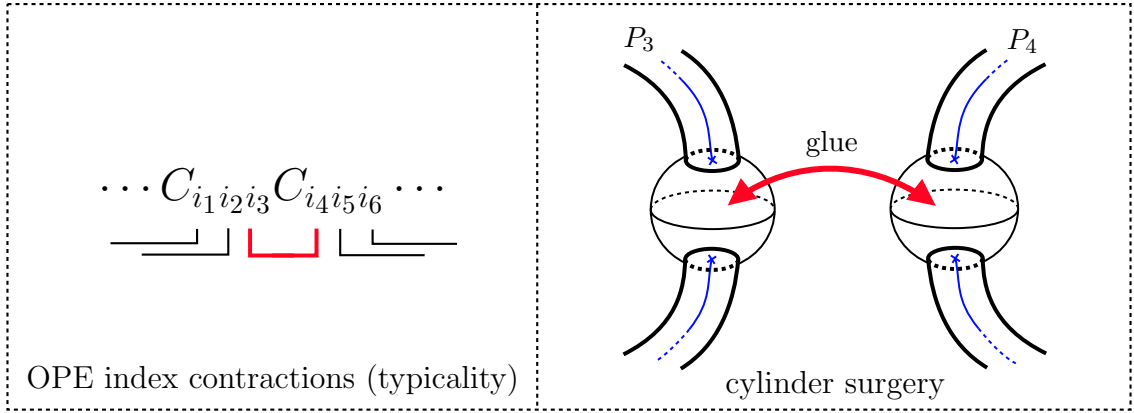

    \centering
    \tikzset{every picture/.style={line width=0.8pt}} %set default line width to 0.75pt
    \tikzset{
pattern size/.store in=\mcSize, 
pattern size = 5pt,
pattern thickness/.store in=\mcThickness, 
pattern thickness = 0.3pt,
pattern radius/.store in=\mcRadius, 
pattern radius = 1pt}
\makeatletter
\pgfutil@ifundefined{pgf@pattern@name@_7jrft2iun}{
\pgfdeclarepatternformonly[\mcThickness,\mcSize]{_7jrft2iun}
{\pgfqpoint{0pt}{0pt}}
{\pgfpoint{\mcSize+\mcThickness}{\mcSize+\mcThickness}}
{\pgfpoint{\mcSize}{\mcSize}}
{
\pgfsetcolor{\tikz@pattern@color}
\pgfsetlinewidth{\mcThickness}
\pgfpathmoveto{\pgfqpoint{0pt}{0pt}}
\pgfpathlineto{\pgfpoint{\mcSize+\mcThickness}{\mcSize+\mcThickness}}
\pgfusepath{stroke}
}}
\makeatother

% Pattern Info
 
\tikzset{
pattern size/.store in=\mcSize, 
pattern size = 5pt,
pattern thickness/.store in=\mcThickness, 
pattern thickness = 0.3pt,
pattern radius/.store in=\mcRadius, 
pattern radius = 1pt}
\makeatletter
\pgfutil@ifundefined{pgf@pattern@name@_expa5zgm6}{
\pgfdeclarepatternformonly[\mcThickness,\mcSize]{_expa5zgm6}
{\pgfqpoint{0pt}{0pt}}
{\pgfpoint{\mcSize+\mcThickness}{\mcSize+\mcThickness}}
{\pgfpoint{\mcSize}{\mcSize}}
{
\pgfsetcolor{\tikz@pattern@color}
\pgfsetlinewidth{\mcThickness}
\pgfpathmoveto{\pgfqpoint{0pt}{0pt}}
\pgfpathlineto{\pgfpoint{\mcSize+\mcThickness}{\mcSize+\mcThickness}}
\pgfusepath{stroke}
}}
\makeatother

% Pattern Info
 
\tikzset{
pattern size/.store in=\mcSize, 
pattern size = 5pt,
pattern thickness/.store in=\mcThickness, 
pattern thickness = 0.3pt,
pattern radius/.store in=\mcRadius, 
pattern radius = 1pt}
\makeatletter
\pgfutil@ifundefined{pgf@pattern@name@_g54laaovo}{
\pgfdeclarepatternformonly[\mcThickness,\mcSize]{_g54laaovo}
{\pgfqpoint{0pt}{0pt}}
{\pgfpoint{\mcSize+\mcThickness}{\mcSize+\mcThickness}}
{\pgfpoint{\mcSize}{\mcSize}}
{
\pgfsetcolor{\tikz@pattern@color}
\pgfsetlinewidth{\mcThickness}
\pgfpathmoveto{\pgfqpoint{0pt}{0pt}}
\pgfpathlineto{\pgfpoint{\mcSize+\mcThickness}{\mcSize+\mcThickness}}
\pgfusepath{stroke}
}}
\makeatother

% Pattern Info
 
\tikzset{
pattern size/.store in=\mcSize, 
pattern size = 5pt,
pattern thickness/.store in=\mcThickness, 
pattern thickness = 0.3pt,
pattern radius/.store in=\mcRadius, 
pattern radius = 1pt}
\makeatletter
\pgfutil@ifundefined{pgf@pattern@name@_cc1njbb37}{
\pgfdeclarepatternformonly[\mcThickness,\mcSize]{_cc1njbb37}
{\pgfqpoint{0pt}{0pt}}
{\pgfpoint{\mcSize+\mcThickness}{\mcSize+\mcThickness}}
{\pgfpoint{\mcSize}{\mcSize}}
{
\pgfsetcolor{\tikz@pattern@color}
\pgfsetlinewidth{\mcThickness}
\pgfpathmoveto{\pgfqpoint{0pt}{0pt}}
\pgfpathlineto{\pgfpoint{\mcSize+\mcThickness}{\mcSize+\mcThickness}}
\pgfusepath{stroke}
}}
\makeatother

    \caption{In section \ref{sec:minimal completion}, we show that index contractions, in a statistical ensemble of heavy OPE coefficients, correspond to 3-manifold surgery where the gluing surface is a cylinder. Combining this surgery with crossing transformations generates many non-handlebodies.}
    \label{fig:machine1}
\end{figure}
\paragraph{Non-handlebodies from statistics.} In section \ref{sec:minimal completion}, we demonstrate that including non-handlebodies in the sum over topologies becomes necessary when, in addition to crossing symmetry, we require the ensemble of CFT$_2$ data to satisfy \emph{typicality}. This concept is familiar from the Eigenstate Thermalization Hypothesis \cite{Deutsch1991,Srednicki1994,Foini:2018sdb,Pappalardi:2022aaz,Jafferis:2022uhu} and can be generalized to chaotic 2d CFTs \cite{Belin:2021ryy, Belin:2021ibv, deBoer:2024mqg}, where it amounts to approximate local rotational invariance in microcanonical windows around primary states of high conformal dimension. Importantly, typicality predicts the structure of index contractions of random OPE coefficients, which in the bulk translates to precise \emph{surgery} operations on 3-manifolds, as illustrated in figure \ref{fig:machine1}. We show that iteratively applying these surgery moves and combining them with crossing transformations generates a large class of non-handlebody topologies. 

We will refer to this inductive procedure for generating topologies from properties of the statistical ensemble of CFT$_2$ data as the ``gravitational machine''. This machine takes as input any handlebody (or multiple handlebodies) and repeatedly acts on it with three basic surgery operations, which will be specified in section \ref{subsec:general gravitational machine}. The resulting set of 3-manifold topologies should be seen as the \emph{minimal} completion of the handlebody sum that is compatible with typicality, crossing symmetry and consistency between higher and lower statistical moments.\footnote{However, this is not the end of the story: in this paper, we focus on OPE statistics and ignore spectral correlations. The latter are expected to arise from off-shell topologies \cite{Cotler:2020ugk,Maxfield:2020ale, DiUbaldo:2023qli, Haehl:2023tkr,deBoer:2025rct,Boruch:2025ilr} that further correct the dictionary \eqref{eq:sumovertopologies}. We will have more to say on this in the \hyperref[sec:discussion]{discussion}.}

\begin{figure}
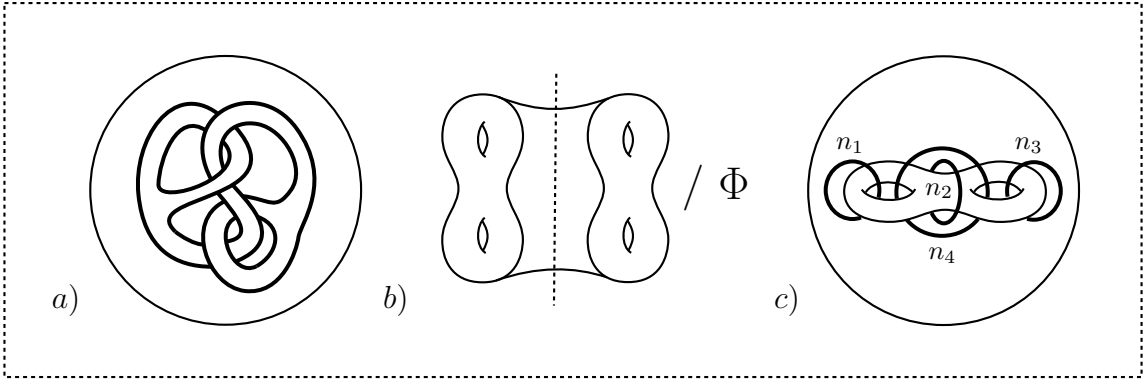

    \centering
    \tikzset{every picture/.style={line width=0.8pt}} %set default line width to 0.75pt 

    \caption{Examples of non-handlebodies with genus-2 boundary. These non-handlebodies are all necessary in the sum over topologies, in order for the statistical description of the CFT to be consistent. $a)$ Complement of a handlebody-knot in $\mathrm{S}^3$. $b)$ Twisted $I$-bundle. $c)$ Dehn surgery on the complement of a handlebody embedded in $\mathrm{S}^3$. }
    \label{fig:nonhandlebodies}
\end{figure}
The main results of this paper are to study and rigorously prove certain appealing properties of the gravitational machine, which we summarize here:
\begin{itemize}
    \item \textbf{The machine generates non-handlebodies.} We will explicitly show that the examples illustrated in figure \ref{fig:nonhandlebodies} are generated by the machine. Moreover, we show that they are indeed non-handlebodies by analyzing their fundamental groups. As a simple class of examples of non-handlebodies, we similarly analyze all genus-2 handlebody-knots with fewer than 6 crossings (see table \ref{table:handlebodyknots}). More generally, the machine applies for arbitrary genus $g\geq 2$ of the boundary manifold and also works for disconnected boundaries. 
    \item \textbf{The machine is hyperbolic.} In section \ref{subsec:machine is hyperbolic}, we prove that starting from any hyperbolic 3-manifold, the action of the gravitational machine only produces 3-manifolds that are also hyperbolic. The proof follows from studying how the fundamental group changes under the operations of the machine, making use of Thurston's hyperbolization theorem \cite{Thurston:1982} and geometric group theory. This key result makes precise the intuition that OPE statistics in the boundary description only requires on-shell saddles in the bulk. 
    \item \textbf{But not all hyperbolic non-handlebodies are generated.} This is perhaps the most interesting outcome of our analysis: in section \ref{subsec:what 3-manifolds are generated}, we show that all manifolds generated by the machine are so-called \emph{cylindrical} 3-manifolds, which contain embedded essential cylinders. However, there are many topologies that are not cylindrical; as an example of a hyperbolic, acylindrical manifold with genus-2 boundary, we study Thurston's so-called `tripus' manifold \cite{thurston_book}. We will discuss the implications of this fact for the sum over topologies below.
\end{itemize}

Throughout the paper, we demonstrate that the above results, which are mostly about topology rather than gravity, can be implemented directly on the level of the gravitational path integral by using Virasoro TQFT \cite{Collier:2023fwi}. This formalism gives the exact quantum gravity partition (i.e.~at finite $c$), topology by topology, so we do not rely on semiclassical arguments. Virasoro TQFT also allows us to straightforwardly extend our results to 3-manifolds with conical defects: for example, when $\Sigma = \Sigma_{0,4}$ is the 4-punctured sphere, the analogue of the handlebody sum is the `conformal block Farey tale' of \cite{Maloney:2016kee} and the analogue of a non-handlebody is a non-rational tangle.

\paragraph{Is 3d gravity non-minimal?}
Let us end this introduction by contemplating what the above results imply for 3d gravity and the sum over topologies. Although we have managed to find a boundary interpretation for a large class of non-handlebodies, there are still infinitely many on-shell manifolds not generated by our gravitational machine. The main source of such counterexamples are the \emph{acylindrical} 3-manifolds described in the third point above, which can be characterized by admitting a finite-volume hyperbolic metric with totally geodesic boundary \cite{Thurston1986HyperbolicI}. Apparently, these topologies are not strictly necessary in order for the statistical ensemble to obey crossing and typicality, leaving the intriguing possibility that the full sum over topologies in 3d gravity is not a minimal solution of the statistical bootstrap. 

In fact, it seems perfectly consistent to take any on-shell and acylindrical `seed', consider its orbit under the action of the machine, and decide to add this orbit to the sum over topologies or not. This gives an infinite number of choices, signaling a large redundancy in the possible consistent statistical descriptions for 3d gravity. This seems contrary to our expectation that pure 3d gravity would, in some sense, be the simplest theory satisfying the statistical bootstrap.

We could turn this question around and instead ask: What are the minimal properties imposed on a statistical ensemble of CFT$_2$ data that uniquely select 3d gravity, including the full sum over topologies? It is logically possible that adding conditions \emph{beyond} those we have imposed (such as a version of the moment problem \cite{Janssen:2021stl}, the existence of a matrix-tensor model \cite{Belin:2023efa, Jafferis:2025vyp}, or the presence of random matrix universality in the CFT spectrum \cite{Perlmutter:2025ngj}) requires more topologies for these additional conditions to be satisfied. We come back to this important open question in the discussion.

\paragraph{Plan of the paper.}
In section \ref{sec:2}, we will define the conditions that any statistical theory of CFT$_2$ data dual to pure 3d gravity must satisfy. Then, in section \ref{sec:3} we show that the condition of crossing symmetry (at all genera) is solved by summing over handlebodies, explaining in particular the relation to non-Gaussianities in the statistics of OPE coefficients. Although this is mainly a review, it sets up a lot of the formalism that will be needed in the following sections. In section \ref{sec:minimal completion}, we show how the condition of typicality, combined with crossing and consistency conditions between higher and lower moments, leads to non-handlebodies in the bulk. The gravitational machine is introduced as a way to systematize these non-handlebody contributions, and its main properties are proven. In section \ref{sec:5}, we apply the machine to produce a variety of examples of non-handlebodies, and compute their partition functions. We end with a discussion and an outlook for future work. Finally, in the appendices we have collected many technical results about 3-manifolds, their fundamental groups, and Virasoro TQFT that are needed in the main text.

%%%%%%%%%%%%%%%%%%%%%%%%%%%%%%%%%%%%
\section{The statistical bootstrap}
\label{sec:2}

The goal of this section is to define a statistical theory that models compact irrational CFT$_2$'s with a pure Einstein gravity dual at large central charge $c$ \cite{Hartman:2014oaa}. Such CFT$_2$'s have no extended chiral algebra beyond Virasoro symmetry, so we will organize the spectrum of $(L_0,\bar L_0)$ into Virasoro primaries and descendants. The descendant states are fixed by symmetry, so they should not be treated as random variables. On the other hand, the conformal primaries will be treated statistically. 

Let us imagine there exists a probability distribution $\mu(\{h_i, \bar h_i, C_{ijk}\})$ that we can use as a statistical model for the dynamical CFT$_2$ data, which consist of the infinite list of primary conformal dimensions $\Delta_i = h_i+\bar h_i$, integer spins $J_i = h_i-\bar h_i$ and primary OPE coefficients $C_{ijk} \equiv C_{h_ih_jh_k}$. We package the spectral data in terms of the density of primary states,
\begin{equation}
    \rho(h,\bar h) = \sum_i \delta(h_i-h)\delta(\bar h_i - \bar h),
\end{equation}
where the sum is over the primary spectrum,
with a similar formula for the density of OPE coefficients (as in \cite{Collier:2019weq}).
The average of any function of OPE coefficients and conformal weights is then defined by integrating it against the measure $\mu$:\vspace{2mm}
\begin{equation}
    \overline{F(\{h_i,\bar h_i,C_{ijk}\})} = \int \d L_0 \d \bar L_0 \d C \,\mu(\{h_i, \bar h_i, C_{ijk}\})F(\{h_i,\bar h_i,C_{ijk}\}).\vspace{2mm}
\end{equation}
Here $\d C$ is a volume element on the space of rank-3 tensors.

Crucially, we will impose a number of conditions on the moments of this probability distribution $\mu$, such that the statistical model is a good approximation of a holographic CFT$_2$ in the large-$c$ limit.\footnote{In particular, we do not assume that the average is taken over an ensemble of `true' CFT's. For an exploration of probability distributions on the space of 2d CFT's, see the recent work \cite{Belin:2025qjm}. Also, for a more in-depth discussion about the notion of an `approximate CFT' we refer the reader to section 2 of \cite{Belin:2023efa}.}  The goal of this section is to specify these conditions, which we will refer to as the ``statistical bootstrap''. We will mostly assume that the spectrum of $(L_0,\bar L_0)$ has a twist gap all the way up to the black hole threshold, meaning that $h_i,\bar h_i$ is either zero (the vacuum, denoted by $\mathds{1}$) or bigger than $\frac{c-1}{24}$. However, the material presented in this section can also be extended to incorporate states below the threshold in the conical defect regime \cite{Benjamin:2020mfz}, as will be discussed in section \ref{sec:3}.

Let us emphasize that although we take the conditions considered in this section to be \emph{necessary}, they need not be sufficient: we have not exhausted all constraints on the distribution $\mu$ that one may want to impose. In the discussion, we comment on the possibility of imposing additional constraints on the moments, that arise (for example) from positivity of $\mu$ and the Stieltjes/Hamburger moment problem. 
 
\subsection{Kinematic constraints}
First, we state the constraints that we will impose on each member of the ensemble---that is, not just on average. For each $i$, we demand the spin $J_i$ to be an integer, as is appropriate for a bosonic CFT$_2$. Note that although this is a natural property to demand, it was not imposed in the original ensemble of \cite{Chandra:2022bqq}. In our case, spin quantization will follow as a special case of the stronger constraint of crossing symmetry and modular invariance that we impose in section \ref{sec:crossing}. 

Next, the OPE coefficients must satisfy the following properties under complex conjugation and permutation of the indices (here $\sigma \in S_3$): 
\begin{equation}\label{eq:OPEpermutation}
    C_{ijk}^* = C_{kji}, \quad C_{\sigma(i)\sigma(j)\sigma(k)} = \text{sgn}(\sigma)^{J_i + J_j + J_k}C_{ijk}\,.
\end{equation}
Combining these constraints implies that $C_{ijk}$ is purely real when the sum of spins is even, and purely imaginary when odd. When two operators coincide, the second constraint also implies that $C_{jjk} = (-1)^{J_k}C_{jjk}$, hence the OPE coefficient $C_{jjk}$ can only be nonzero if $J_k$ is even (and moreover $C_{jjk}$ is real). Lastly, we impose the selection rule when one of the operators is the identity, $
    C_{ij\mathds{1}} = \delta_{ij},$
where $\delta_{ij}$ is the Kronecker delta.

To account for the structure of descendant states, we can expand any genus-$g$ partition function into Virasoro conformal blocks \cite{Belavin:1984vu}. Loosely speaking, the blocks resum all contributions to the partition function from descendants, so that the remaining sums are only over primary states and OPE coefficients. More precisely, the conformal block decomposition requires us to pick a channel $\mathcal{C}$, so that\vspace{1mm}
\begin{equation}\label{eq:block_decomposition}
    \overline{Z[\Sigma_g;\Omega,\bar\Omega]} = \int \d^{3g-3} h \,\d^{3g-3} \bar h\,\,\overline{\prod_{e_i\in \Gamma_\mathcal{C}}\!\!\rho(h_i,\bar h_i)\!\prod_{v_{ijk}\in \Gamma_\mathcal{C}}\!\!\!C_{ijk}} \,\,\mathcal{F}_g^{\,\mathcal{C}}(\bm{h};\Omega) \bar{\mathcal{F}}_g^{\,\mathcal{C}}(\bar{\bm{h}};\bar\Omega).
\end{equation}
Let us explain the notation used. $\mathcal{F}_g^{\,\mathcal{C}}$ denotes the genus-$g$ Virasoro conformal block in the channel $\mathcal{C}$. The choice of channel involves a pair-of-pants decomposition of the surface $\Sigma_g$, which gives a set of $3g-3$ moduli, to which we collectively refer as $\Omega$. Similarly, the right-moving conformal block $\bar{\mathcal{F}}_g^{\,\mathcal{C}}$ depends on $3g-3$ independent right-moving moduli $\bar\Omega$. The pair-of-pants decomposition defines a dual trivalent graph $\Gamma_{\mathcal{C}}$, whose edges are labeled by the conformal weights $\bm{h} = (h_1,\dots, h_{3g-3})$ and $\bar{\bm{h}} = (\bar h_1,\dots, \bar h_{3g-3})$, and whose vertices are labeled by the OPE coefficients $C_{ijk}$. This graph specifies the products appearing in the average \eqref{eq:block_decomposition}: each edge $e_i$ contributes a factor of the density of states and each vertex $v_{ijk}$ a factor of $C_{ijk}$, where the pattern of OPE indices is determined by the topology of the graph $\Gamma_{\mathcal{C}}$.

In equation \eqref{eq:block_decomposition}, we used linearity to write the average of $Z[\Sigma_g]$ in terms of the average of the primary data encoded in the $\rho(h_i,\bar h_i)$'s and the OPE coefficients. Recall that this average is defined with respect to the measure $\mu(\{h_i,\bar h_i,C_{ijk}\})$. After averaging, we expect the moment $\overline{\prod \rho \prod C}$ to be a smooth function of the conformal weights $\bm{h},\bar{\bm{h}}$, at least for the states above the black hole threshold---more generally, it is a distribution, with discrete support on the light spectrum and continuous support on the heavy spectrum.
In section \ref{sec:3}, we will see how these moments are related to gravitational partition functions in the bulk.

As an example of the above formalism, let us draw the `sunset' and `dumbbell' channels for the genus-2 partition function, with the corresponding index structures:
\begin{equation}\label{eq:genus2graphs}
     \vcenter{\hbox{

    }}
\rightarrow \, C_{iij}C_{jkk}.
\end{equation}
We read the cyclic ordering of the indices of the structure constants in a counter-clockwise sense of the three outgoing lines at each vertex.
There are of course also the same diagrams where some of the indices are permuted.
At genus 3, we have the following inequivalent graph topologies:
\be 
\vcenter{\hbox{%
}}\  \label{eq:genus3graphs}
\ee
where the first diagram corresponds to the index structure $C_{112} C_{234} C_{345} C_{566}$, the second to $C_{123}C_{124} C_{345}C_{566}$, and so forth.
In general, at genus $g$ we get $2g-2$ OPE coefficients, giving $(3(2g-2)-1)!!$ pairwise index contractions (many of which are related by graph automorphisms).

Note that these are all diagrams that can be drawn on the level of index structures of the OPE coefficients. However, the conformal blocks encode more data than just the graph topology: they are fully specified by the pair-of-pants decomposition and the braiding phases at each vertex of the dual graph (see the figures in \cite{Eberhardt:2023mrq} for the complete graphical representations of the blocks). There are an infinite number of such channels (where the trivalent graphs can be knotted), which are obtained from the basic graphs in \eqref{eq:genus2graphs} and \eqref{eq:genus3graphs} by acting with crossing transformations, also called Moore-Seiberg moves \cite{HatcherThurston1980, LegoTeichmuller,Moore:1988qv}. These crossing transformations will be the topic of the next subsection.

\subsection{Crossing symmetry and modular invariance}\label{sec:crossing}
\begin{figure}
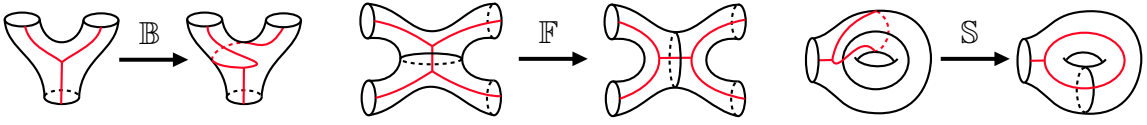

    \centering
    \tikzset{every picture/.style={line width=0.8pt}} 

    \caption{The three basic Moore-Seiberg moves, acting on embedded bordered surfaces in a given pair-of-pants decomposition of a Riemann surface. From left to right, these are the braiding transformation, the fusion move and the modular S-transform.}
    \label{fig:MS_moves}
\end{figure}

The next set of consistency conditions that we impose on the ensemble are modular invariance of the torus partition function and crossing symmetry of all higher-genus partition functions. We will impose these constraints on the average, instead of on each member of the ensemble. That is, we demand that
\begin{equation}\label{eq:crossinginvariance}
    \overline{Z\!\left[\Sigma_g;\gamma\!\cdot \Omega,\gamma\!\cdot\bar\Omega\right]} = \overline{Z\!\left[\Sigma_g; \Omega, \bar\Omega\right]} \quad \text{for all }\gamma\in \text{MCG}(\Sigma_g) \text{ and } g\geq 1.
\end{equation}
Since we are not taking an average over an ensemble of exact CFT's, this is a non-trivial constraint on the measure $\mu(\{h_i,\bar h_i,C_{ijk}\})$. 
 When $g=1$, $\gamma$ is an element of the modular group $\text{MCG}(\Sigma_1) \cong \text{PSL}(2,\mathbb{Z})$, which acts on the torus moduli $\tau,\bar\tau$ by the usual fractional linear transformations. For $g\geq 1$, $\gamma$ is a mapping class group transformation, which can be represented as a composition of the basic generating moves of the Moore-Seiberg groupoid $\text{MS}_g$ shown in figure \ref{fig:MS_moves}. These moves are allowed to act on all subsurfaces $\Sigma_{0,3},\Sigma_{0,4}$ and $\Sigma_{1,1}$ that are embedded in $\Sigma_g$.\footnote{The Moore-Seiberg groupoid can be used to represent the mapping class group $\text{MCG}(\Sigma_g) = \text{Diff}^+(\Sigma_g)/\text{Diff}_0(\Sigma_g)$ of large diffeomorphisms of $\Sigma_g$. This can be seen by composing two braids $\mathbb{B}$ to obtain a Dehn twist \cite{Eberhardt:2023mrq} and using that the mapping class group of $\Sigma_g$ is generated by Dehn twists along $2g+1$ standard $a$- and $b$-cycles (the so-called Humphries generators \cite{Lickorish_1964,Humphries}). Each of these cycles can be reached by acting with $\mathbb{F}$ and $\mathbb{S}$ on a given pair-of-pants decomposition.}

As an illustration, consider again the genus-2 crossing transformation shown in figure \ref{fig:crossing} from the introduction. To see how it can be obtained by composing the basic Moore-Seiberg moves, we make the following sequence of crossing transformations:
\begin{equation}\label{eq:sequence1}
    \tikzset{every picture/.style={line width=1.2pt}} %set default line width to 0.75pt        

\end{equation}
Here we only drew the dual graph, which can be drawn on a genus-2 surface in such a way that each edge intersects a cycle in the pair-of-pants decomposition once. In particular, the pair-of-pants decomposition on the right of figure \ref{fig:crossing} corresponds to the dual graph on the right of equation \eqref{eq:sequence1}. 

Note that the constraint in \eqref{eq:crossinginvariance} implies that crossing symmetry is also enforced for averages of \emph{products} of partition functions, such as $\overline{Z[\Sigma_{g_1}]Z[\Sigma_{g_2}]}$. To see this, we just have to take a higher-genus Riemann surface $\Sigma_g$, such that pinching it along a separating geodesic gives two disjoint `daughter' surfaces $\Sigma_{g_1}$ and $\Sigma_{g_2}$. The mapping class group of the `parent' $\Sigma_g$ contains as a strict subset the elements of the form $\gamma = \gamma_1\circ \gamma_2$, where $\gamma_1 \in \text{MCG}(\Sigma_{g_1})$ and $\gamma_2 \in \text{MCG}(\Sigma_{g_2})$, such that the separating geodesic is fixed pointwise by $\gamma$. Pictorially, we have the following example at $g=4$:
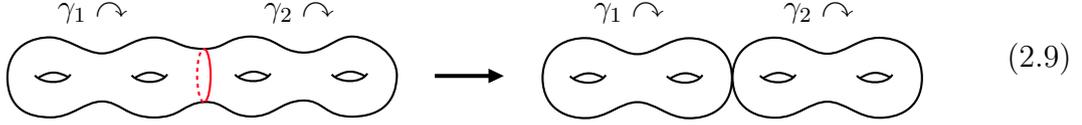
\begin{equation}
    \tikzset{every picture/.style={line width=0.8pt}}    

\end{equation}
On the level of the partition functions, we can implement the pinching limit by degenerating  the modulus $\omega$ corresponding to the separating geodesic, with associated plumbing coordinate $q$, and multiplying by an appropriate factor of the conformal anomaly \cite{Cardy:2017qhl}. Then we have, schematically (suppressing the dependence on the right-moving moduli in the notation): 
\begin{align}
    \overline{Z[\Sigma_{g_1};\gamma_1\cdot \Omega_1]Z[\Sigma_{g_2};\gamma_2\cdot \Omega_2]} &= \lim_{q,\bar q \to 0} q^{\frac{c}{24}}\bar q^{\frac{c}{24}} \,\overline{Z[\Sigma_g;\gamma_1\circ \gamma_2\cdot \Omega]}\\[1em]
    &=\lim_{q,\bar q \to 0} q^{\frac{c}{24}}\bar q^{\frac{c}{24}} \,\overline{Z[\Sigma_g; \Omega]} = \overline{Z[\Sigma_{g_1}; \Omega_1]Z[\Sigma_{g_2}; \Omega_2]}\ ,
\end{align}
where in the second line we used crossing symmetry \eqref{eq:crossinginvariance} of the average higher-genus partition function. So we see that imposing crossing symmetry of the average partition function at all genera implies, as a special case, crossing symmetry of all higher statistical moments of $Z[\Sigma]$.\footnote{This is to be contrasted to the approach taken in the tensor/matrix model of \cite{Belin:2023efa, Jafferis:2025vyp}, in which the square of the genus-2 crossing equation is enforced by a highly constraining potential, whose width is set by a tolerance parameter $\hbar$.} This is a non-trivial conclusion because, in general, the average of a product is not the product of the averages.

\paragraph{Block decomposition and Virasoro crossing kernels.} We now want to see what the constraints of crossing symmetry and modular invariance imply for the measure $\mu$ and its moments. To do so, let us substitute the conformal block decomposition \eqref{eq:block_decomposition} on both the left- and right-hand side of equation \eqref{eq:crossinginvariance}. We can then use the fact that Virasoro conformal blocks with $c>1$ transform under crossing transformations by integrating them against \emph{Virasoro crossing kernels} \cite{Ponsot:1999uf,Ponsot:2000mt,Teschner:2012em,Eberhardt:2023mrq}. These crossing kernels form a projective unitary representation of the Moore-Seiberg groupoid (and thus in particular of the mapping class group) of $\Sigma_g$, acting on the Hilbert space of conformal blocks by integral transformations:
\begin{equation}
    \mathcal{F}_g^{\,\mathcal{C}}(\bm{h};\gamma\cdot\Omega) = \mathcal{F}_g^{\,\gamma\cdot\mathcal{C}}(\bm{h};\Omega) =\int_{\frac{c-1}{24}}^\infty \d^{3g-3} h' \,\,\mathbb{K}^\gamma_{\bm{hh'}}\, \mathcal{F}_g^{\,\mathcal{C}}(\bm{h}';\Omega).
\end{equation}
A similar formula holds for the right-moving conformal blocks. The crossing kernel $\mathbb{K}^\gamma_{\bm{hh'}}$ is an analytic function of the conformal weights with $h_i\geq \frac{c-1}{24}$ and can always be expressed as a composition of the basic crossing kernels $\mathbb{B}, \mathbb{F}$ and $\mathbb{S}$ that represent the three Moore-Seiberg moves in figure \ref{fig:MS_moves}. For an overview of the analytic and structural properties of these kernels, we refer the reader to \cite{Ponsot:1999uf,Teschner:2012em, Teschner:2013tqy,Eberhardt:2023mrq}.

Having expressed the left- and right-hand side of \eqref{eq:crossinginvariance} as superpositions of Virasoro conformal blocks in the same channel, we can use the fact that the above-threshold non-degenerate blocks span a complete orthogonal basis to obtain the following constraint equation:\vspace{1mm}
\begin{equation}\label{eq:crossingOPE}
    \int \d^{3g-3}h\,\d^{3g-3}\bar h\,\overline{\prod_{e_i\in \Gamma_{\gamma\cdot\mathcal{C}}}\!\!\!\rho(h_i,\bar h_i)\!\!\!\prod_{v_{ijk}\in \Gamma_{\gamma\cdot\mathcal{C}}}\!\!\!\!C_{ijk}} \,\big |\mathbb{K}_{\bm{hh'}}^\gamma \big |^2 = \overline{\prod_{e_i'\in \Gamma_\mathcal{C}}\!\!\rho(h_i',\bar h_i')\!\!\!\prod_{v_{i'j'k'}\in \Gamma_\mathcal{C}}\!\!\!\!C_{i'j'k'}}
\end{equation}
for all $\gamma\in \text{MS}_g$ and $g\geq 1$. Here we used the notation $|\cdot |^2$ to denote the product of left- and right-moving crossing kernels. These constraint equations are equivalent to the constraints on the `canonical' partition function in \eqref{eq:crossinginvariance}, but they have been translated into a `fixed energy' language as integral equations for the moments of the ensemble of CFT$_2$ data. This strategy, of stripping off the Virasoro conformal blocks from the bootstrap constraints using crossing kernels, was first used in \cite{Collier:2019weq} to derive universal OPE statistics from crossing symmetry combined with identity dominance, and was extended to higher genus in \cite{Belin:2021ryy,Anous:2021caj,deBoer:2024mqg}. 

There is yet another equivalent way of expressing the constraints of modular invariance and crossing symmetry that will be useful later. This uses the fact that the crossing kernels form a representation of $\text{MS}_g$, so if \eqref{eq:crossingOPE} holds for the generators $\mathbb{B},\mathbb{F}$ and $\mathbb{S}$, it automatically holds for all $\gamma$. Let us first see how this works at genus $g=1$, where \eqref{eq:crossingOPE} is equivalent to invariance of $\bar\rho$ under the generators $S$ and $T$:
\begin{align}
    |\mathbb{T}_h|^2\,\overline{\rho(h,\bar h)} &=  \overline{\rho(h,\bar h)} \label{eq:Tinvariance}\\
    \int \d h \d\bar h \,\overline{\rho(h,\bar h)} \,|\mathbb{S}_{hh'}[\mathds{1}]|^2 &= \overline{\rho(h',\bar h')}.\label{eq:Sinvariance} 
\end{align}
The $T$-transform is represented by the multiplicative phase $\mathbb{T}_h = \e^{-2\pi i (h-\frac{c}{24})}$ acting on the torus block (i.e.~the Virasoro character). Combined with the right-moving phase, equation \eqref{eq:Tinvariance} is just the quantization of spin, since $|\mathbb{T}_h|^2 = \e^{2\pi i (\bar h-h)} = 1$ implies $J = h-\bar h \in \mathbb{Z}$. The non-trivial constraint is the invariance under $S:\tau \to -\frac{1}{\tau}$, which is represented by the modular $S$-kernel $\mathbb{S}_{hh'}[\mathds{1}]$. Since any $\gamma \in \text{PSL}(2,\mathbb{Z})$ can be written as a word $\gamma = ST^{n_1}ST^{n_2}\cdots ST^{n_k}$ in the generators, iteratively applying \eqref{eq:Tinvariance} and \eqref{eq:Sinvariance} implies the invariance of $\overline{\rho(h,\bar h)}$ under the general modular transformation\footnote{We can express this general modular kernel in the form given in \cite{Benjamin:2020mfz} by using the continued fraction expansion of the rational number $\frac{p}{q}$ and the Gaussian integral identity (C9) in \cite{deBoer:2025rct}.}  
\begin{equation}
    \mathbb{K}^\gamma_{hh'} = \int \prod_{i=1}^{k-1}\d h_i\,\, \mathbb{S}_{hh_1}[\mathds{1}] \,\mathbb{T}_{h_1}^{n_1} \cdots \mathbb{S}_{h_{k-1} h'}[\mathds{1}]\,\mathbb{T}_{h'}^{n_k}\, \quad (g=1).
\end{equation}

 We can follow a similar logic for $g>1$. Now the general crossing kernel $ \mathbb{K}^\gamma_{hh'}$ is composed of the fusion kernel, the one-point S-kernel and the braiding phase \cite{Eberhardt:2023mrq}. Thus, the crossing equation \eqref{eq:crossingOPE} is equivalent to the following set of constraints:
\begin{subequations}
\begin{align}
    \overline{C_{123}X} &=  \big| \mathbb{B}^{h_1}_{h_2h_3} \big |^2 \,\overline{C_{132}X} \label{eq:Binvariance}\\[1em]
     \overline{\rho(h_2,\bar h_2)C_{223}X}&= 
     \int \d h_1 \d \bar h_1\,\, \overline{\rho(h_1,\bar h_1)C_{113}X} \,\,\big|\mathbb{S}_{h_1h_2}[h_3]\big|^2\label{eq:S1invariance}\\[1em]
     \overline{\rho(h_3,\bar h_3)C_{123}C_{345}X}&=
     \int \d h_3' \d \bar h_3' \,\,\overline{\rho(h_3',\bar h_3')C_{13'5}C_{43'2}X}\,\,\left| \fker{h_3'}{h_3}{h_1}{h_2\,}{h_4\,}{h_5}\right |^2 \,.\label{eq:Finvariance}
\end{align}
\end{subequations}
In the above equations, $X$ is a placeholder for all the other factors of $\rho(h_i,\bar h_i)$ and $C_{ijk}$ that can appear inside the average in \eqref{eq:crossingOPE}. (In particular, all OPE indices come with their own factors of $\rho$, but the symbol $X$ makes the notation lighter.) One should think of the above constraints as holding inside every graph $\Gamma_{\mathcal{C}}$, where we have `zoomed in' to any portion of the graph that looks like one of the diagrams in figure \ref{fig:MS_moves}. Within each such piece of the graph we can then locally apply braiding, fusion or modular transformations.  

The first constraint \eqref{eq:Binvariance} is trivially satisfied in our ensemble because we already assumed the permutation property \eqref{eq:OPEpermutation}. The braiding phase times its right-moving counterpart is given by $\big| \mathbb{B}^{h_1}_{h_2h_3} \big |^2 = \e^{\pi i (h_2+h_3-h_1)}\e^{-\pi i (\bar h_2+\bar h_3-\bar h_1)},$ which cancels the factor of $(-1)^{J_1+J_2+J_3}$ coming from the permutation of the indices of $C_{123}$ (this follows because the spins are integers). We can combine the braiding phase with the fusion and modular kernels in \eqref{eq:S1invariance} and \eqref{eq:Finvariance} to obtain the general crossing kernel $\mathbb{K}^\gamma_{\bm{hh'}}$, which implies the general crossing symmetry constraint \eqref{eq:crossingOPE}.

\subsection{Typicality and index contractions}\label{sec:typicality}
In the previous subsection, we have stated the requirements of crossing symmetry and modular invariance on the moments of CFT data, which were clearly based on the same requirements as in the regular CFT bootstrap. Now we will state an additional assumption on the ensemble, to which we refer as \emph{typicality}, which has its origin in the theory of quantum chaos.

In chaotic many-body quantum systems, the Eigenstate Thermalization Hypothesis (ETH) is a statistical ansatz to model the matrix elements of simple operators $\mathcal{O}_{ij} = \bra{E_i}\mathcal{O}\ket{E_j}$ in high-energy eigenstates \cite{Deutsch1991,Srednicki1994,Foini:2018sdb,Pappalardi:2022aaz,Jafferis:2022uhu}.  Instead of giving the exact, microscopic matrix elements $\mathcal{O}_{ij}$, the ETH posits a smoothly varying envelope for the diagonal matrix elements, and exponentially small erratic noise in the off-diagonal matrix elements that can effectively be treated as random variables.  As shown in \cite{Foini:2018sdb}, the statistics of these random variables are approximately Gaussian, but important non-Gaussianities can contribute at leading order to higher point thermal correlators \cite{Murthy:2019fgs} (in particular out-of-time-ordered ones). The structure of these non-Gaussian moments is fixed by assuming local unitary invariance in narrow microcanonical windows around high energy eigenstates:
\begin{equation}\label{eq:typicality}
    \overline{\mathcal{O}^U \cdots \mathcal{O}^U}\approx \overline{\mathcal{O}\cdots\mathcal{O}}\,.
\end{equation}
Here $\mathcal{O}^U = U\mathcal{O}U^\dagger$, with $U$ a block-diagonal unitary matrix. This block-diagonal structure is such that $U$ acts non-trivially in small microcanonical windows (that nevertheless contain many energy levels), so as to keep the overall band structure of the Hamiltonian intact. In the context of ETH, the assumption \eqref{eq:typicality} is referred to as typicality. Just like level repulsion, the intuition behind typicality stems from random matrix universality: if the Hamiltonian in a narrow energy band resembles a random matrix, then the relative basis change between the eigenbasis of simple operators and that of the Hamiltonian should be a random unitary. By averaging \eqref{eq:typicality} over $U$ in the unitary group, we immediately infer the index structure for the moments $\overline{\mathcal{O}_{a_1b_1}\cdots \mathcal{O}_{a_nb_n}}$ evaluated in energy eigenstates. Namely, for the moment to be non-zero, each index $a_i$ should be paired with an index $a_j$, and each $b_k$ with some $b_l$. Diagrammatically, the allowed index contractions are visualized by cyclic, cactus and non-cactus diagrams (see \cite{Pappalardi:2022aaz}). 

This statistical approach to chaos and thermalization, formalized in the ETH, is of course very similar to our point of view on OPE statistics and, in fact, a sharp analogy can be made for OPE coefficients of 1 light and 2 heavy operators \cite{deBoer:2024mqg,Belin:2021ryy}. More generally, for theories with a gravitational dual, $C_{ijk}$ can be treated as a random variable as long as at least one index is in the heavy part of the spectrum \cite{Belin:2020hea,Chandra:2022bqq}. 

If the statistics of these random variables were assumed to be Gaussian,\footnote{Here, by Gaussian we mean that the probability distribution $\mu(\{h_i,\bar h_i,C_{ijk}\})$ for the heavy OPE coefficients is the form $\exp(-\sum_{i,j,k} f(h_i,\bar h_i,h_j,\bar h_j, h_k,\bar h_k)|C_{ijk}|^2)$, where $f$ is some smooth function of the conformal weights.} the index structure of some statistical moment of OPE coefficients would be entirely fixed by Wick contractions. Although this Gaussian ansatz reproduces many bulk computations \cite{Chandra:2022bqq}, non-Gaussianities are necessary to match the ensemble prediction with gravitational calculations of multiboundary wormhole topologies \cite{deBoer:2024mqg,Collier:2024mgv} and topologies with higher-genus boundaries \cite{Belin:2021ryy, Collier:2023fwi, Hartman:2025ula}. Apart from bulk considerations, these non-Gaussianities are also necessary in order to satisfy the full crossing symmetry requirements discussed in the previous section \cite{Belin:2021ryy,Anous:2021caj}. Hence, we can no longer resort to Wick contractions.

This means we need a different principle to fix the index structure of a given moment of OPE coefficients. As in the ETH, the full index structure can be fixed by appealing to a version of typicality, now tailored to 2d CFT.\footnote{This notion of typicality has also been discussed in \cite{Belin:2021ryy,Belin:2021ibv,deBoer:2024mqg}.} We will focus on theories with a gap up to the black hole threshold, so we are interested in moments of $C_{ijk}$ where all indices correspond to heavy primaries (when one or more indices is the vacuum state, the index structure is trivially fixed by the selection rule $C_{ij\mathds{1}} = \delta_{ij}$). For these moments, typicality is the assumption that the average is approximately invariant under orthogonal transformations in microcanonical windows:
\begin{equation}\label{eq:typicalityCFT}
     \overline{C^T_{i_1i_2i_3}\cdots C^T_{i_{n-2}i_{n-1}i_n}}\,\approx\,\overline{C_{i_1i_2i_3}\cdots C_{i_{n-2}i_{n-1}i_n}} , \quad C^T_{ijk} \coloneqq \sum_{a,b,c} T_{ia}T_{jb}T_{kc} C_{abc}\,.
\end{equation}
Here $T$ is a block-diagonal orthogonal matrix, with each block acting in a microcanonical Hilbert space of dimension $\e^{S(\Delta_i)}$ centered at energy $\Delta_i$ with fixed spin $J_i$. The fact that $T$ is real orthogonal, instead of unitary, follows from the reality condition \eqref{eq:OPEpermutation} (see also section 2 of \cite{Jafferis:2025vyp}). As before, the typicality assumption fixes the index structure by taking the Haar average over the local rotations $T$: the moment of structure constants in \eqref{eq:typicalityCFT} is non-zero only if all indices are repeated pairwise. For example, for the second moment, typicality implies the following structure: 
\begin{multline}\label{eq:genus2typicality}
    \overline{C_{123}C^*_{1'2'3'}} = \sum_{\sigma \in S_3}\delta_{1\sigma(1')}\delta_{2\sigma(2')}\delta_{3\sigma(3')} \,(\text{sgn}(\sigma))^{J_1+J_2+J_3} \,\mathcal{G}_S\!\left[\vcenter{\hbox{
    \begin{tikzpicture}[yscale=-0.5,xscale=0.5]
        \draw[very thick, wilsonred] (0,0) ellipse (1.5 and 1);
        \draw[very thick, wilsonred] (0,1) -- (0,-1);
        \node at (-1.2,0) {\footnotesize $1$};
        \node at (0.29,0) {\footnotesize $2$};
        \node at (1.2,0) {\footnotesize $3$};
    \end{tikzpicture}
    }}\right] \\ 
    + \sum_{\tau,\sigma \in A_3} \delta_{\tau(1)\sigma(1')}\delta_{\tau(2) \tau(3)}\delta_{\sigma(2')\sigma(3')} \,\mathcal{G}_D\!\left[\vcenter{\hbox{
    \begin{tikzpicture}[yscale=-0.85,xscale=0.85]
        \draw[very thick, wilsonred] (-1,0) circle (0.5 and 0.5);
        \draw[very thick, wilsonred] (1,0) circle (0.5 and 0.5);
        \draw[very thick, wilsonred] (-0.5,0) -- (0.5,0);
        \node at (-1,-0.15) {\scriptsize $\tau(2)$};
        \node at (0,0.3)  {\scriptsize $\tau(1)$};
        \node at (1,-0.15)  {\scriptsize $\sigma(2')$};
    \end{tikzpicture}
    }}\right]\,.
\end{multline}
This is a sum over all pairings of 6 indices, with in total $(6-1)!! = 15$ terms, where we have split the sum into the two distinct contraction patterns that can appear for two OPE coefficients (recall eq.~\eqref{eq:genus2graphs}). The first sum, corresponding to the `sunset' contraction $C_{123}C_{123}^*$, is over the symmetric group $S_3$, with sign factors that account for the permutation property \eqref{eq:OPEpermutation}. The function $\mathcal{G}_S$ is a real-valued, totally symmetric smooth function of the conformal weights $h_{1,2,3},\bar h_{1,2,3}$, whose functional form will be discussed below. The second term is a double sum over the alternating group $A_3 = \{\text{Id},(123),(132)\}$, corresponding to the contractions of `dumbbell' type $C_{\tau(1)\tau(2)\tau(2)}C_{\tau(1)\sigma(2')\sigma(2')}$. The real function $\mathcal{G}_D$ also depends smoothly on the weights $\{h_{a},\bar h_{a}\}_{a\in\{\tau(1),\tau(2),\sigma(2')\}}$ in the heavy regime and vanishes when $J_{\tau(1)}$ is odd.\footnote{In each term of \eqref{eq:genus2typicality}, $\mathcal{G}_D$ is the same function, but evaluated on different conformal weights, as specified by the even permutations $\tau$ and $\sigma$.} Because the functions multiplying the Kronecker delta's are smooth, they are approximately constant over a narrow microcanonical window in which $T$ acts, which explains why the structure \eqref{eq:genus2typicality} indeed obeys the typicality condition \eqref{eq:typicalityCFT}.

For the higher statistical moments, the index structure is similarly fixed to be a sum over pairwise index contractions. Because each OPE coefficient has three indices, this automatically implies that all odd moments vanish.\footnote{The fact that odd moments vanish also already follows from the fact that the canonical normalization of the CFT operator $\mathcal{O}_i$
is unchanged under $\mathcal{O}_i\to-\mathcal{O}_i$, so there is no canonical definition of the overall sign of a moment containing an odd number of appearances of $\mathcal{O}_i$. Let us also remark that the statement that odd moments vanish holds in the `pure' gravity ensemble, where all non-vacuum states are above the black hole threshold. If we allow states in the conical defect regime, we can get non-vanishing odd moments: for a contribution to the cubic moment, see \cite{deBoer:2024mqg}.} The non-zero index contractions can be visualized again by trivalent graphs $\Gamma$, such as the ones depicted in \eqref{eq:genus3graphs}, so that the general ansatz compatible with typicality is\vspace{1mm}
\begin{equation}\label{eq:generaltypicality}
    \overline{C_{i_1i_2i_3}\cdots C_{i_{n-2}i_{n-1}i_n}} = \sum_{\text{pairings } P} \delta^P_{i_1\cdots i_n} \,\mathrm{sgn}(P)\,\mathcal{G}_{\Gamma}\big[\{h_a,\bar h_a\}_{e_a\in \Gamma}\big]\,.
\end{equation}
The sum is over the set of partitions of $n$ elements into $n/2$ pairs (where $n$ is even). To each pairing $P$ we associate a trivalent graph $\Gamma$ (really an equivalence class up to relabellings of the edges) and a product of Kronecker delta's $\delta^P\coloneqq \prod_{(a,b)\in P}\delta_{i_ai_b}$. We define the sign of $P$ by labeling the edges of $\Gamma$ and choosing an orientation for each vertex; then we define $\mathrm{sgn}(P) \coloneqq \prod_{v_{abc}\in \Gamma}\mathrm{sgn}(v_{abc})^{J_{i_a}+J_{i_b}+J_{i_c}} $ where $\mathrm{sgn}(v_{abc})$ is 1 if the index labels $abc$ are ordered clockwise; and $-1$ if they are ordered anti-clockwise. When two index labels coincide, we also define $\mathrm{sgn}(v_{abb}) = 1$. These $\pm$ signs are necessary for compatibility of the ansatz with the permutation property \eqref{eq:OPEpermutation} of the OPE coefficients.

There are a few things to note about the ansatz \eqref{eq:generaltypicality} predicted by typicality. First, there are more allowed contractions in \eqref{eq:generaltypicality} than in the ETH. This is because the Haar average is taken over orthogonal matrices, whereas in `standard' ETH the average is over the unitary group.\footnote{Straightforward extensions of the ETH to the GOE have also been discussed, see \cite{DAlessio:2015qtq}.} Hence, the ETH ansatz for $\mathcal{O}_{a_ib_i}$ always pairs an $a_i$-index with an $a_j$-index, and $b_i$ with $b_j$, while in the CFT case \eqref{eq:generaltypicality} any index can be paired to any other index. This means that, for example, the analog of the dumbbell contraction $\overline{C_{122}C_{133}}$ is absent in ETH, because in ETH the connected part of $\overline{\mathcal{O}_{22}\mathcal{O}_{33}}$ is zero for $E_2\neq E_3$.

Second, the non-zero contraction patterns in \eqref{eq:generaltypicality} are naturally encoded in averages of genus-$g$ partition functions $Z[\Sigma_g]$, expanded in Virasoro conformal blocks as in \eqref{eq:block_decomposition}. However, as explained in the previous section, although the sum over pairings is finite (with $(2n-1)!!$ terms), there is an infinite number of conformal blocks. So, for each graph topology $\Gamma$, the function $\mathcal{G}_{\Gamma}$ can itself be an infinite sum of contributions. In gravity, these contributions consist of bulk topologies with a boundary given by a marked Riemann surface, schematically $\mathcal{G}_\Gamma = \sum_{M_\Gamma} Z_\text{grav}[M_\Gamma]$, as will be explained in section \ref{sec:3}. However, so far we have not assumed a bulk dual, and hence the functions $\mathcal{G}_\Gamma$ only need to be sufficiently smooth for \eqref{eq:generaltypicality} to obey the typicality assumption \eqref{eq:typicalityCFT}.

Third, in the literature on OPE statistics it is customary, by a slight abuse of notation (also perpetrated by some of the present authors), to refer to the functions $\mathcal{G}_\Gamma$ by the corresponding moment $\overline{\prod_{v_{ijk}\in \Gamma}C_{ijk}}$. For example, we could write
\begin{equation}
    \overline{C_{123}C_{123}^*} = \mathcal{G}_S(h_{1,2,3},\bar h_{1,2,3})\,.
\end{equation}
This is not an abuse of notation as long as all indices are different, because only a single term in the sum of \eqref{eq:genus2typicality} will be non-zero. However, when some indices coincide, this is no longer entirely correct: for example,  $\overline{C_{122}C_{122}}$ receives contributions from both the sunset and dumbbell contractions, $\mathcal{G}_S$ and $\mathcal{G}_D$ (see below). This observation may seem like a minor detail, but it is at the root of answering our question why non-handlebodies appear in the sum over topologies! Saving a full explanation of this fact for section \ref{sec:minimal completion}, let us now briefly discuss the structure of the ansatz \eqref{eq:generaltypicality} when a subset of the indices is repeated.

\paragraph{Implications of typicality for repeated indices.} Typicality implies that moments of OPE coefficients are non-zero only when all indices are paired. This includes contractions in which indices are repeated \emph{more than once}. For example, for the OPE variance, the typicality ansatz \eqref{eq:typicalityCFT} for $\mathcal{O}_1\neq \mathcal{O}_2$ reads:\vspace{1mm}
\begin{equation}\label{eq:repeated indices a}
    \overline{C_{122}C_{122}} = \left(1+(-1)^{J_1}\right) \mathcal{G}_S\!\left[\vcenter{\hbox{
    \begin{tikzpicture}[yscale=-0.5,xscale=0.5]
        \draw[very thick, wilsonred] (0,0) ellipse (1.5 and 1);
        \draw[very thick, wilsonred] (0,1) -- (0,-1);
        \node at (-1.2,0) {\footnotesize $1$};
        \node at (0.29,0) {\footnotesize $2$};
        \node at (1.2,0) {\footnotesize $2$};
    \end{tikzpicture}
    }}\right] + \mathcal{G}_D\!\left[\vcenter{\hbox{
    \begin{tikzpicture}[yscale=-0.85,xscale=0.85]
        \draw[very thick, wilsonred] (-1,0) circle (0.5 and 0.5);
        \draw[very thick, wilsonred] (1,0) circle (0.5 and 0.5);
        \draw[very thick, wilsonred] (-0.5,0) -- (0.5,0);
        \node at (-1,-0.15) {\footnotesize $2$};
        \node at (0,0.3)  {\footnotesize $1$};
        \node at (1,-0.15)  {\footnotesize $2$};
    \end{tikzpicture}
    }}\right]\,.\vspace{1mm}
\end{equation}
Note that the factor $\left(1+(-1)^{J_1}\right)$ in the first term projects onto even spin $J_1$, in accordance with the fact that $C_{122}$ vanishes when $J_1$ is odd.\footnote{The function $\mathcal{G}_D$ is non-zero only for even $J_1$, which follows from \eqref{eq:OPEpermutation} and \eqref{eq:genus2typicality} evaluated on the dumbbell contraction $\overline{C_{122}C_{133}}$ with $\mathcal{O}_2 \neq \mathcal{O}_3$.} 

Similarly, when all indices coincide, all contractions contribute:
\begin{equation}\label{eq:repeated indices b}
    \overline{C_{111}C_{111}} = 3\left(1+(-1)^{J_1}\right) \mathcal{G}_S\!\left[\vcenter{\hbox{
    \begin{tikzpicture}[yscale=-0.5,xscale=0.5]
        \draw[very thick, wilsonred] (0,0) ellipse (1.5 and 1);
        \draw[very thick, wilsonred] (0,1) -- (0,-1);
        \node at (-1.2,0) {\footnotesize $1$};
        \node at (0.29,0) {\footnotesize $1$};
        \node at (1.2,0) {\footnotesize $1$};
    \end{tikzpicture}
    }}\right] + 9\,\mathcal{G}_D\!\left[\vcenter{\hbox{
    \begin{tikzpicture}[yscale=-0.85,xscale=0.85]
        \draw[very thick, wilsonred] (-1,0) circle (0.5 and 0.5);
        \draw[very thick, wilsonred] (1,0) circle (0.5 and 0.5);
        \draw[very thick, wilsonred] (-0.5,0) -- (0.5,0);
        \node at (-1,-0.15) {\footnotesize $1$};
        \node at (0,0.3)  {\footnotesize $1$};
        \node at (1,-0.15)  {\footnotesize $1$};
    \end{tikzpicture}
    }}\right]\,.
\end{equation}
These combinatorial factors (or `degeneracies') in the statistical moments of OPE coefficients will be important when we match to gravity in sections \ref{sec:minimal completion} and \ref{sec:5}. 
An entirely analogous observation holds for higher moments of OPE coefficients: when a subset of indices coincide, there are more terms in the ansatz \eqref{eq:generaltypicality} that can contribute, which can originate from different graph topologies.

One interesting point to observe here is that index repetition is not a notion that is invariant under crossing. For example, consider the contribution to $\overline{C_{113}C_{223}}$ from the index contraction $\delta_{12}$. Let us rewrite the discrete Kronicker delta using the continuous Dirac delta $\delta^2(h_1-h_2) \equiv \delta(h_1-h_2)\delta(\bar h_1-\bar h_2)$, multiplied by appropriate densities of states $\rho_i\equiv \rho(h_i,\bar h_i)$. Under the one-point S-transform, this index contraction gets mapped to:
\begin{equation}\label{eq:repeated indices and crossing}
   \int \d h_1\d \bar h_1 \, \delta^2(h_1-h_2)\,\overline{\rho_2\rho_3C_{113}C_{223}} \,|\mathbb{S}_{h_1h_4}[h_3]|^2 =  \overline{\rho_2\rho_3C_{223}C_{223}}\,|\sker{h_2}{h_4}{h_3}|^2.
\end{equation}
By the crossing equation \eqref{eq:S1invariance}, the RHS now contributes to $\overline{\rho_2\rho_{3}\rho_4C_{223}C_{443}}$, and is non-zero even when $h_2$ and $h_{4}$ do not coincide. So we see that crossing symmetry relates moments with repeated pairs of indices to moments with distinct (non-repeated) pairs of indices. This interplay between crossing symmetry and typicality will be at the heart of the mechanism that, on the bulk side, will be responsible for generating a large class of non-handlebodies in the sum over topologies.

\subsection{Spectral statistics} \label{sec:spectral_statistics}

In the previous subsections, we have described the constraints imposed by crossing symmetry and typicality on the ensemble. We will now make one further assumption on our ensemble, again motivated by the theory of quantum chaos. Namely, we will assume that, to a good approximation, we can ignore the effects of non-trivial spectral statistics, treating $\rho(h,\bar h)$ fully classically. 

In chaotic many-body quantum systems, which are characterized by a ramp in the spectral form factor after the Thouless time \cite{Altland:2020ccq,Saad:2018bqo,Cotler:2016fpe}, this is indeed a valid assumption for early times, because the spectral variance  is exponentially suppressed in the entropy compared to the leading factorized contribution:
\begin{equation}
    \overline{\rho(E_i)\rho(E_j)} = \overline{\rho(E_i)}\,\, \overline{\rho(E_j)} \,\Big[1+ \mathcal{O}(\e^{-2S})\Big]\,.
\end{equation}
Indeed, in most treatments of the Eigenstate Thermalization Hypothesis it is a standard assumption that discrete sums can be replaced by integrals over products of smooth densities of states \cite{DAlessio:2015qtq}, ignoring any fluctuations around the average $\bar\rho$, because the time scales for thermalization are much shorter than the post-Thouless time scales where spectral rigidity becomes important.  

For chaotic 2d CFT's, one can also study the spectral variance in fixed spin sectors\footnote{For discussions of spectral statistics and the relation to 3d gravity, see \cite{Cotler:2020hgz,DiUbaldo:2023qli, Haehl:2023xys,Haehl:2023mhf, Haehl:2023tkr, deBoer:2025rct, Boruch:2025ilr,Perlmutter:2025ngj,Jafferis:2025jle}.} and formulate a version of late-time random matrix universality that is compatible with modular invariance. Under a suitable coarse-graining procedure, akin to Berry's diagonal approximation \cite{Berry1985SpectralRigidity}, the 2d CFT analog of the spectral form factor displays a linear ramp  \cite{DiUbaldo:2023qli}, which matches to the gravitational calculation of an off-shell torus wormhole $T^2\times I$ by Cotler and Jensen \cite{Cotler:2020ugk}. Recently, \cite{Perlmutter:2025ngj} made the `simple extreme conjecture', which formulates a precise statement of random matrix universality for the primary spectrum of Virasoro CFT's with $c>1$, in fixed spin sectors, in the asymptotic limit of large conformal dimension $\Delta$.

In this paper we do not account for the fluctuations of the primary-state density around its mean. In the discussion, we comment on corrections from higher statistical moments of 
$\rho(h,\bar h)$, and interpret these corrections gravitationally using the “RMT surgery’’ construction of \cite{deBoer:2025rct}. In the main text, however, we make the replacement:
\begin{equation}\label{eq:rho_factorized}
    \overline{\prod_{e_i\in \Gamma_\mathcal{C}}\rho(h_i,\bar h_i)\!\!\!\prod_{v_{ijk}\in \Gamma_\mathcal{C}}\!\!\!\!C_{ijk}} \,\,\boldsymbol{\longrightarrow} \prod_{e_i\in \Gamma_\mathcal{C}}\!\!\overline{\rho(h_i,\bar h_i)}\times\overline{\prod_{v_{ijk}\in \Gamma_\mathcal{C}}\!\!\!\!C_{ijk}}\,.
\end{equation}
That is, we factorize moments of the density of states into products of the mean density $\bar\rho$,
 while retaining the full (non-factorized) OPE statistics.

The above simplification parallels that of \cite{Chandra:2022bqq}, with the only difference that in their CFT ensemble, the mean spectral density was further approximated by $\overline{\rho(h,\bar h)} \approx \rho_0(h)\rho_0(\bar h)$, where $\rho_0$ is the universal Cardy density of states \cite{Collier:2019weq}. By contrast, in this paper we subject the mean $\bar\rho$ to the constraints of modular invariance \eqref{eq:Tinvariance}-\eqref{eq:Sinvariance}. This refinement of the approach of \cite{Chandra:2022bqq} will be essential for identifying some non-handlebody saddles in section \ref{sec:minimal completion}.

The simplification \eqref{eq:rho_factorized} can also be motivated from the bulk. Replacing the statistical moments of $\rho$ by products of means plus \emph{connected} spectral correlations effectively separates out contributions from different bulk topologies. The connected terms---which we neglect---arise exclusively from \emph{off-shell} bulk geometries. As shown by Maloney and Witten \cite{Maloney:2007ud}, any 3-manifold with more than one asymptotic torus boundary is necessarily off-shell (i.e.~non-hyperbolic). Since higher connected moments of $\rho$ are generated by precisely such multi-boundary torus wormholes, all spectral statistics beyond the mean $\bar\rho$ correspond to off-shell topologies. The RMT surgery framework of \cite{deBoer:2025rct} provides a systematic way to build some of these off-shell contributions by `gluing in' torus wormholes. 

Although a complete ensemble description of 3d gravity must ultimately include such off-shell effects, here we focus solely on the contributions from on-shell topologies. Even at this level, the resulting bulk sum includes many nontrivial 3-manifolds. In the next section we begin assembling the sum over topologies, and see to what extent they solve the boundary constraints imposed in this section.

%%%%%%%%%%%%%%%%%%%%%%%%%%%%%%%%%%%%
\section{Summing over handlebodies}
\label{sec:3}
%%%%%%%%%%%%%%%%%%%%%%%%%%%%%%%%%%%%

In this section, we will start to build up the ensemble from the boundary. As we proceed, we will start to recover the gravitational sum over 3-manifolds. Here we will only be concerned with the constraints of crossing symmetry and modular invariance, leaving the additional assumption of typicality for section \ref{sec:minimal completion}. The simplest class of topologies that, when summed over, solve crossing and modular invariance are the handlebody topologies, which we study in this section.

\subsection{The mapping class group}

In general, a 3-dimensional handlebody $M$ of genus $g$ is defined by gluing $g$ 1-handles to a 3-ball. Its boundary is a genus-$g$ Riemann surface $\Sigma_g$, with a collection of $3g-3$ non-intersecting closed curves that bound disks in $M$. These essential curves are also called contractible cycles, or meridians. An example at genus 2 is shown below:
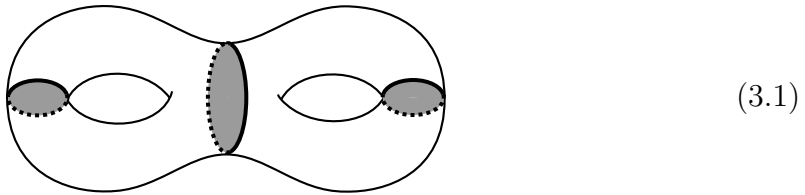
\begin{equation}\label{eq:genus2handlebody}
\tikzset{every picture/.style={line width=0.8pt}}

\end{equation}
where the compressible disks chosen in this example have been shaded gray. 
The contractible cycles are associated to the Teichm\"uller parameters that specify the conformal structure moduli at the boundary. If we fix the values of these moduli, the handlebody carries a hyperbolic metric (i.e.~of constant negative curvature) with infinite volume, which is unique up to isometry \cite{Sullivan1981ErgodicInfinity}.  

There are three important groups associated to a handlebody. The first is the fundamental group $\pi_1(M)$, which for a handlebody is isomorphic to the free group on $g$ generators. The second is the mapping class group of the boundary surface:
\begin{equation}
    \text{MCG}(\Sigma_g) = \frac{\text{Diff}^+(\Sigma_g)}{\text{Diff}_0(\Sigma_g)}
\end{equation}
where the numerator is the group of orientation preserving diffeomorphisms of $\Sigma_g$ and the denominator is the subgroup of diffeomorphisms that are continuously connected to the identity. A famous theorem by Lickorish states that the mapping class group is generated by a finite collection of Dehn twists \cite{Lickorish_1964}. A Dehn twist $T$ is a local operation on a cylinder neighborhood of a closed curve $C$, where the cylinder is cut along $C$ and reglued after a rotation by $2\pi$:\vspace{2mm}
\begin{equation}\label{eq:Dehntwist}
T:\quad\tikzset{every picture/.style={line width=0.8pt}}
 \,.
\end{equation}
The minimal number of Dehn twist generators for $\text{MCG}(\Sigma_{g})$ with $g\geq 2$ is $2g+1$ \cite{Humphries} (while for $g=1$ the minimum number is 2). A canonical choice of cycles, whose corresponding Dehn twists generate the mapping class group, is shown in figure \ref{fig:humphries}. Although finitely generated, it is not a free group: there are relations among the Dehn twist generators. These relations can be used to reduce the number of generators to 2 elements \cite{Wajnryb1996}, which consist of products of Dehn twists. 

\begin{figure}
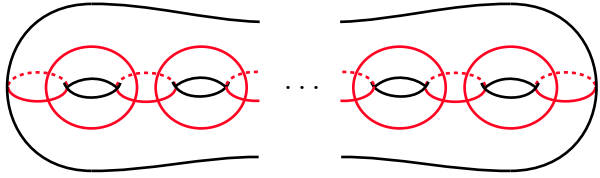

    \centering
    \tikzset{every picture/.style={line width=1pt}}

    \caption{For a closed Riemann surface of genus $g\geq 2$, the mapping class group $\text{MCG}(\Sigma_g)$ is generated by Dehn twists along the $2g+1$ cycles drawn in red.}
    \label{fig:humphries}
\end{figure}

The third important group associated to a handlebody is the so-called \emph{handlebody group} $\mathcal{H}_g$ of $M$. To describe it, consider the \emph{bulk} mapping class group $\text{MCG}(M)$, which consists of all large diffeomorphisms of $M$. Then we have the injection 
\begin{equation}
    \text{MCG}(M) \hookrightarrow \text{MCG}(\Sigma_g)
\end{equation}
by restricting the diffeomorphisms of $M$ to its boundary.
The image of this map is precisely the handlebody group. It is an infinite index (and non-normal) subgroup of the boundary MCG, which is rather complicated to describe explicitly \cite{Wajnryb1998}. An equivalent criterion for a mapping class $\gamma$ to be an element of $\mathcal{H}_g$ is that for every contractible cycle $C$, the image $\gamma(C)$ is also a contractible cycle \cite{Hensel2018APO}. Hence, the set of inequivalent handlebodies is parametrized by the coset 
\begin{equation}\label{eq:coset}
    \text{MCG}(\Sigma_g)/\mathcal{H}_g\,.
\end{equation} 

For genus 1, the above groups are easy to describe. The genus-1 handlebodies are the solid tori described by Maloney and Witten \cite{Maloney:2007ud} (consisting of thermal AdS$_3$ and an infinite family of Euclidean BTZ black holes), with fundamental group $\pi_1(M)=\mathbb{Z}$. The mapping class group of the boundary torus is $\text{MCG}(\Sigma_1) \cong \mathrm{PSL}(2,\mathbb{Z})$ and the handlebody group is
$
    \mathcal{H}_1 \cong \mathbb{Z}.
$
This follows from the fact that, at genus 1, there is only one contractible cycle $C$, hence any $\gamma \in \mathcal{H}_1$ that maps $C$ to itself is a power of a Dehn twist $T$ along $C$, i.e.~$\gamma = T^n$ for some $n\in \mathbb{Z}$. 

For genus $g\geq 2$, the handlebody group also contains powers of Dehn twists along contractible cycles, but the group is strictly larger than the subgroup generated by these simple twists. As an example of a distinct type of element in $\mathcal{H}_g$, consider any triple of closed curves $(\alpha,\beta,\delta)$ that bound a pair of pants. Then, if $\delta$ is a contractible cycle, the product of Dehn twists  $T_\alpha T_\beta^{-1}$ is in $\mathcal{H}_g$ \cite{Hensel2018APO}.

\subsection{Handlebodies as conformal blocks} Let us now discuss how handlebodies contribute to the path integral of 3d gravity. As explained in \cite{Collier:2023fwi}, 3d gravity with a negative cosmological constant on a hyperbolic 3-manifold is related to two copies of Virasoro TQFT (VTQFT). The gravitational partition function is just the product 
\begin{equation}\label{eq:grav_to_VTQFT}
    Z_\text{grav}[M;\Omega,\bar\Omega] = Z_{\text{Vir}}[M;\Omega]Z^*_{\text{Vir}}[M;\bar\Omega].
\end{equation}
Here $\Omega,\bar\Omega$ are complex structure moduli fixed at the conformal boundary. For a genus-$g$ handlebody $M^{\mathcal{C}}$ with conformal boundary $\Sigma_g$, the VTQFT amplitude is given by the vacuum Virasoro conformal block:
\begin{equation}\label{eq:vir_block}
    Z_{\text{Vir}}[M^{\mathcal{C}};\Omega] = \mathcal{F}_g^{\,\mathcal{C}}(\mathds{1};\Omega).
\end{equation}
Here $\mathcal{C}$ denotes a choice of contractible cycles (i.e.~an identity channel) that is part of the data specifying the handlebody $M^{\mathcal{C}}$. In general, the genus-$g$ Virasoro conformal blocks are defined as solutions of the conformal Ward identities \cite{Belavin:1984vu} but for $g>1$ and $c>1$ they are not known analytically (although recursive representations exist in special cases \cite{Zamolodchikov:1985ie,Hadasz:2009db,Cho:2017oxl}, using the plumbing construction). In \cite{Collier:2024mgv}, the relation between the gravitational path integral on a handlebody and the square of the identity block was verified, for $g=2$, in the semiclassical ($c\to \infty$) limit, including the one-loop determinant (in a perturbative expansion around a degenerating locus of moduli space). See also \cite{Yin:2007gv,Headrick:2015gba}. In this match, one uses the Brown-Hennaux relation \cite{Brown:1986nw}
\begin{equation}
    c \sim \frac{3\ell_{\text{AdS}}}{2G_N}
\end{equation}
between the gravitational coupling and the central charge. However, we stress that the identification between the gravitational path integral and the identity block \eqref{eq:vir_block} is exact in $c$ and does not rely on a semiclassical approximation. This is striking, because it shows that the gravitational answer is fully determined by the representation theory of the Virasoro algebra, which is universal to every 2d CFT. 

\paragraph{Action of the mapping class group.}
The fact that 3d gravity partition functions are given by representation-theoretic objects holds more generally: in VTQFT, the partition function on any  3-manifold with boundary $\Sigma_g$ can be expressed as a state in the Hilbert space of Virasoro conformal blocks. This Hilbert space is spanned by non-degenerate blocks, which have internal conformal weights $h_i \geq \frac{c-1}{24}$. The identity module is a degenerate representation of the Virasoro algebra with $c>1$, but the identity block can be expressed in the basis of non-degenerate blocks in a dual channel through the crossing kernels introduced in section \ref{sec:2}:
\begin{equation}\label{eq:identitykernel}
    \mathcal{F}_g^{\,\gamma\cdot\mathcal{C}}(\mathds{1};\Omega) =\int_{0}^\infty \d^{3g-3} P \,\,\mathbb{K}^\gamma_{\mathds{1}\boldsymbol{P}}\, \mathcal{F}_g^{\,\mathcal{C}}(\boldsymbol{P};\Omega).
\end{equation}
Here we introduced the convenient Liouville parametrization $
    h_i= \frac{c-1}{24} + P_i^2,
$ where the Jacobian factor is implicit in $\mathbb{K}^\gamma_{\id \boldsymbol{P}}$, and used the notation
$\boldsymbol{P} = (P_1,\dots, P_{3g-3})$. This corresponds to a non-normalizable state in the VTQFT Hilbert space. So the handlebody partition functions, labeled by the mapping classes $\gamma$, are particular (non-normalizable) states in the Hilbert space of Virasoro conformal blocks whose expansion coefficients are the crossing kernels $\mathbb{K}^\gamma$. Here the relevant kernels are the \emph{identity} kernels $\mathbb{K}^\gamma_{\mathds{1}\boldsymbol{P}}$, whose definition requires the subtraction of null states by hand to account for the fact that the vacuum module is a degenerate representation \cite{Eberhardt:2023mrq}. 

Through equation \eqref{eq:identitykernel}, these identity kernels implement the action of the mapping class group on the set of handlebodies \eqref{eq:coset}. We can find the action of a Dehn twist generator, twisting along some non-contractible cycle $C$, by crossing to the channel where $T_C$ is diagonalized (acting on the conformal block as multiplication by a phase), and then cross back to the original channel. For example, consider the genus-2 block, with $\mathcal{C}$ the dumbbell channel: 
\begin{equation}
    \mathcal{F}_{g=2}^{\,\mathcal{C}}(\boldsymbol{P},\Omega) =  \mathcal{F}_{g=2}^{\,\mathcal{C}}\left[\tikzset{every picture/.style={line width=0.8pt}} %set default line width to 0.75pt        
\right]\,.
\end{equation}
We have drawn the dual graph $\Gamma_\mathcal{C}$ in gray and labelled the edges by Liouville momenta $P_1,P_2,P_3$. Now choose $\gamma$ to be following product of Dehn twists, where we twist along the non-contractible cycles $(B_1,B_2,A)$ highlighted in red:
\begin{equation}\label{eq:dehntwistmappingclass}
    \gamma = T_{B_1}T_{B_2}T_{A} :\quad \tikzset{every picture/.style={line width=0.8pt}} %set default line width to 0.75pt        
%
\,.
\end{equation}
 One can convince oneself that $\gamma$ acts on the contractible cycles as shown in figure \ref{fig:crossing}. The corresponding kernel $\mathbb{K}^\gamma_{\mathds{1}\boldsymbol{P}}$ can be found by using $\mathbb{F}$ and $\mathbb{S}$ kernels to cross to the dual channels, where $T$ acts by a phase $\mathbb{T}_P = \e^{-2\pi i h_P}$, giving:
\begin{equation}\label{eq:dehntwistkernel}
    \mathbb{K}^\gamma_{\mathds{1}P_1P_2P_3} = (\mathbb{S}\mathbb{T}\mathbb{S})_{\mathds{1}P_1}(\mathbb{S}\mathbb{T}\mathbb{S})_{\mathds{1}P_2} (\mathbb{F}\mathbb{T}\mathbb{F})_{\mathds{1}P_3}\,.
\end{equation}
More explicitly, we have:
\begin{align}
    (\mathbb{F}\mathbb{T}\mathbb{F})_{\mathds{1}P_3} &= \int_0^\infty \d P \,\fker{\mathds{1}}{P}{P_2}{P_1}{P_1}{P_2}\,\mathbb{T}_P\,\fker{P}{P_3}{P_1}{P_1}{P_2}{P_2}, \\
    (\mathbb{S}\mathbb{T}\mathbb{S})_{\mathds{1}P_1} &= \int_0^\infty \d P\,\sker{\mathds{1}}{P}{\mathds{1}}\,\mathbb{T}_{P}\,\sker{P}{P_1}{\mathds{1}} = \sker{\mathds{1}}{P_1}{\mathds{1}}\,\mathbb{T}_{P_1}^*\,,
\end{align}
where in the second line we used the $\text{PSL}(2,\mathbb{Z})$ identity $ \mathbb{S}\mathbb{T}\mathbb{S} = \mathbb{T}^*\mathbb{S}\mathbb{T}^*$ and $\mathbb{T}_{\mathds{1}} = 1$.
This example shows how Dehn twists are represented by crossing kernels acting on identity blocks. The expressions in terms of Dehn twists can always be re-expressed in terms of the Moore-Seiberg generators $\mathbb{B},\mathbb{F},\mathbb{S}$, as we will see below. 

\paragraph{Action of the handlebody group.} 
We can ask when a mapping class, represented by $\mathbb{K}^\gamma$, is in the handlebody group $\mathcal{H}_g$.  On the level of conformal blocks, a handlebody group element leaves the vacuum block invariant up to a phase:
\begin{equation}\label{eq:H_group_invariance}
\mathcal{F}^{\,\gamma\cdot\mathcal{C}}_g(\mathds{1};\Omega) = \e^{i\theta_\gamma}\mathcal{F}^{\,\mathcal{C}}_g(\mathds{1};\Omega), \quad \gamma \in \mathcal{H}_g.
\end{equation}
The overall phase is a manifestation of the framing anomaly familiar from Chern Simons theory.
Since the gravitational partition function on the handlebody is the absolute value squared of the identity block (recall \eqref{eq:grav_to_VTQFT}), this phase factor drops out. Hence, any two  handlebodies that are diffeomorphic have the same gravitational partition function, as they should. 

On the level of the crossing kernels, the condition in equation \eqref{eq:H_group_invariance} can be translated using the block decomposition \eqref{eq:identitykernel} to the statement that for $\gamma \in \mathcal{H}_g$, the associated crossing kernel is the Dirac delta up to an overall $\bm{h}$-independent phase:
\begin{equation}
    \mathbb{K}^\gamma_{\mathds{1}\bm{h}} = \e^{i\theta_\gamma} \delta(\bm{h})\,, \quad \gamma \in \mathcal{H}_g.
\end{equation}
We can be more explicit in the genus-$1$ case. There, the vacuum block is simply the Virasoro vacuum character $\chi_{\mathds{1}}(\tau) = q^{-\frac{c-1}{24}}\frac{1}{\eta(\tau)}(1-q)$, and the handlebody group consists of the Dehn twists around the contractible cycle $\tau \to \tau+n$, where $n\in \mathbb{Z}$. These act on the vacuum character as 
multiplication by a phase:  
\begin{equation}
    \chi_{\mathds{1}}(\tau+1) = \e^{-2\pi i \frac{c}{24}}\chi_{\mathds{1}}(\tau),
\end{equation}
where the phase is just the conformal anomaly determined by the central charge $c$.

\subsection{Fixed-\texorpdfstring{$P$}{P} partition functions}
\label{sec:fixedP}
Instead of working with the `canonical' partition functions, with fixed values of the complex structure moduli at the asymptotic boundary, in Virasoro TQFT it is useful to define the corresponding `fixed-$P$' partition function $Z^{\,\mathcal{C}}_{\text{Vir}}[M;\boldsymbol{P}]$. We define it by stripping off the non-degenerate blocks and including a normalization factor:
\begin{equation}\label{eq:fixed_P}
    Z_{\text{Vir}}[M;\Omega] = \int_0^\infty \d^{3g-3} P \,\rho^{\,\mathcal{C}}_g(\boldsymbol{P}) \,Z^{\,\mathcal{C}}_{\text{Vir}}[M;\boldsymbol{P}] \,\mathcal{F}_g^{\,\mathcal{C}}(\boldsymbol{P};\Omega).
\end{equation}
This equation is analogous to the gluing of trumpets to fixed-length partition functions (i.e.~Weil-Petersson volumes) in JT-gravity \cite{Saad:2018bqo}.
Let us explain the notation used. 
First, the integral in \eqref{eq:fixed_P} is only over the heavy part of the spectrum, which we parametrized by the Liouville momenta $\boldsymbol{P}$ as before. In other words, we have expanded in a basis of non-degenerate conformal blocks in the channel $\mathcal{C}$. (This heavy channel should not be confused with the identity channel that defined the contractible cycles of the handlebodies discussed above.) 

Second, the normalization factor is defined as:
\begin{equation}
    \rho^{\,\mathcal{C}}_g(\boldsymbol{P}) = \prod_{e_a\in \Gamma_{\mathcal{C}}}\rho_0(P_a)\prod_{v_{ijk}\in\Gamma_{\mathcal{C}}}C_0(P_i,P_j,P_k).
\end{equation}
The product is over all edges and vertices of the dual graph $\Gamma_{\mathcal{C}}$, with known functions $\rho_0$ and $C_0$ that are related to the modular S-kernel and fusion kernel (see App.~\ref{app:technology}). This normalization appears in the inner product on the space of conformal blocks derived in \cite{Collier:2023fwi} and will be convenient later. Indeed, one may interpret \eqref{eq:fixed_P} as the resolution of the identity in the VTQFT Hilbert space associated to the genus-$g$ boundary, which implements the change of basis from fixed-moduli to fixed-$P$:
\begin{equation}
    \braket{\Omega|Z_\text{Vir}(M)} = \int_0^\infty \d^{3g-3}P\,  \rho^{\,\mathcal{C}}_g(\boldsymbol{P}) \braket{\mathcal{F}^{\,\mathcal{C}}_g(\boldsymbol{P})|Z_\text{Vir}(M)} \braket{\Omega|\mathcal{F}^{\,\mathcal{C}}_g(\boldsymbol{P})}\,.
\end{equation}
 
 Concretely, in the case that $M$ is a handlebody $M^\gamma$, labeled by some crossing transformation $\gamma$, the fixed-$P$ partition function can be read off from  \eqref{eq:identitykernel} to be:
\begin{equation}\label{eq:def_fixedP}
    Z_{\text{Vir}}^{\,\mathcal{C}}[M^{\gamma};\boldsymbol{P}] = \frac{1}{\rho^{\,\mathcal{C}}_g(\boldsymbol{P})} \,\mathbb{K}^\gamma_{\mathds{1}\boldsymbol{P}}\,.
\end{equation}
Plugging this into \eqref{eq:fixed_P}, we see that the corresponding identity channel is  
$
    \gamma\cdot \mathcal{C}.
$ We stress that the choice of $\mathcal{C}$ is arbitrary once we sum over all $\gamma$'s in section \ref{sec:3.3}. In practice, we will choose the basis $\mathcal{C}$ to be `simple’, in the sense that the dual graph $\Gamma_{\mathcal{C}}$ has the minimal number crossings. At genus 2 and 3, this will be one of the planar graphs in equations \eqref{eq:genus2graphs} and \eqref{eq:genus3graphs}.\footnote{However, at genus 4 and higher there are contraction patterns that cannot be made planar: an example is the contraction in equation (3.46) of \cite{Collier:2024mgv}. That is why we define $\mathcal{C}$ by requiring the minimal number of crossings of $\Gamma_{\mathcal{C}}$. These simple bases are in 1-1 correspondence with the distinct contraction patterns of OPE coefficients introduced in section \ref{sec:2}. } With respect to this simple choice of basis $\mathcal{C}$, the complexity of the handlebody $M^\gamma$ is visualized by a VTQFT diagram, which represents the \emph{embedding} of $\Gamma_{\mathcal{C}}$ in a closed 3-manifold, as we will explain below. 

Again, an example will clarify the above formalism. At genus 2, consider the crossing transformation $\gamma = \mathbb{S}\circ \mathbb{S}\circ \mathbb{F}$, where one S-kernel acts on the left dumbbell and the other on the right.  
The corresponding dual graph $\Gamma_{\mathcal{C}}$ is the sunset diagram. Using the definition \eqref{eq:def_fixedP}, its fixed-$P$ partition function evaluates to: 
\begin{align}\label{eq:theta graph ZVir}
    Z_{\rm Vir}\!\left[\!\vcenter{\hbox{
    \begin{tikzpicture}[yscale=-0.7,xscale=0.7]
        \draw[very thick, wilsonred] (0,0) ellipse (1.5 and 1);
        \draw[very thick, wilsonred] (0,1) -- (0,-1);
        \node at (-1.15,0) {\footnotesize $P_1$};
        \node at (0.31,0) {\footnotesize $P_2$};
        \node at (1.15,0) {\footnotesize $P_3$};
    \end{tikzpicture}
    }}\!\right] &= \frac{\mathbb{S}_{\mathds{1}P_1}[\mathds{1}]\,\mathbb{S}_{\mathds{1}P_3}[\mathds{1}]\,\fker{\mathds{1}}{P_2}{P_3}{P_1}{P_1}{P_3}}{\rho_0(P_1)\rho_0(P_2)\rho_0(P_3)C_0(P_1,P_2,P_3)C_0(P_1,P_2,P_3)} \\
    &= \frac{1}{C_0(P_1,P_2,P_3)}\, .
\end{align}
In the second line, we used that $\rho_0(P) = \mathbb{S}_{\mathds{1}P}[\mathds{1}]$, as well as the relation:
\begin{equation}
    \fker{\mathds{1}}{P_2}{P_3}{P_1}{P_1}{P_3} = \rho_0(P_2)C_0(P_1,P_2,P_3).
\end{equation}
Now we can find the corresponding identity channel by acting with $\gamma$ on the sunset diagram. Doing so, we see that the identity channel $\gamma\cdot\mathcal{C}$ is the dumbbell channel:
\begin{multline}
    \int_0^\infty \d^3 P\, \sker{\mathds{1}}{P_1}{\mathds{1}}\,\sker{\mathds{1}}{P_3}{\mathds{1}} \,\fker{\mathds{1}}{P_2}{P_3}{P_1}{P_1}{P_3}\mathcal{F}^{\,\mathcal{C}}_{g=2}\!\left[\!\!\vcenter{\hbox{
}}\!\!\right]\,,
\end{multline}
where the `bells' (sub-tori) are in the S-dual channel. The contribution of this handlebody to the statistical moment is just the universal density studied in \cite{Chandra:2022bqq}:
\begin{equation}\label{eq:C0 contribution}
    \overline{\rho_1\rho_2\rho_3C_{123}C_{123}^*} \supset |\rho_0(h_1)\rho(h_2)\rho_0(h_3)C_0(h_1,h_2,h_3)|^2.
\end{equation}
As before, we used the notation $\supset$ to indicate that the left-hand side is given by a full sum over topologies, and the right-hand side is only one term in that sum. We also used the notation $|\cdot |^2$ as a shorthand for the product of left- and right-moving contributions, as in equation \eqref{eq:grav_to_VTQFT}. Furthermore, we used the shorthand $\rho_i \equiv \rho(h_i,\bar h_i)$ and changed variables from Liouville momenta back to conformal weights $h_i$.

As a second slightly more involved example, consider 
the crossing transformation $\gamma$ that was illustrated in \eqref{eq:sequence1}. Additionally, we add two S-moves acting on the bells of the dumbbell. So the total mapping class is:
\begin{equation}\label{eq:crossingtrans}
    \gamma = \mathbb{S}\circ \mathbb{S}\circ \mathbb{F}\circ \mathbb{B}^2\circ \mathbb{F} \ ,
\end{equation}
where the two S-kernels act on the two dumbbells and $\mathbb{F} \circ \mathbb{B}^2 \circ \mathbb{F}$ only acts on the centrally embedded 4-punctured sphere.
The associated VTQFT graph is the linked dumbbell channel, where the two bells have been linked as in \eqref{eq:sequence1}. We can find the fixed-$P$ partition function associated to this handlebody by representing \eqref{eq:crossingtrans} by its genus-2 crossing kernel:
\begin{equation}
    \mathbb{K}^\gamma_{\mathds{1}P_1P_2P_3} = \sker{\mathds{1}}{P_1}{\mathds{1}}\,\sker{\mathds{1}}{P_2}{\mathds{1}}\int_0^\infty \d P \,\fker{\mathds{1}}{P}{P_2}{P_1}{P_1}{P_2}\,\left(\mathbb{B}^P_{P_1P_2}\right)^2\,\fker{P}{P_3}{P_1}{P_1}{P_2}{P_2}\,.
\end{equation}
Using the expression for the braiding phase \eqref{eq:appbraid}, we note that this kernel is precisely equal to the Dehn twist representation \eqref{eq:dehntwistkernel}, and indeed $\gamma$ defines the same mapping class as \eqref{eq:dehntwistmappingclass}. Furthermore,
we note that the $P$-integral can be expressed in terms of the one-point S-kernel by using the integral identity \eqref{eq:Shat and 6j}. Dividing by the normalization factor $\rho_{g=2}^{\mathcal{C}}$, we thus obtain:
\begin{align}
Z_{\text{Vir}}\left[
    \tikzset{every picture/.style={line width=1.2pt}} 
\right] &= \frac{\mathbb{K}^\gamma_{\mathds{1}P_1P_2P_3}}{\rho_0(P_1)\rho_0(P_2)\rho_0(P_3)C_0(P_1,P_1,P_3)C_0(P_2,P_2,P_3)} \\ 
&= \frac{\sker{P_1}{P_2}{P_3}}{\rho_0(P_2)C_0(P_2,P_2,P_3)} = \frac{1}{\sqrt{\mathsf{C}_{113}\mathsf{C}_{223}}} \,\skerhat{P_1}{P_2}{P_3}\,.\label{eq:Z_linkedhandcuff} 
\end{align}
In the last equality we rewrote the fixed-$P$ partition function in terms of the symmetric S-kernel, cf.~\eqref{eq:Shat_def}, in the so-called Racah-Wigner normalization \cite{Post:2024itb}. The prefactors are useful shorthands for the $C_0$ function:
\begin{equation}
    \mathsf{C}_{ijk} \equiv C_0(P_i,P_j,P_k)\, .
\end{equation}
In this example, $\mathcal{C}$ is the dumbbell channel, and we have represented the non-trivial topology of the handlebody $M^\gamma$ by the VTQFT diagram where the bells are linked. To find the corresponding identity channel $\gamma\cdot\mathcal{C}$, we follow the sequence of moves in \eqref{eq:sequence1} and apply the additional S-moves. The resulting identity channel is also the linked dumbbell diagram:
\begin{equation}\label{eq:identitylinkeddumbbell}
    \int_0^\infty \d^3 P \,\mathbb{K}^\gamma_{\mathds{1}P_1P_2P_3}\,\mathcal{F}_{g=2}^{\,\mathcal{C}}\left[\tikzset{every picture/.style={line width=1.2pt}} %set default line width to 0.75pt        
\!\right]\,.
\end{equation}
Here we used the prime $'$ to indicate that the bells are now in the S-dual channel.

The linked dumbbell in \eqref{eq:Z_linkedhandcuff} has a counterpart where the over-under crossings of the bells are reversed \cite{Hartman:2025cyj}. Their VTQFT partition functions differ by a complex conjugation, which leads to a phase $\skerhat{P_1}{P_2}{P_3}^* = \e^{-i\pi h_3}\skerhat{P_1}{P_2}{P_3}$. The sum of these two distinct handlebodies projects onto even spin $J_3$, compatible with the permutation property \eqref{eq:OPEpermutation}, and contributes as the following non-Gaussian moment:
\begin{equation}\label{eq:cherryOPE}
\overline{\rho_1\rho_2\rho_3C_{113}C_{223}} \supset 
(1+(-1)^{J_3}) \left| \rho_0(h_1)\rho_0(h_2)\rho_0(h_3)
\sqrt{\mathsf{C}_{113}\mathsf{C}_{223}}\,
\skerhat{h_1}{h_2}{h_3}
\right|^2. 
\end{equation}
As an aside, we note that the linked dumbbell diagram is also known as the `knotted handcuff' in the literature \cite{Hartman:2025cyj,Hartman:2025ula}, and it is often depicted as such: 
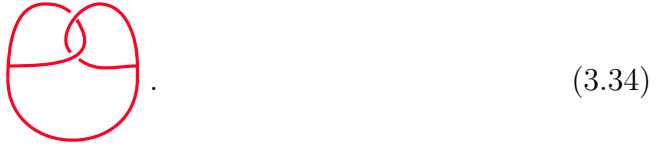
\begin{equation}
\label{eq:cherry}
\vcenter{
\hbox{

}
}.
\end{equation}

As a third and last example of a more complicated handlebody, we study a diagram that is in the mapping class group orbit of the sunset diagram. We can obtain it from the linked dumbbell by acting with a braiding and a fusion move:
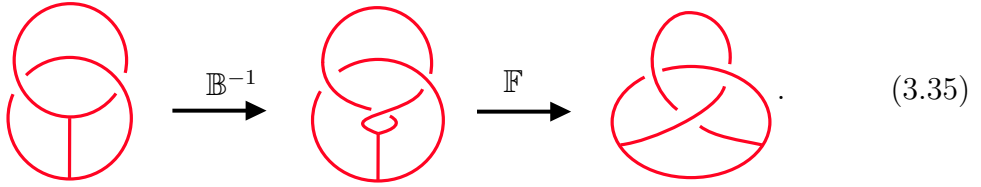
\begin{equation}
    \tikzset{every picture/.style={line width=1.2pt}} %set default line width to 0.75pt        
%
\,.\vspace{1mm}
\end{equation}
The right-hand side indeed has the same graph type as the sunset in \eqref{eq:theta graph ZVir}, but one of the edges `loops' around one of the other edges. We can also view this diagram as a particular fragmentation of the trefoil knot, in the sense of \cite{Chandra:2025fef}. The associated genus-2 handlebody has a fixed-$P$ partition function that we compute as before: first, we write down the crossing kernel $\mathbb{K}^\gamma$ that represents $\gamma = \mathbb{S}\circ \mathbb{S}\circ \mathbb{F}\circ \mathbb{B}^2\circ \mathbb{F} \circ \mathbb{B}^{-1} \circ \mathbb{F}$, and then we divide by the sunset normalization factor. This gives:
\begin{equation}
    Z_\text{Vir}\!\left[\tikzset{every picture/.style={line width=1.2pt}} %set default line width to 0.75pt        
\begin{tikzpicture}[x=0.75pt,y=0.75pt,yscale=-1.3,xscale=1.3,baseline={([yshift=-0.5ex]current bounding box.center)}]
%Shape: Arc [id:dp2979398283370033] 
\draw  [draw opacity=0] (350.19,107.95) .. controls (353.46,107.13) and (357,106.68) .. (360.69,106.68) .. controls (377.37,106.68) and (390.89,115.78) .. (390.89,127.01) .. controls (390.89,138.23) and (377.37,147.33) .. (360.69,147.33) .. controls (344.02,147.33) and (330.5,138.23) .. (330.5,127.01) .. controls (330.5,120.07) and (335.66,113.95) .. (343.54,110.28) -- (360.69,127.01) -- cycle ; \draw  [color=wilsonred  ,draw opacity=1 ] (350.19,107.95) .. controls (353.46,107.13) and (357,106.68) .. (360.69,106.68) .. controls (377.37,106.68) and (390.89,115.78) .. (390.89,127.01) .. controls (390.89,138.23) and (377.37,147.33) .. (360.69,147.33) .. controls (344.02,147.33) and (330.5,138.23) .. (330.5,127.01) .. controls (330.5,120.07) and (335.66,113.95) .. (343.54,110.28) ;  
%Curve Lines [id:da17480498866159244] 
\draw [color=wilsonred  ,draw opacity=1 ]   (333.68,135.35) .. controls (345.72,134.07) and (371.15,122.94) .. (373.49,113.96) ;
%Curve Lines [id:da5494768527268993] 
\draw [color=wilsonred  ,draw opacity=1 ]   (355.43,122.52) .. controls (337.69,108.82) and (347.73,87) .. (359.77,87) .. controls (371.82,87) and (376.84,97.7) .. (375.5,105.83) ;
%Curve Lines [id:da7380808724993313] 
\draw [color=wilsonred  ,draw opacity=1 ]   (364.12,128.51) .. controls (369.42,132.43) and (379.51,133.64) .. (387.88,135.35) ;
% Text Node
\draw (318.89,110.07) node [anchor=north west][inner sep=0.75pt]  [font=\footnotesize]  {$P_{1}$};
% Text Node
\draw (354.56,134.73) node [anchor=north west][inner sep=0.75pt]  [font=\footnotesize]  {$P_{3}$};
% Text Node
\draw (377.89,88.73) node [anchor=north west][inner sep=0.75pt]  [font=\footnotesize]  {$P_{2}$};
\end{tikzpicture}\,\,\,\right] = \frac{1}{\mathsf{C}_{123}}\int_0^\infty \d P \, \rho_0(P)\,\skerhat{P_1}{P_2}{P}\,\mathbb{B}_{P}^{P_2P_2} \sixj{P_1}{P_2}{P_3}{P_2}{P_1}{P} .
\end{equation}
Here $\mathbb{B}_P^{P_2P_2}$ is the complex conjugate of the braiding phase $\mathbb{B}^P_{P_2P_2}$, and we have expressed the fusion kernel in terms of the Virasoro $6j$ symbol using the relation \eqref{eq: relation btw fusion and modular kernels}. In this example, $\mathcal{C}$ is the sunset channel, and acting on it with $\gamma$ one can show that the corresponding identity channel $\gamma\cdot\mathcal{C}$ is again a linked dumbbell as in \eqref{eq:identitylinkeddumbbell}, with the only difference that now one bell is in the S-dual channel and one bell is acted on by $\mathbb{S}\mathbb{T}$. Multiplying by the normalization factor and taking the absolute value square as in \eqref{eq:grav_to_VTQFT}, this handlebody is also a contribution to the statistical moment $\overline{\rho_1\rho_2\rho_3C_{123}C_{123}^*}$.

From the above examples, we see that the fixed-$P$ VTQFT partition functions are fully expressed in terms of crossing kernels only. We stress that, contrary to conformal blocks, these kernels are known analytically as integrals over known special functions \cite{Eberhardt:2023mrq}. Although the integrals may be hard to compute in general, they could in principle be evaluated numerically, for any value of $c$. What's more, since we stripped off the conformal blocks, the fixed-$P$ partition functions are what actually contribute to OPE statistics.

\paragraph{Geometric interpretation of the fixed-$P$ partition function.} There are several ways to interpret the VTQFT diagrams that appear in the fixed-$P$ partition functions calculated above. First, one can view the fixed-$P$ VTQFT graph as being embedded in the 3-sphere $\mathrm{S}^3$, and obtain a 3-manifold with boundary by taking the \emph{complement} of a regular neighborhood of the graph. Here is a pictorial example:
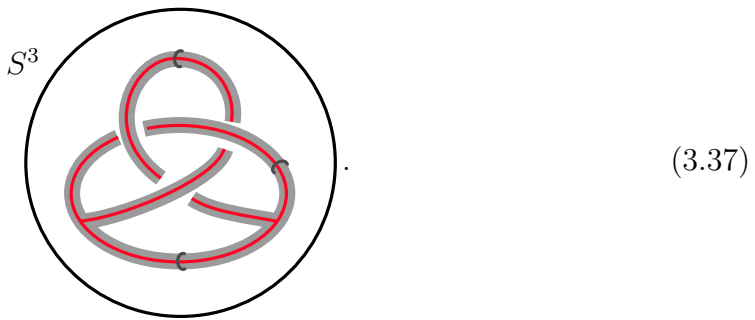
\begin{equation}\label{eq:loopedtheta}
    \tikzset{every picture/.style={line width=1.2pt}} %set default line width to 0.75pt        
\,.
\end{equation}
This is similar to the construction of a knot complement, but instead of embedding a link (thickened into a union of solid tori), we have embedded a trivalent graph (i.e.~a network of Wilson lines). If this graph is connected, the boundary of the graph complement is a genus-$g$ surface. The cycles in $\mathcal{C}$ are now non-contractible cycles in the ambient 3-manifold, for complicated enough $\gamma$. 

More generally, instead of $\mathrm{S}^3$ we could use any other closed 3-manifold as our embedding manifold. For example, our notation of the prime in \eqref{eq:identitylinkeddumbbell} to indicate an S-transform should really be thought of as an embedding of the dumbbell graph in $\mathrm{S}^2\times \mathrm{S}^1$ (or connected sums thereof, for multiple S-transforms). We will come back to this point when we discuss Dehn surgery in section \ref{sec:3.3}. 

Geometrically, viewing the fixed-$P$ diagrams as graph complements can be understood using the theory of \emph{pleated surfaces}. Namely, for a handlebody $M^\gamma$ the complete hyperbolic metric is always of infinite volume, and `flares out' to the conformal boundary \cite{Schlenker:2022dyo}. $M^\gamma$ does not support any smooth finite-volume hyperbolic metrics with totally geodesic boundary. However, we can put a finite-volume hyperbolic metric on $M^\gamma$ if we allow the geometry to end on a boundary surface that is pleated. This means that the boundary surface has vanishing extrinsic curvature everywhere except along a collection of geodesics, where the surface is allowed to have corners. As explained in \cite{Hartman:2025ula}, the fixed-$P$ partition function can be understood geometrically as the AdS$_3$ gravitational path integral with precisely such a boundary condition, in which the fixed Liouville momenta $P_i$ are related to the lengths $\ell_i$ of the geodesics around which the pleated surface bends by the formula
$
    \ell_i = 4\pi b P_i.
$\footnote{Here $b$ is the Liouville coupling, which parametrizes the central charge as $c = 1+ 6(b+1/b)^2$.}

We remark that such a geometric interpretation does not always exist, for a number of `simple’ handlebodies. This happens if a closed curve $C_i$ dual to an edge $P_i$ bounds a compressible disk in the graph complement. An example is given by the complement of the dumbbell graph, in which the middle cycle can be contracted in the ambient $\mathrm{S}^3$:
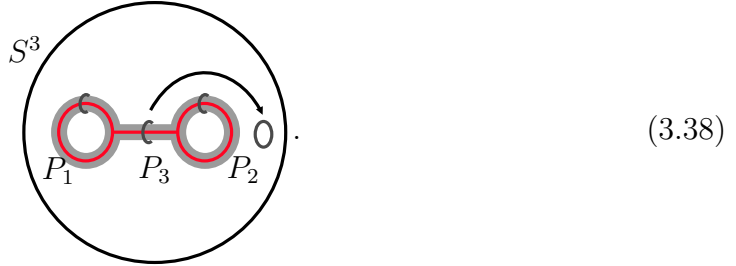
\begin{equation}\label{eq:contractible dumbbell}
\tikzset{every picture/.style={line width=1.2pt}} %set default line width to 0.75pt        
\,.
\end{equation}
This is reflected by the fact that the VTQFT partition function has support only on the identity for the weight $P_3$:\footnote{Here $\mathds{1}$ means $P_3 = \frac{iQ}{2}$, which is equivalent to the real delta function setting $h_3 = 0$.} 
\begin{equation}
Z_{\text{Vir}}\!\left[\tikzset{every picture/.style={line width=1.2pt}} %set default line width to 0.75pt        
%
\right] = \frac{\delta(P_3 - \mathds{1})}{\rho_0(P_3)C_0(P_1,P_1,P_3)C_0(P_2,P_2,P_3)}\,.
\end{equation}
Since this is a distribution, it clearly does not have an interpretation as a classical saddlepoint to a fixed-length path integral. For a more complicated handlebody, like the \emph{linked} dumbbell \eqref{eq:Z_linkedhandcuff}, the $P_3$ cycle is non-contractible, and indeed in that case the fixed-length path integral exists and was analyzed directly in gravity in \cite{Hartman:2025ula}. More generally, if the mapping class $\gamma$ that defines the handlebody is complicated enough (technically, a sufficient condition is that $\gamma$ is ``pseudo-Anosov'' \cite{Thurston:1988}), then the fixed-length path integral exists, and the above geometric understanding of the fixed-$P$ partition function makes sense.

\paragraph{Fixed-$P$ partition functions, wormholes, and OPE statistics.}
The fixed-$P$ partition functions that have been the subject of this subsection straightforwardly encode the moments of structure constants in the dual statistical ensemble of CFT data.
The precise relation is made transparent by noting that there is a second geometric interpretation of the VTQFT Wilson line diagrams: as Euclidean wormhole geometries with boundaries that are thrice-punctured spheres. That is, instead of excising a regular neighborhood of the full graph embedded in $\mathrm{S}^3$, we only excise a spherical neighborhood around each vertex. In replacing trivalent vertices by 3-punctured spheres, we normalize each boundary component with a factor of $C_0$:
\begin{equation}\label{eq:vertex replacement rule}
    \vcenter{\hbox{
        \begin{tikzpicture}[scale=0.75]
            \draw[very thick,wilsonred] (0,0) to (0,1);
            \draw[very thick,wilsonred] (0,0) to (-.866,-.5);
            \draw[very thick,wilsonred] (0,0) to (.866,-0.5);
            \node[above] at (0,1) {$P_1$};
            \node[left] at (-.866,-.7) {$P_2$};
            \node[right] at (.866,-0.7) {$P_3$};
        \end{tikzpicture}
    }}
    = \,\,\frac{1}{\mathsf{C}_{123}}
    \vcenter{\hbox{
    \begin{tikzpicture}[scale=0.75]
        \draw[very thick] (0,0) circle (3/4);
        \draw[very thick] (-3/4,0) to[out=-90,in=-90,looseness=.8] (3/4,0);
        \draw[thick,densely dashed] (-3/4,0) to[out=90,in=90,looseness=.8] (3/4,0);
        \draw[very thick,wilsonred] (0,1/2) to (0,1);
        \draw[very thick, wilsonred] (-.433,-.25) to (-.866,-.5); 
        \draw[very thick, wilsonred] (.433,-.25) to (.866,-.5); 
        \node[above] at (0,1) {$P_1$};
        \node[left] at (-.866,-.7) {$P_2$};
        \node[right] at (.866,-0.7) {$P_3$};
    \end{tikzpicture}
    }}\, .
\end{equation} 
This turns our trivalent graph into a Euclidean wormhole with 3-punctured
sphere boundaries and a (possibly knotted) network of Wilson lines in the bulk, where the bulk Wilson lines have fixed holonomies parametrized by $P_i$. 

When the conformal weight $h_i$ is in the conical defect regime (defined e.g. in \cite{Benjamin:2020mfz}), the Wilson line has a semiclassical interpretation as creating a conical defect in the bulk geometry. As an example, we see that the sunset graph is equivalent to the Maldacena-Maoz wormhole:
\begin{equation}\label{eq:theta graph MM wormhole}
    \vcenter{\hbox{

    }}\, .
\end{equation}
This was also the perspective taken in section 3.3 of \cite{Collier:2024mgv}. The two perspectives on the fixed-$P$  partition functions are both useful to keep in mind. They interpolate between each other by taking the conformal weights of the Wilson lines above the black hole threshold, in which case the boundary of the graph complement becomes a connected Riemann surface \cite{Chandra:2022bqq}. We then effectively `trade’ the Wilson lines for tunnels.\footnote{See also \cite{Abajian:2023bqv} for a geometric description of the Euclidean wormholes in the heavy regime.} Likewise, the geometric interpretation of the conformal blocks that we integrate against is that of `trumpets’, or flares, which trade the fixed momentum labels $\boldsymbol{P}$ for complex structure moduli $\Omega$ at the conformal boundary.

Thus in order to extract the statistics of the structure constants in the boundary ensemble it is sufficient to consider trivalent networks of Wilson lines in VTQFT. While these are most naturally associated with contributions to the gravitational path integral with a single genus-$g$ boundary, they are also straightforwardly related to multi-boundary sphere 3-point wormholes which more directly encode the OPE statistics without the additional baggage of conformal blocks. 
To extract the contributions to the OPE statistics from the VTQFT Wilson line configurations we simply replace the trivalent vertices with 3-punctured spheres via the replacement rule (\ref{eq:vertex replacement rule}) and square the associated VTQFT partition function. This gives:
\begin{equation}\label{eq:ZVirOPE}
    \overline{\prod_{e_i\in \Gamma_{\mathcal{C}}}\!\!\rho(P_i,\bar P_i)\!\!\!\prod_{v_{ijk}\in \Gamma_{\mathcal{C}}}\!\!\!C_{ijk}} \,\,\,\supset \,\,\,\Big|\,\rho_g^{\mathcal{C}}(\boldsymbol{P})\, Z^{\,\mathcal{C}}_{\rm Vir}[M^{\gamma};\boldsymbol{P}]\,\Big|^2\, .
\end{equation}
 Crucially, the RHS of \eqref{eq:ZVirOPE} is generically a non-Gaussian contribution to the statistical moment on the LHS, which does not factorize into a product of variances.

\paragraph{How to tell a handlebody from a non-handlebody.}

 The fixed-$P$ formulation of Virasoro TQFT is useful for computations, because we have all of the tools of 3d topological quantum field theory at our disposal to evaluate the partition function. A list of useful calculation rules is collected in appendix \ref{app:technology}. However, one drawback of this formalism is that, given some random knotted trivalent graph,  it is in general hard to distinguish a handlebody from a non-handlebody. This problem comes down to finding a Moore-Seiberg move $\gamma$ such that the diagram can be untangled to a simple  diagram.\footnote{Another equivalent way of saying this is that in order to show that $M$ is a handlebody, one has to show that the VTQFT graph and its complement form a Heegaard splitting of $\mathrm{S}^3$.} This $\gamma$ could be a very long word in the generators $\mathbb{S}, \mathbb{F}$ and $\mathbb{B}$. For instance, in the example below, the left diagram corresponds to a handlebody and the right to a non-handlebody: 
\begin{equation}
\vcenter{\hbox{

}}.
\end{equation}
The left is just the looped version of the sunset diagram \eqref{eq:loopedtheta} that we studied above. The right, on the other hand, corresponds to a known non-handlebody that we will study in section \ref{sec:5}, called $\mathbf{5}_2$ in our table \ref{table:handlebodyknots}. The two diagrams only differ by one under-over crossing. Moreover, 
both diagrams result from a fragmentation of the trefoil knot \cite{Chandra:2025fef}: on the left, the edge $3$ fragments the trefoil, and on the right the edge $1$ fragments the (mirror of the) trefoil. 

This example illustrates that the fragmentation procedure of \cite{Chandra:2025fef} obscures the handlebodiness of a given 3-manifold in VTQFT. This is not a problem if one’s perspective is to sum over \emph{all} 3-manifolds; but from the perspective of crossing symmetry, we want to be able to distinguish contributions from the sum over identity blocks from those that cannot be written as a mapping class group image of an identity block. 
Fortunately, for trivalent graphs embedded in $\mathrm{S}^3$, there is a simple test for handlebodiness, which we explain in appendix \ref{subapp:tricolorability}. A second method of determining handlebodiness is to study the fundamental group using Wirtinger's presentation, and running Whitehead's algorithm as explained in appendix \ref{subapp:fundamental group}.

\subsection{The handlebody sum is crossing symmetric}\label{sec:3.3}

Now that we have reviewed the topology of handlebodies and their description in Virasoro TQFT, we will study the sum over handlebodies and its implications for the statistics of OPE coefficients. This handlebody sum is the simplest sum over a subset of topologies that is compatible with crossing symmetry.

Crossing symmetry is most clear on the level of the canonical partition function. Namely, by the previous section, the handlebody sum is just a sum over mapping class group images of some `seed’ vacuum block:\vspace{1mm} \begin{equation}\label{eq:handlebodysum3}
        \overline{Z[\Sigma_{g};\Omega,\bar\Omega]} \quad \supset \,\sum_{\gamma \in \text{MCG}(\Sigma_g)/\mathcal{H}_g} \left |  \mathcal{F}_g^{\,\gamma\cdot\mathcal{C}_0}(\mathds{1};\Omega) \right |^2\,.
    \end{equation}
This is clearly invariant under the action of the mapping class group $\Omega \mapsto \gamma’\cdot\Omega$, $\bar\Omega \mapsto \gamma’\cdot\bar\Omega$, simply by relabelling the sum $\gamma \circ \gamma’ \to \gamma$. The choice of seed $\mathcal{C}_0$ is irrelevant, as the mapping class group coset \eqref{eq:coset} acts transitively on the full set of handlebodies. Our convention is to pick $\mathcal{C}_0$ to be the genus-$g$ generalization of the dumbbell channel (called the `necklace' in \cite{Belin:2021ryy}):
\begin{equation}\label{eq:choiceofseed}
\mathcal{C}_0 = \vcenter{\hbox{
\begin{tikzpicture}
    \draw[very thick, wilsonred] (-2,0) circle (0.5);
    \draw[very thick, wilsonred] (-1.5,0) -- (-1,0);
    \draw[very thick, wilsonred] (-0.5,0) circle (0.5);
    \draw[very thick, wilsonred] (0,0) -- (0.15,0);
    \draw[very thick, wilsonred, dotted] (0.15,0)--(0.8,0);
    \draw[very thick, wilsonred] (0.8,0)--(1,0);
    \draw[very thick, wilsonred] (1.5,0) circle (0.5);
\end{tikzpicture}
}}.
\end{equation}

Each vacuum block in the sum \eqref{eq:handlebodysum3} is the leading saddle (in the semiclassical limit) in some region of moduli space. This is because the classical on-shell action is proportional to the renormalized volume of the handlebody, which goes to $-\infty$ in the limit where we pinch a compressible disk \cite{Schlenker:2022dyo} (which can be thought of as a high-temperature limit in a corner of moduli space). Thus, the handlebody sum gives the leading large-$c$ contribution to the average $\overline{Z[\Sigma_g]}$.

Next, let us translate the handlebody sum \eqref{eq:handlebodysum3} to the level of OPE coefficients.  We do so by expanding the LHS of \eqref{eq:handlebodysum3} in conformal blocks in some channel $\mathcal{C}$, and using the identity crossing kernels to rewrite the RHS in the same channel. We can then strip off the conformal blocks, so that we are left with:
\begin{equation}\label{eq:handlebody sum statistics}
    \overline{\prod_{e_i\in \Gamma_{\mathcal{C}}}\!\!\rho(h_i,\bar h_i)\!\!\!\prod_{v_{ijk}\in \Gamma_{\mathcal{C}}}\!\!\!C_{ijk}} \,\,\,\supset \sum_{\gamma\in \text{MCG}(\Sigma_g)/\mathcal{H}_g} \big |\mathbb{K}^{\,\gamma\cdot\gamma_\mathcal{C}}_{\mathds{1}\bm{h}} \big |^2\,.
\end{equation}
Here we have defined $\gamma_\mathcal{C}$ to be the crossing transformation that maps 
 $\gamma_\mathcal{C} \cdot \mathcal{C} = \mathcal{C}_0$, where $\mathcal{C}_0$ was our choice of seed in \eqref{eq:choiceofseed}. Its corresponding crossing kernel can be written (in terms of the Liouville momenta) as:
 \begin{equation}
     \mathbb{K}^{\,\gamma\cdot\gamma_\mathcal{C}}_{\mathds{1}\boldsymbol{P}} = \int_0^\infty \d^{3g-3} P' \,\mathbb{K}^{\,\gamma}_{\mathds{1}\boldsymbol{P}'}\, \mathbb{K}^{\gamma_\mathcal{C}}_{\boldsymbol{P'P}}\,.
 \end{equation}
 The crossing transformation $\gamma_\mathcal{C}$ parametrizes the different mapping class group orbits that arise in \eqref{eq:handlebody sum statistics}: for each choice of channel $\mathcal{C}$ in which we decide to expand the partition function, there is a corresponding orbit $\{\gamma\cdot\gamma_\mathcal{C}\}_{\gamma\in \text{MCG}(\Sigma_g)/\mathcal{H}_g}$. There are only finitely many orbits,\footnote{Technically, there are finitely many diffeomorphism classes of markings of $\Sigma_g$. One way to see this is to note that any given $\mathcal{C}$ is related by finitely many MS moves to a channel where all contractible cycles are non-separating, and then invoke the fact that the mapping class group acts transitively on the set of non-separating simple closed curves \cite{HatcherThurston1980}.} which are in one-to-one correspondence with the patterns of OPE index contractions that we discussed in section \ref{sec:2}. We highlight the distinction between $\gamma$ and $\gamma_{\mathcal{C}}$ because a mapping class $\gamma$ does not change the topology of the graph $\Gamma_{\mathcal{C}}$ (instead,  what changes is the embedding of $\Gamma_{\mathcal{C}}$ in the Riemann surface). On the other hand, $\gamma_{\mathcal{C}}$ can be any crossing transformation (in the Moore-Seiberg groupoid), which may change the graph topology of $\Gamma_{\mathcal{C}}$. 

\begin{figure}
    \centering
    $Z_\text{Vir}\Bigg[$\!\tikzset{every picture/.style={line width=1.2pt}} %set default line width to 0.75pt        
$\,\Bigg]$
    \caption{Two VTQFT diagrams that describe the same handlebody: one is in the sunset basis (left), the other in the dumbbell basis (right).}
    \label{fig:twopartitionfunctions}
\end{figure}

An example at genus 2 will help to understand this point. In the dumbbell channel, the handlebody sum \eqref{eq:handlebody sum statistics} should include the contribution from the mapping class $\gamma = \mathbb{S}\circ \mathbb{S}\circ \mathbb{F}\circ \mathbb{B}^2 \circ \mathbb{F}$,
whose VTQFT diagram is the linked dumbbell shown in figure \ref{fig:twopartitionfunctions} (right). This is indeed a MCG transformation, and $\gamma_\mathcal{C} = 1$. In the sunset channel, on the other hand, we have $\gamma_\mathcal{C} = \mathbb{F}$, so the same handlebody is now represented by $\gamma\circ \gamma_\mathcal{C} = \mathbb{S}\circ \mathbb{S}\circ \mathbb{F}\circ \mathbb{B}^2$, where we used that $\mathbb{F}^2 =1$. The corresponding VTQFT diagram is shown in figure \ref{fig:twopartitionfunctions} (left). The two representations of the same handlebody indeed lead to the same identity block:
\begin{multline}
  \int_0^\infty \d^3 P \,\sker{\mathds{1}}{P_1}{\mathds{1}}\sker{\mathds{1}}{P_2}{\mathds{1}}\fker{\mathds{1}}{P_3}{P_2}{P_1}{P_1}{P_2}\big(\mathbb{B}^{P_3}_{P_1P_2}\big)^2\mathcal{F}_{g=2}\!\left[\!\vcenter{\hbox{
\!\right] .
\end{multline}
This reproduces the answer from our previous calculation \eqref{eq:identitylinkeddumbbell} in the dumbbell channel. The two representations of the same handlebody are related by an overall fusion move, which we can think of as a change of basis.

\paragraph{Spin quantization.} One immediate corollary of 
equation \eqref{eq:handlebody sum statistics}, which expresses a statistical moment of CFT data in terms of a mapping class group orbit, is that the spins $J_i = h_i-\bar h_i$ are manifestly quantized. To see this, note that the sum over the mapping class group contains a sub-sum where we precompose with powers of Dehn twists $T_i^{n_i}$ acting on the cycles in the channel $\mathcal{C}$. Writing $\gamma = \tilde\gamma \circ T^{n_1}\circ \dots \circ T^{n_{3g-3}}$, the corresponding partial sum over crossing kernels takes the form:
\begin{equation}
    \sum_{\bm{n}\in \mathbb{Z}^{3g-3}} \Big |\mathbb{K}^{\,\tilde \gamma}_{\mathds{1}\bm{h}}\, \mathbb{T}_{h_1}^{\,n_1}\cdots \mathbb{T}_{h_{3g-3}}^{\,n_{3g-3}}\Big |^2 = \left |\mathbb{K}^{\,\tilde \gamma}_{\mathds{1}\bm{h}}\right |^2 \prod_{i=1}^{3g-3}\delta_{\mathbb{Z}}(J_i),
\end{equation}
where we used the Poisson resummation formula to rewrite the sum over Dehn twists as the Dirac comb, $\sum_{n\in \mathbb{Z}} \e^{2\pi i n (h_i-\bar h_i)} = \delta_{\mathbb{Z}}(h_i-\bar h_i)$.
Note that this holds when all weights $h_i>0$ are in a heavy channel: for the vacuum contributions ($h_i,\bar h_i = 0$), spin is trivially quantized ($J_i=0$), and we do not sum over Dehn twists because $[T_{\mathds{1}}^n] = [1]$ in the handlebody group $\mathcal{H}_g$. So, spin quantization naturally follows from the sum over topologies, as was also stressed in \cite{deBoer:2025rct}.

\paragraph{Reduction to punctured surfaces of lower genus.} Another immediate consequence of  \eqref{eq:handlebody sum statistics} is that the sum over the genus-$g$ MCG contains, as a special case, the contributions from crossing in identity channels of punctured surfaces with lower genus. This can be seen by cutting the surface along non-separating geodesics, or, in terms of VTQFT diagrams, along edges that leave the VTQFT graph connected. The subset of mapping classes acting only on the cut surface gives rise to a sub-sum of  \eqref{eq:handlebody sum statistics}, which implements crossing symmetry of the lower-genus correlation functions of CFT operators below the black hole threshold. 

As an example at genus 2, consider the subset of mapping classes of the form $\gamma= \gamma_1 \circ \gamma_2\circ \gamma_{0,4}$, where $\gamma_1$ and $\gamma_2$ are modular transformations on the two dumbbell handles and $\gamma_{0,4}$ only acts on the 4-punctured sphere obtained by cutting the two bells. The corresponding crossing kernel simplifies to:
\be 
\mathbb{K}^{\gamma}_{\id P_1P_2P_3}= \mathbb{K}_{\id P_1}^{\gamma_1}\, \mathbb{K}_{\id P_2}^{\gamma_2}\, \mathbb{K}^{\gamma_{0,4}}_{\id P_3}\!\begin{bmatrix} P_1 & P_2 \\ P_1 & P_2 \end{bmatrix}\,  , \label{eq:genus 2 to genus0,4 reduction}
\ee
where $\mathbb{K}^{\gamma_{0,4}}_{\id P_3}$ only acts on the embedded 4-punctured sphere conformal block with internal momentum $P_3$ and pairwise identical external momenta $P_1,P_2$. The sum over $\gamma_{0,4}\in \text{MCG}(\Sigma_{0,4})$ implements crossing symmetry of the averaged 4-point function $\langle \mathcal{O}_1 \mathcal{O}_1 \mathcal{O}_2 \mathcal{O}_2\rangle_{\mathrm{S}^2}$. The only difference is that in \eqref{eq:genus 2 to genus0,4 reduction}, the conformal weights $h_1,h_2$ are continued above the black hole threshold. Moreover, summing over $\gamma_1$ and $\gamma_2$ in the modular group leads to two copies of the modular invariant spectral density at genus 1:
\be \label{eq:MWKdensity}
\rho_\text{MWK}(P,\bar P)=\sum_{\gamma \in \text{PSL}(2,\mathbb{Z})/\ZZ} \left|\mathbb{K}_{\id P}^{\gamma}\, \right|^2,
\ee
where MWK stands for Maloney-Witten-Keller \cite{Maloney:2007ud,Keller:2014xba}.

Since the full genus-$g$ mapping class group is bigger than the product of mapping class groups of its subsurfaces, crossing symmetry of the genus-$g$ partition function imposes a stronger constraint on the CFT ensemble than just crossing symmetry of the bordered surfaces embedded in $\Sigma_g$.
The moments of OPE coefficients with light external states arise as special cases of the handlebody sum \eqref{eq:handlebody sum statistics} (appropriately analytically continued) for restricted crossing transformations such as \eqref{eq:genus 2 to genus0,4 reduction}.

\paragraph{Identity contributions and support of $\mathbb{K}^\gamma$.} The sum over crossing kernels \eqref{eq:handlebody sum statistics} has a subtlety that has to do with the contributions from the vacuum state $h_i = \bar h_i =0$. Namely, we should understand both sides of equation \eqref{eq:handlebody sum statistics} as distributions, to be integrated against conformal blocks. For moments with $h_i,\bar h_i$ all heavy, the crossing kernels contributing to the handlebody sum are smooth meromorphic functions. However,  they may have delta function support on identity contributions. 

The simplest example where this occurs is at genus one, where the  handlebody sum is the MWK density \eqref{eq:MWKdensity}. Transforming this density from Liouville variables to conformal weights, $\rho_{\text{MWK}}(P,\bar P)\d P \d \bar P = \rho_{\text{MWK}}(h,\bar h)\d h \d \bar h$, we can write the modular sum as:
\begin{equation}\label{eq:MWK2}
     \rho_{\text{MWK}}(h,\bar h) = \delta(h)\delta(\bar h) + \sum_{\gamma \neq I} \left|\mathbb{K}_{\id h}^{\gamma} \right|^2\,,
\end{equation}
where we separated out the contribution from the unit matrix $\gamma = I$. The modular kernel of the unit matrix is just the Dirac delta $\mathbb{K}_{\id h}^{\gamma= I} = \delta(h)$, with support only on $h=0$, whereas the modular kernels of the other elements of $\text{PSL}(2,\mathbb{Z})/\mathbb{Z}$ are smooth functions of $h$ (written, for example, in \cite{Benjamin:2020mfz}). The identity contribution $\delta(h)\delta(\bar h)$, besides being physically relevant for the spectrum of the CFT, is also important for crossing symmetry: under the modular S-transform in \eqref{eq:Sinvariance}, the term with $\gamma = S$ in \eqref{eq:MWK2} gets mapped to the identity $\gamma = I$, because $S$ squares to one. The fact that this holds as distributions is best understood by integrating against a Virasoro character and carefully analyzing the integration contour. 

At genus 2 and higher, there are more identity contributions, again encoded in the analytic properties of the crossing kernels. For example, in the dumbbell channel $\overline{\rho_1\rho_2\rho_3C_{113}C_{223}}$, the identity contributions come from simple mapping classes:
\begin{equation}
    \mathbb{K}^{\gamma = I}_{\id \bm{h}} = \prod_{i=1}^3 \delta(h_i), \quad \mathbb{K}^{\gamma = S}_{\id \bm{h}} = \sker{\id}{h_1}{\id}\delta(h_2)\delta(h_3), \quad \mathbb{K}^{\gamma = S\circ S}_{\id \bm{h}} = \sker{\id}{h_1}{\id}\sker{\id}{h_2}{\id}\delta(h_3), \dots
\end{equation}
and more generally any kernel that acts trivially on the middle edge of the dumbbell. As remarked above, this generates a subgroup isomorphic to $\text{PSL}(2,\mathbb{Z})\times \text{PSL}(2,\mathbb{Z})$ acting on the two subtori. In the sunset channel $\overline{\rho_1\rho_2\rho_3C_{123}C^*_{123}}$, the identity contributions come from $\gamma = \mathbb{F}$ and $\gamma = \mathbb{S}\circ \mathbb{F}$:
\begin{align}
   \mathbb{K}^{\gamma = F}_{\id \bm{h}} &= \delta(h_1)\delta(h_2) \fker{\id}{h_3}{\id}{\id}{\id}{\id} = \prod_{i=1}^3 \delta(h_i) ,\\
    \mathbb{K}^{\gamma = S\circ F}_{\id \bm{h}} &= \delta(h_1)\sker{\id}{h_2}{\id}\fker{\id}{h_3}{h_2}{\id}{\id}{h_2} = \delta(h_1)\rho_0(h_2)\delta(h_3-h_2),
\end{align}
and more generally any element of the form $\gamma_{p,q}\circ \mathbb{F}$, where $\gamma_{p,q} \in \text{PSL}(2,\mathbb{Z})$.
Here we used that the fusion kernel reduces to a delta function when one or more of the external operators is the identity \cite{Eberhardt:2023mrq}. This reflects the fact that $F$ and $I$ are identified under the handlebody group, i.e.~they lead to same handlebody:
\begin{equation}
  \tikzset{every picture/.style={line width=0.8pt}} %set default line width to 0.75pt        
\,.
\end{equation}
Note that the above identity contributions in the handlebody sum correctly implement the selection rule $C_{\id 23} = \delta_{23}$ for the statistical moment \eqref{eq:handlebody sum statistics}, as a distribution.

\paragraph{Crossing from handlebody surgery.}

So far, we discussed crossing symmetry of the handlebody sum on the level of the crossing kernels $\mathbb{K}^\gamma$. However, it is also useful to understand how crossing symmetry is manifest on the level of the fixed-$P$ partition functions $Z_{\text{Vir}}[M;P]$. This provides a diagrammatic interpretation of the terms in the sum. Namely, translating the handlebody sum in \eqref{eq:handlebody sum statistics} to VTQFT diagrams, we see that we only need to sum over fixed-$P$ partition functions on Wilson line networks with the same graph topology as $\Gamma_{\mathcal{C}}$. This is also clear from the interpretation of OPE coefficients as 3-punctured sphere boundaries discussed above equation \eqref{eq:ZVirOPE}.
This reformulation allows us to check that the handlebody sum \eqref{eq:handlebody sum statistics} satisfies the fixed-$P$ versions of the crossing symmetry constraints \eqref{eq:Binvariance}-\eqref{eq:Finvariance}. 

Namely, take any element in the handlebody sum and consider its corresponding VTQFT diagram, contributing as \eqref{eq:ZVirOPE}. Then, around any vertex of the diagram (embedded in $\mathrm{S}^3$), we can cut along a 3-punctured sphere and apply a braiding move:
\begin{equation}\label{eq:Z_vir braiding}
    \mathbb{B}^{P_1}_{P_2P_3} \,Z_\text{Vir}\!\left[\tikzset{every picture/.style={line width=0.8pt}} %set default line width to 0.75pt        

\right]\,.
\end{equation}
We labeled the Liouville momenta by $i\equiv P_i$. The dashed lines indicate that this vertex is part of the larger VTQFT diagram. We can view this as a local surgery operation, which sends handlebodies to handlebodies. If we then multiply by the anti-holomorphic counterpart, the LHS contributes to $\overline{C_{123}X}$, and the RHS to $\overline{C_{132}X}$; hence, the sum over handlebodies satisfies the first constraint equation \eqref{eq:Binvariance}. 

Similarly, given some VTQFT diagram, we can take any \emph{pair} of vertices that are connected by an edge, and cut along a 4-punctured sphere surrounding only these vertices and one edge. We then apply an F-move inside this 4-punctured sphere:
\begin{align}
\rho_0(P_3)\mathsf{C}_{123}&\mathsf{C}_{345} \, Z_\text{Vir}\!\left[\tikzset{every picture/.style={line width=0.8pt}} %set default line width to 0.75pt        

\right]\,.\label{eq:RHScrossing}
\end{align}
Here we used that the fixed-$P$ partition function is multiplied by the normalization factor \eqref{eq:ZVirOPE}, and we have written only the part of this factor that involves the vertices that have been singled out. In the last line, these normalization factors have nicely combined with the fusion kernel to swap the internal indices of $\mathbb{F}_{P_3P_3'}$, using the formula \eqref{eq:fusioninverse}. If we now multiply the LHS and RHS by the remaining normalization factors of \eqref{eq:ZVirOPE}, and then take the absolute value square on both sides, we see that the LHS contributes to $\overline{\rho(h_3,\bar h_3)C_{123}C_{345}X}$ (where $X$ represents the rest of the diagram) and the integrand of \eqref{eq:RHScrossing} contributes to $\overline{\rho(h_3',\bar h_3')C_{13'5}C_{43'2}X}$. Since the above 
surgery move sends handlebodies to handlebodies, and has a well-defined inverse $\mathbb{F}^{-1}$, we conclude that the handlebody sum satisfies the second constraint equation \eqref{eq:Binvariance}.

Finally, let us discuss invariance under the S-transform $\mathbb{S}_{P_1P_2}[P_3]$. Unlike the fusion move, $\mathbb{S}$ is a genuine mapping class group transformation and does not change the graph $\Gamma_{\mathcal{C}}$. Since the handlebody sum is a mapping class group orbit, it is clearly invariant under $\mathbb{S}$. Still, the S-transform has the interesting property that it changes the embedding manifold of the VTQFT graph $\Gamma_{\mathcal{C}}$, as we will now explain. Namely, take any VTQFT diagram with a loop, contributing to $\overline{C_{113}X}$. Regardless of how this loop is linked to the rest of the diagram, we can always surround it by a once-punctured torus, and apply an S-transform on the torus, giving:
\begin{multline}
   Z_\text{Vir}\!\left[\tikzset{every picture/.style={line width=0.8pt}} %set default line width to 0.75pt        

\right].\label{eq:omegaloop}
\end{multline}
In the second line we used the integral identity \eqref{eq:Verlinde lasso} (the ``Verlinde lasso''). In the final equality we recognized the VTQFT version of the $\Omega$-loop \cite{Hartman:2025cyj}, which implements a Dehn surgery with a relative S-transform. This surgery changes the manifold in which the VTQFT graph is embedded. For example, if the loop is embedded trivially in the ambient $\mathrm{S}^3$, the addition of the $\Omega$-loop in \eqref{eq:omegaloop} turns the embedding manifold into $\mathrm{S}^2\times \mathrm{S}^1$, which is a closed manifold obtained by gluing the boundaries of two solid tori with the identity map. Note, however, that in a generic tangled VTQFT diagram, there will be Wilson lines passing through the $P_1$ loop on the LHS of \eqref{eq:omegaloop}, 
so that the $\Omega$-loop acts non-trivially on the topology of the background manifold. We will see a number examples of this in section \ref{sec:5}.

\paragraph{Convergence.} One final obvious point that we have not addressed so far is the question of convergence of the handlebody sum \eqref{eq:handlebodysum3}. In the genus-1 case, this sum is known to be divergent, but it can be regulated as an analytic continuation of a convergent Eisenstein series \cite{Maloney:2007ud,Keller:2014xba}. For $g>1$, regulating the handlebody sum is still an open problem. First, as we saw in \eqref{eq:MWKdensity}, the genus-$g$ MCG contains $\text{PSL}(2,\mathbb{Z})$ as a subgroup, which leads to a divergence that can be regularized similarly to \cite{Maloney:2007ud,Keller:2014xba}. Second, there is a canonical map from $\text{MCG}(\Sigma_g)$ to $\mathrm{Sp}(2g,\mathbb{Z})$ by reducing to homology. However, this map has a non-trivial kernel (the Torelli group), and moreover the coset $\text{MCG}(\Sigma_g)/\mathcal{H}_g$ is strictly bigger than $\mathrm{Sp}(2g,\mathbb{Z})/\Gamma_\infty$, with $\Gamma_\infty$ the parabolic subgroup \cite{Collier:2023fwi}. 
In \cite{Maloney:2020nni,Collier:2021rsn}, a divergent sum over $\mathrm{Sp}(2g,\mathbb{Z})/\Gamma_\infty$ was regulated by a higher degree Eisenstein series, so potentially a similar method can be used for our MCG coset. However, this is technically challenging, first and foremost because we do not know the behavior of conformal blocks under mapping classes of high complexity. We leave this as a question for future work.

\section{Minimal completion of the handlebody sum}\label{sec:minimal completion}

Although the sum over handlebodies solves the constraints of crossing symmetry and modular invariance (and consequently spin quantization and OPE selection rules), there are still constraints imposed in section \ref{sec:2} that are not satisfied by the handlebody sum alone.
In this section, we will iteratively construct a complete set of topologies that does solve our constraints in a minimal way. 

The inclusion of these additional topologies can be systematized by introducing three basic surgery operations that can be applied to a given `seed' topology. By iterating these operations, we generate a family of 3-manifolds that forms the minimal set needed to reproduce the properties of the CFT ensemble laid out in section \ref{sec:2}. We call the process of generating new topologies from these basic gluing conditions the `gravitational machine'. The goal of this section is to prove the basic  properties of the gravitational machine, as summarized in the introduction \ref{sec:intro}.

\subsection{Handlebodies are not sufficient}
\label{subsec:gravitational machine 1}
Before introducing the gravitational machine in its full generality, we will just point out two examples of conditions imposed in section \ref{sec:2} that fail to be satisfied by the handlebody sum alone, showing that the handlebody sum is indeed insufficient.

\paragraph{Consistency between higher and lower genus.}
In section \ref{sec:3.3} we have seen that the handlebody sum at genus $g$ contains information about the handlebody sum at lower genus. For example, at genus 2 the sum over handlebodies (in the sunset channel) contains as a subsum the elements of the form $\gamma_{p_1,q_1}\circ \gamma_{p_2,q_2}\circ \mathbb{F}$, recall \eqref{eq:genus 2 to genus0,4 reduction}. Here $\gamma_{p,q}$ is a modular transformation at genus 1, labeled by a pair of co-prime integers that uniquely specify the $\mathrm{PSL}(2,\mathbb{Z})/\mathbb{Z}$ element. The sum over this subclass of crossing transformations contributes as:
\begin{equation}\label{eq:partialMWKs}
    \overline{\rho_1\rho_2\rho_3C_{123}C_{123}^*} \,\supset \,\rho_{\text{MWK}}(h_1,\bar h_1)\rho_{\text{MWK}}(h_2,\bar h_2) \,\big|\rho_0(h_3)C_0(h_1,h_2,h_3)\big|^2\,.
\end{equation}
Here $\rho_{\text{MWK}}$ is the genus-1 handlebody sum \eqref{eq:MWKdensity}, which is the full sum over on-shell topologies contributing to the mean spectral density $\bar \rho$. As discussed in section \ref{sec:spectral_statistics}, if we ignore spectral correlations (associated to off-shell topologies in the bulk), the above statistical moment should factorize as:
\begin{equation}\label{eq:factorizegenus2}
    \overline{\rho_1\rho_2\rho_3C_{123}C_{123}^*} = \overline{\rho_1}\,\overline{\rho_2}\,\overline{\rho_3}\,\overline{C_{123}C_{123}^*}\,.
\end{equation}
Comparing to eq.~\eqref{eq:partialMWKs}, we see that the handlebody (sub-)sum correctly reproduces the means $\bar\rho_1 = \rho_{\text{MWK}}(h_1,\bar h_1)$ and $\bar\rho_2 = \rho_{\text{MWK}}(h_2,\bar h_2)$, but fails to reproduce the third mean spectral density $\bar\rho_3$, giving only the leading Cardy density $\rho_0(h_3)\rho_0(\bar h_3)$. Intuitively, the reason is that the genus-2 mapping class group is too small to accommodate three commuting copies of $\text{PSL}(2,\mathbb{Z})$. 

The set of topologies that \emph{does} factorize into lower moments as in \eqref{eq:factorizegenus2} is rather easy to describe, and it necessarily includes non-handlebodies. We describe this set using Dehn surgery: given a VTQFT diagram $\Gamma$ (i.e.~a network of Wilson lines), and a regular neighborhood $N(\Gamma)$ embedded in a closed 3-manifold $M_E$, take an edge dual to a non-contractible cycle in $M_E\setminus N(\Gamma)$. Then carve out a solid torus enclosing this edge and re-glue the solid torus after a modular transformation $\gamma_{p,q}$. Pictorially:
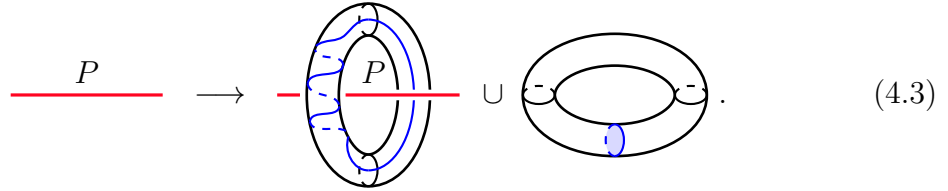
\begin{equation}\label{eq:dehn surgery picture}
\vcenter{\hbox{
\begin{tikzpicture}
    \draw[very thick, wilsonred] (-1,0) to node[above, black] {$P$} node[below, black] {$\phantom{P}$} (1,0);
\end{tikzpicture}
}}\  \longrightarrow\
\vcenter{\hbox{
\begin{tikzpicture}
    \draw[line width=13, black] (0,0) ellipse (0.6 and 1);
    \draw[line width=11, white] (0,0) ellipse (0.6 and 1);
    \draw[thick] (0,0.79) arc (-90:90:0.12 and 0.21);
    \draw[thick, dashed] (0,0.79) arc (270:90:0.12 and 0.21);
    \draw[thick] (0,-0.79) arc (90:-90:0.12 and 0.21);
    \draw[thick, dashed] (0,-0.79) arc (90:270:0.12 and 0.21);
    \draw[thick, blue] (-0.8,0.15) to[out = 80, in = -100] (-0.33,0.4);
    \draw[thick, blue] (-0.38,-0.15) to[out = -100, in = 80] (-0.77,-0.35);
    \draw[thick, blue] (-0.26,-0.55) to[out = -110, in =180] (0,-1);
    \draw[thick, blue] (0,-1) arc (-90:90:0.6 and 1);
    \draw[thick, blue] (-0.69,0.65) to[out = 40, in = -135] (-0.3,0.8) to[out=45, in=180] (0,1);
    \draw[thick, blue,dashed] (-0.33,0.4) to[out = 100, in = -80] (-0.7,0.6);
    \draw[thick, blue,dashed] (-0.8,0.15) to[out = -80, in = 100] (-0.38,-0.15);
    \draw[thick, blue,dashed] (-0.77,-0.35) to[out = -80, in = 100] (-0.25,-0.55);
    \draw[line width=4, white] (-0.3,0) to (1.2,0);
    \draw[very thick, wilsonred] (-0.3,0) to node[above, black] {$P\quad\;\;\;$} (1.2,0);
    \draw[very thick, wilsonred] (-1.2,0) to (-0.9,0);
\end{tikzpicture}
}} \cup
\vcenter{\hbox{
\begin{tikzpicture}[rotate = -90]
\draw[line width=13, black] (0,0) ellipse (0.6 and 1);
    \draw[line width=11, white] (0,0) ellipse (0.6 and 1);
    \draw[thick] (0,0.79) arc (-90:90:0.12 and 0.21);
    \draw[thick, dashed] (0,0.79) arc (270:90:0.12 and 0.21);
    \draw[thick] (0,-0.79) arc (90:-90:0.12 and 0.21);
    \draw[thick, dashed] (0,-0.79) arc (90:270:0.12 and 0.21);
    \fill[blue!15] (0.6,0) ellipse (0.21 and 0.12);
    \draw[thick,blue] (0.39,0) arc (180:0:0.21 and 0.12);
    \draw[thick, dashed,blue] (0.39,0) arc (180:360:0.21 and 0.12);
\end{tikzpicture}
}}.
\end{equation}
The element $\gamma_{p,q}$ maps the blue meridian on the right torus to a curve on the left which winds around the meridian $p$ times and around the longitude $q$ times (in the picture $p=3$ and $q=1$). For a curve with no self-intersections $p$ and $q$ are coprime integers, and the ordered pair written as $p/q$ is in one-to-one correspondence with the elements of $\mathrm{PSL}(2,\ZZ)/\ZZ$.

In Virasoro TQFT, the above Dehn surgery procedure is described by the Verlinde loop operator \cite{Collier:2024mgv, deBoer:2025rct}, as follows:
\begin{align}
    Z_\text{Vir}\!\left[\tikzset{every picture/.style={line width=0.8pt}} %set default line width to 0.75pt        

\right]\,.\label{eq:dehnsurg}
\end{align}
Here $\mathbb{K}^{(p,q)}_{\id P}$ is the modular kernel for $\gamma_{p,q}$, implementing the Dehn surgery with surgery coefficient $p/q$. In the second line we used the formula \eqref{eq:verlindeloop} for removing the Verlinde loop, and in the third line we used the fact that $\gamma_{p,q}\circ \mathbb{S} = \gamma_{-q,p}$ under the modular S-transform. The above manipulations can be done locally on any edge of $\Gamma$, indicated as before by the dotted line.

To obtain the gravitational partition function from the above VTQFT computation, we multiply the fixed-$P$ partition function \eqref{eq:dehnsurg} by the normalization factor $\rho^{\,\mathcal{C}}_g(\boldsymbol{P})$---which in particular contains a factor $\rho_0(P_i)$ for the $i^{\text{th}}$ edge---and then we multiply by the right-moving counterpart. Finally, we sum over the co-primes $p,q$. The net effect of this sum over Dehn surgeries is that the Cardy density $\rho_0$ gets modular completed to the MWK spectral density:
\be\label{eq:complete to MWK}
    \rho_0(P_i)\rho_0(\bar P_i)\,\left |\tikzset{every picture/.style={line width=0.8pt}} %set default line width to 0.75pt        

\right |^2.
\ee
Here $\rho'_{\text{MWK}}$ is the MWK handlebody sum \eqref{eq:MWKdensity} without the vacuum contribution, i.e.~$\rho_{\text{MWK}} = \delta(h)\delta(\bar h) + \rho'_{\text{MWK}}$. In other words, we excluded the Dehn surgery with $(p,q) = (0,1)$, which projects $P_i$ to the identity $\id$. This corresponds to the case that the surgery implements the action of the $\Omega$-loop. In this way, the sum over Dehn surgeries upgrades the heavy Cardy density to the heavy part of the MWK density, leaving the identity contributions that are already part of the genus-$g$ handlebody sum. Importantly, for complicated enough Dehn surgery coefficients, the manifolds constructed by our procedure \eqref{eq:dehnsurg} can be non-handlebodies. 

Let us see how this works in the genus-2 example discussed above. Dehn surgery on the three internal lines of the sunset channel gives:\footnote{Here we used the sunset normalization factor $\rho^{\,\mathcal{C}}_{g=2}(\boldsymbol{P}) = \rho_0(P_1)\rho_0(P_2)\rho_0(P_3)C_0(P_1,P_2,P_3)^2$, as well as \eqref{eq:theta graph ZVir} for evaluating the sunset diagram in VTQFT.}
\begin{equation}
    \rho^{\,\mathcal{C}}_{g=2}(\boldsymbol{P})\, Z_{\text{Vir}}\!\left[
\vcenter{\hbox{
\begin{tikzpicture}[yscale=1,xscale=0.75]
\draw[very thick, wilsonred] (0,1) -- (0,-1);
\draw[very thick, wilsonred] (0,0) ellipse (2 and 1);
\draw[very thick] (-0.07,0.2) arc (100:440:0.35 and 0.25)
node[right, black]{{\,$\frac{p_2}{q_2}$}};
\draw[very thick] (-2,0.2) arc (100:440:0.35 and 0.25)
node[right, black]{{\,$\frac{p_1}{q_1}$}};
\draw[very thick] (1.9,0.2) arc (100:440:0.35 and 0.25)
node[right, black]{{\,$\frac{p_3}{q_3}$}};
\draw (0,-0.5) node [anchor=north west][inner sep=0.75pt]  [font=\footnotesize]  {$P_2$};
\draw (-1.4,-0.9) node [anchor=north west][inner sep=0.75pt]  [font=\footnotesize]  {$P_1$};
\draw (1,-0.9) node [anchor=north west][inner sep=0.75pt]  [font=\footnotesize]  {$P_3$};
\end{tikzpicture}
}}
\right] = \mathbb{K}_{\id P_1}^{(-q_1,p_1)} \mathbb{K}_{\id P_2}^{(-q_2,p_2)} \mathbb{K}_{\id P_3}^{(-q_3,p_3)} C_0(P_1,P_2,P_3).
\end{equation}
In section \ref{sec:5}, we will identify the topology of the filling $(\frac{p_1}{q_1},\frac{p_2}{q_2},\frac{p_3}{q_3}) = (2,2,2)$ with a non-handlebody known as the twisted $I$-bundle \cite{Yin:2007at}, and  $(\frac{p_1}{q_1},\frac{p_2}{q_2},\frac{p_3}{q_3}) = (2,2,3)$ with another non-handlebody called the $\mathbf{5}_2$ handlebody-knot. In total, the sum over all the surgery coefficients $(p_i,q_i)$ contributes as:
\begin{equation}\label{eq: sunset surgery}
    \overline{\rho_1\rho_2\rho_3C_{123}C_{123}^*} \,\supset\, \rho_{\text{MWK}}(h_1,\bar h_1)\rho_{\text{MWK}}(h_2,\bar h_2) \rho_{\text{MWK}}(h_3,\bar h_3)\,\big|C_0(h_1,h_2,h_3)\big|^2\,,
\end{equation}
which is indeed of the factorized form
\eqref{eq:factorizegenus2}. Although compatible with genus-1 modular invariance, the sum over Dehn surgeries itself still has to be completed by a sum over the genus-2 mapping class group, which acts via crossing transformations on the sunset diagram. The non-trivial interplay between crossing and the factorization \eqref{eq:factorizegenus2} into lower moments will be explained in more depth in section \ref{subsec:general gravitational machine}.

\paragraph{Typicality.}
As a second constraint that is not satisfied by the handlebody sum, we give an example of a violation of the typicality condition discussed in section \ref{sec:typicality}. As explained there, typicality predicts extra terms contributing to statistical moments with repeated pairs of indices, such as $\overline{C_{113}C_{113}}$. On the level of distributions, this means there are delta-function contributions such as:
\begin{multline}\label{eq:typicality again}
    \overline{\rho_1\rho_2\rho_3C_{123}C_{123}^*} \, \supset \\ \delta^2(h_1-h_2)\, \overline{\rho_1}\,\overline{\rho_3} \,\left[(1+(-1)^{J_3})\,\mathcal{G}_S\!\left[\vcenter{\hbox{
    \begin{tikzpicture}[yscale=-0.5,xscale=0.5]
        \draw[very thick, wilsonred] (0,0) ellipse (1.5 and 1);
        \draw[very thick, wilsonred] (0,1) -- (0,-1);
        \node at (-1.2,0) {\footnotesize $1$};
        \node at (0.29,0) {\footnotesize $1$};
        \node at (1.2,0) {\footnotesize $3$};
    \end{tikzpicture}
    }}\right] + \mathcal{G}_D\!\left[\vcenter{\hbox{
    \begin{tikzpicture}[yscale=-0.85,xscale=0.85]
        \draw[very thick, wilsonred] (-1,0) circle (0.5 and 0.5);
        \draw[very thick, wilsonred] (1,0) circle (0.5 and 0.5);
        \draw[very thick, wilsonred] (-0.5,0) -- (0.5,0);
        \node at (-1,-0.15) {\footnotesize $1$};
        \node at (0,0.3)  {\footnotesize $3$};
        \node at (1,-0.15)  {\footnotesize $1$};
    \end{tikzpicture}
    }}\right] \right].
\end{multline}
This is an example of a term for the index contraction $\delta_{12}$, which, since we are integrating over densities, has been represented by the Dirac delta $\delta^2(h_1-h_2) \equiv \delta(h_1-h_2)\delta(\bar h_1-\bar h_2)$. 
In some cases, these delta-function contributions may be accounted for in the bulk by handlebodies, but in general we require non-handlebodies to account for all index contractions. 

For example, \eqref{eq:typicality again} receives a contribution from the following handlebody:
\begin{equation}\label{eq:handlebody contribution}
\tikzset{every picture/.style={line width=1.2pt}} %set default line width to 0.75pt        
.
\end{equation}
 Here we drew the linked dumbbell from the previous section (with the opposite over-under crossing), where one of the bells is in the S-dual channel as indicated by the prime. We can check that this indeed produces the desired delta function by using the $\Omega$-loop defined in \eqref{eq:omegaloop} to implement the S-transform:
\begin{equation}
    Z_\text{Vir}\!\left[\!\!\!\!\!\tikzset{every picture/.style={line width=1.2pt}} %set default line width to 0.75pt        
%
\right].
\end{equation}
Here we used the VTQFT identity \eqref{eq:omega loop on pair of lines} for the $\Omega$-loop encircling two Wilson lines. We can also directly use the VTQFT partition function for the linked dumbbell computed in \eqref{eq:Z_linkedhandcuff}, together with the fact that $\mathbb{S}$ is unitary:
\begin{align}
    \int_0^\infty \d P\, \sker{P_1}{P}{P_3} Z_\text{Vir}\!\left[\!\!\!\!\!\tikzset{every picture/.style={line width=1.2pt}} %set default line width to 0.75pt        
%
\right] &= \int_0^\infty \d P\, \sker{P_1}{P}{P_3} \frac{\sker{P}{P_2}{P_3}^*}{\rho_0(P_2)C_0(P_2,P_2,P_3)} \\&=  \frac{\delta(P_1-P_2)}{\rho_0(P_2)C_0(P_2,P_2,P_3)}. 
\end{align}
Multiplying by the normalization factor $\rho_{g=2}(\boldsymbol{P})$ and the anti-holomorphic counterpart, we see that this handlebody contributes in the following way:
\begin{equation}
\overline{\rho_1\rho_2\rho_3C_{123}C_{123}^*} \, \supset \delta^2(h_1-h_2)\, |\rho_0(h_1)\rho_0(h_3)C_0(h_1,h_1,h_3)|^2\,.
\end{equation}
There is also the handlebody with the opposite over-under crossing of the linked handles, which gives an extra phase factor $\e^{-\pi i h_3}$, as already remarked in \eqref{eq:cherryOPE}. This accounts for the $(-1)^{J_3}$ contribution to the moment \eqref{eq:typicality again}. 

Even though handlebodies account for some of the contractions predicted by typicality, not all of the contractions can be explained by the handlebody sum alone. For example, the enhancement in \eqref{eq:typicality again} by the dumbbell terms $\mathcal{G}_D$ consists of non-handlebody contributions, as we will see in section \ref{sec:5}. A simpler example is given by the double contraction $\delta_{12}\delta_{23}$, recall \eqref{eq:repeated indices b}, which contributes to the second moment when all indices coincide, $\overline{C_{iii}C_{iii}}$. When $h_i\neq \id$, the only contributions to this moment come from non-handlebodies. In fact, the simplest contribution comes from a so-called \emph{handlebody-knot}, whose VTQFT diagram can be drawn as:
 \begin{equation}\label{eq:41prime}
\tikzset{every picture/.style={line width=0.75pt}} %set default line width to 0.75pt        
 \,.
\end{equation}
If we take this knotted graph and embed it in $\mathrm{S}^3$, its complement is a non-handlebody with genus-2 boundary. It is the handlebody-knot with the least number of over-under crossings, called the $\mathbf{4}_1$ handlebody-knot in the classification of  \cite{ISHII:2012}. The primes indicate that we additionally perform an S-transform on the two bells. Let us check that this manifold produces the desired delta functions for the double contraction: 
\begin{align}
    Z_\text{Vir}\!\left[\tikzset{every picture/.style={line width=0.75pt}} %set default line width to 0.75pt        

\right].\label{eq:41prime self-gluing}
\end{align}
 Multiplying by the normalization factor and the anti-holomorphic counterpart, we confirm that this non-handlebody is the leading contribution to the double index contraction $\delta_{12}\delta_{23}$ at genus two:
 \begin{equation}\label{eq:C_111C_111}
\overline{\rho_1\rho_2\rho_3C_{123}C_{123}^*} \, \supset \delta^2(h_1-h_2)\,\delta^2(h_2-h_3) \,|\rho_0(h_1)C_0(h_1,h_1,h_1)|^2\,.
\end{equation}
This example shows that non-handlebodies like $\mathbf{4}_1$ are needed for consistency with typicality.
Since a non-handlebody always remains a non-handlebody after crossing transformations, all of the mapping class group images of \eqref{eq:41prime} (which are needed to complete \eqref{eq:C_111C_111} into a crossing symmetric density) are also non-handlebodies. Moreover, it may be further modified by Dehn surgery to upgrade $\rho_0$ to $\rho_{\text{MWK}}$. 

In the next subsection, we will take the two examples studied above as the basis for an algorithm for generating 3-manifolds, which we call the `gravitational machine'. 

\subsection{The gravitational machine} \label{subsec:general gravitational machine}

The examples discussed above suggest a general procedure that systematically constructs a minimal set of 3-manifolds that should be summed over in the bulk, in order to satisfy the constraints imposed in section \ref{sec:2} on the moments of the CFT data. In this section, we will introduce this algorithm---the gravitational machine---in full generality. It will generalize both the Dehn surgery and the $\Omega$-loop computations of the previous section and works for manifolds with arbitrary number of boundary components of any genus $g\geq 1$. We will formulate the rules on the level of the fixed-$P$ partition functions, i.e.~for networks of Wilson lines in VTQFT. It is simple to change the formulation into the canonical ensemble by thickening the Wilson lines and considering the drilled out manifold with the asymptotic conformal boundary. 

\paragraph{Cylinder surgery.}
The main topological operation in the gravitational machine is what we will call \emph{cylinder surgery}. This procedure can be described as follows. Given a (possibly disconnected) network of Wilson lines in VTQFT, surround one of the edges by a 2-punctured sphere, and carve out its interior to obtain a new manifold with a 2-punctured sphere boundary. Do the same with another edge, and glue the two 2-punctured spheres to each other with the opposite orientation:
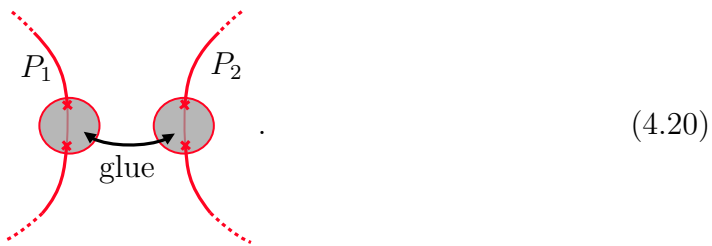
\begin{equation}\label{eq:cylinder surgery}
    \tikzset{every picture/.style={line width=0.75pt}} %set default line width to 0.75pt        
\,.
\end{equation}
We call this `cylinder surgery', because the 3-manifolds in VTQFT are the complement of a regular neighborhood of the Wilson line network embedded in $M_E$, so that the 2-punctured sphere along which we glue becomes a cylinder. This was illustrated in the introduction in figure \ref{fig:machine1}.

When $P_1$ and $P_2$ belong to two disconnected networks of Wilson lines, corresponding to disjoint manifolds $M_1$ and $M_2$, 
the above construction is similar to a connected sum of 3-manifolds, in which one removes two empty balls and identifies the resulting $\mathrm{S}^2$ boundaries. In this case, cylinder surgery is a generalization of the connected sum of $M_1$ and $M_2$, now including Wilson lines. We will denote it by
\be 
M_1 \#_{P_1 \sim P_2} M_2\ , \label{eq:generalized connected sum}
\ee
with $P_1$ and $P_2$ the Wilson line labels that get identified. 

When the $P_1$ and $P_2$ Wilson lines are part of the same connected VTQFT diagram, with corresponding graph complement $M$, we denote cylinder surgery by
\be
M \#_{P_1 \sim P_2}\,.
\ee 
We will also refer to this as `self-gluing' throughout the rest of the paper.

\paragraph{Reformulating Dehn surgery.} Let us see how the above procedure generalizes the Dehn surgery that we described in section~\ref{subsec:gravitational machine 1}. We can think of the modified manifold in \eqref{eq:dehn surgery picture} as being obtained by applying cylinder surgery to the original manifold $M_1$ (with Wilson line $P_1$), and a genus-1 handlebody $M_2$ (i.e.~one of the $\mathrm{PSL}(2,\mathbb{Z})$-family of solid tori). The fixed-$P$ version of the genus-1 handlebody can be represented in VTQFT as:
\be 
\vcenter{\hbox{
\begin{tikzpicture}[scale=0.8]
    \draw[very thick] (0,-.9) arc (-90:90:.5 and 1) node[black, shift={(.72,-.2)}] {$p/q$};
    \fill[white] (.35,0) rectangle (.65,.12);
    \draw[very thick, wilsonred] (0,1) circle (1.4 and 1);
    \node at (0,1.7) {$P$};
    \fill[white] (-.45,0) rectangle (-.55,.12);
    \draw[very thick] (0,1.1) arc (90:270:.5 and 1);
\end{tikzpicture}
}}
\ee
embedded in $\mathrm{S}^3$.
Thus, we can view the Dehn surgery operation \eqref{eq:dehn surgery picture} on $M_1$ as the result of gluing a genus-1 handlebody to $M_1$ by removing a ball neighborhood around a pair of Wilson lines and identifying the boundaries, that is, $M_1\#_{P_1\sim P_2} M_2$. For example, when $M_1$ is a genus-2 handlebody, we have:
\begin{equation}\label{eq:gluing solid torus}
\vcenter{\hbox{
\begin{tikzpicture}
\draw[very thick, wilsonred] (0,1) -- (0,-1);
\draw[very thick, wilsonred] (0,0) ellipse (2 and 1);
\draw[very thick] (1.9,0.2) arc (100:440:0.35 and 0.25);
\node at (2.3,-0.6) {$p/q$};
\end{tikzpicture}
}}\cong\vcenter{\hbox{
\begin{tikzpicture}
\draw[very thick, wilsonred] (-2,1) -- (-2,-1);
\draw[very thick, wilsonred] (-2,0) ellipse (2 and 1);
\draw[very thick, wilsonred] (2,0) circle (.5);
\draw[very thick] (2.4,0.2) arc (100:440:0.35 and 0.25);
\node at (2.8,-0.6) {$p/q$};
\draw[red, fill=gray, opacity=.75] (0,0) circle (.25);
\draw[red, fill=gray, opacity=.75] (1.5,0) circle (.25);
\draw[very thick,<->,bend left=20] (.2,.2) to node[above] {glue} (1.3,.2);
\end{tikzpicture}
}}\ .
\end{equation}
This requires of course to identify the Wilson line labels of the two Wilson lines that get glued together. On the RHS, before gluing, the genus-1 and genus-2 Wilson line networks are embedded in two disjoint copies of $\mathrm{S}^3$.

\paragraph{Reformulating self-contractions.} Let us also see how cylinder surgery generates the bulk manifolds discussed in section \ref{subsec:gravitational machine 1} that contribute to the index contractions predicted by typicality. In the examples given in \eqref{eq:handlebody contribution} and \eqref{eq:41prime}, we used the $\Omega$-loop encircling a pair of Wilson lines to produce the delta function $\delta(P_1-P_2)$. Topologically, this is equivalent to doing the cylinder surgery described above, see also~\eqref{eq:omega loop on pair of lines}.

For example, the non-handlebody manifold \eqref{eq:41prime} that contributes to the double self-contraction $\overline{C_{111}C_{111}}$ can also be constructed from the following self-gluing: 
\begin{equation}\label{eq:4_1 from surgery}
\tikzset{every picture/.style={line width=1.2pt}} %set default line width to 0.75pt        
\,.
\end{equation}

This can be seen from the figure \eqref{eq:cylinder surgery} by slicing it horizontally through the middle. In the top half of the figure, take the hemispheres punctured by $P_1$ and $P_2$ and puff up the Wilson lines so that the punctured hemispheres becomes annuli. When we glue, these become a torus. Doing the same with the bottom half of the figure, we get two tori that are glued to each other with the identity map, thus making the background manifold $\mathrm{S}^2\times \mathrm{S}^1$. This is precisely what the $\Omega$-loop implements.

\paragraph{Cylinder surgery in VTQFT.}  
In cylinder surgery, we glue along a 2-punctured 2-sphere. The corresponding Hilbert space in VTQFT is one-dimensional  (provided that the labels of the Wilson lines agree). Thus the gravitational path integral in $M_1$ and $M_2$ has to produce the unique state up to proportionality and we obtain:
\be\label{eq:VTQFT generalized connected sum}
Z_\text{Vir}[M_1 \#_{P_1 \sim P_2} M_2]=\frac{\delta(P_1-P_2)}{\rho_0(P_1)}\, Z_\text{Vir}[M_1]Z_\text{Vir}[M_2].
\ee 
The proportionality constant $\rho_0(P_1)^{-1}$ can be understood in two ways. First, after surgery we have one edge less in the VTQFT diagram, so the normalization factor $\rho^{\mathcal{C}}_g(\boldsymbol{P})$ should get one fewer factor of $\rho_0$. Second, 
we note that for $M_1$ and $M_2$ both unknots in $\mathrm{S}^3$, the connected sum is also an unknot in $\mathrm{S}^3$. Using that $Z_\text{Vir}(\text{unknot})=\rho_0(P)$ explains the proportionality factor.\footnote{We may regard (\ref{eq:VTQFT generalized connected sum}) as the VTQFT analog of the identity for the partition function on manifolds constructed as connected sums in ordinary TQFTs (e.g.~based on a modular tensor category). In particular, for $M = M_1 \# M_2$, we have $Z_{\text{TQFT}}(M) = \tfrac{Z_{\text{TQFT}}(M_1)Z_{\text{TQFT}}(M_2)}{Z_{\text{TQFT}}(\mathrm{S}^3)}$. This equation is derived by carving out a three-ball from each of $M_1$ and $M_2$ and gluing them along the resulting $\mathrm{S}^2$ boundaries. This operation is not allowed in VTQFT: the Hilbert space on the naked $\mathrm{S}^2$ is ill-defined, the total quantum dimension badly diverges, and indeed hyperbolic manifolds are prime and so cannot be written as connected sums anyway. However the generalization with Wilson lines \eqref{eq:generalized connected sum} as implemented by (\ref{eq:VTQFT generalized connected sum}) does (under certain conditions) lead to hyperbolic manifolds, as we shall see below.}

For self-gluing, the same reasoning gives:
\be\label{eq:VTQFT self gluing}
Z_\text{Vir}[M \#_{P_1 \sim P_2} ;P_1,P_2,\dots]=\frac{\delta(P_1-P_2)}{\rho_0(P_1)}\, Z_\text{Vir}[M;P_1,P_1,\dots].
\ee 
In the special case that $M = M_1\sqcup M_2$ is a disjoint union, we recover \eqref{eq:VTQFT generalized connected sum}.

\paragraph{Statistical interpretation.} Cylinder surgery has a clear interpretation from the boundary ensemble point of view. Namely, 
consider some moment $\overline{XY}$, where $X$ and $Y$ are products of OPE coefficients and densities of states. $\overline{XY}$ necessarily includes the disconnected moment $\overline{X}\, \overline{Y}$, which is described gravitationally by considering the disconnected union of the set of manifolds computing the lower moments $\overline{X}$ and $\overline{Y}$. Now, the typicality assumption also forces us to include connected contractions of the indices in $X$ and $Y$, which are constructed in the bulk via cylinder surgery. That is, let $M_1$ be a manifold contributing to $\overline{X}$ and $M_2$ a manifold contributing to $\overline{Y}$. Then we should add to $\overline{XY}$ the manifold corresponding to the diagonal contraction by excising a ball around two Wilson lines and identifying their boundaries, i.e.~applying the cylinder surgery $M_1 \#_{P_1 \sim P_2} M_2$ as above.

\paragraph{Adding crossing transformations.} Once we have added a new topology $M$ that was obtained using cylinder surgery, such as the $\mathbf{4}_1$ handlebody-knot in \eqref{eq:4_1 from surgery}, we must also include its sum over mapping class group images in order to preserve crossing symmetry. More precisely, we sum over the boundary mapping class group modulo the bulk diffeomorphisms of $M$, i.e.~the coset \cite{Collier:2023fwi}:
\begin{equation}\label{eq:MCG coset}
\text{MCG}(\partial M) / \text{MCG}(M, \partial M).
\end{equation}
Here $\text{MCG}(M, \partial M)$ is the bulk mapping class group of $M$, containing the large diffeomorphisms of $M$ that preserve the boundary setwise. On the level of the VTQFT diagrams, this means we can act with any of the crossing transformations $\mathbb{B}$, $\mathbb{F}$ and $\mathbb{S}$, as in equations \eqref{eq:Z_vir braiding}, \eqref{eq:RHScrossing} and \eqref{eq:omegaloop}, up to the relations of the Moore-Seiberg groupoid and the identifications by the bulk mapping class group. 

Crucially, these crossing transformations map VTQFT partition functions with delta functions $\delta(P_i-P_j)$ to partition functions without delta functions. For example, recall the $\mathbf{4}_1$ handlebody-knot \eqref{eq:41prime}, which is a contribution to the double contraction $\overline{C_{111}C_{111}}$. If we act with S-transforms on its bells, we get a smooth function of $P_{1,2,3}$ that is a product of two S-kernels, written in \eqref{eq:vir41-2} and further analyzed there. Hence, crossing symmetry combined with typicality predicts a non-handlebody contribution to $\overline{C_{112}C_{233}}$, even when $P_{1,2,3}$ are all different.

Similarly, recall the Dehn surgery \eqref{eq:dehn surgery picture}, which was needed in order to reproduce the correct factors of $\rho_{\text{MWK}}$ as in \eqref{eq: sunset surgery}. Acting with crossing transformations on the VTQFT diagram gives an amplitude that is no longer of the factorized form \eqref{eq:factorizegenus2}; hence a new $\text{PSL}(2,\mathbb{Z})$-family of Dehn surgeries must be added for consistency with modular invariance of the lower moments $\overline{\rho_i}$. To illustrate this, consider the sunset diagram (with Dehn surgery) and act with an $\mathbb{F}$ move:
\begin{equation}
    \vcenter{\hbox{
\,.
\end{equation}
Here $\gamma_i$ specify the Dehn surgery parameters, which we assume to be non-trivial. On the RHS, the VQTFT amplitude is not of the factorized form $\bar \rho_1 \,\bar \rho_3\, \bar \rho_4 \,\overline{C_{114}C_{334}}$. At the same time, the  Dehn filling $\gamma_2$ has resulted in a new non-contractible cycle, namely the loop surrounding the $P_4$ edge. We can restore the factorized form by adding a fourth Dehn surgery around this loop and summing over $\gamma_4\in \PSL(2,\ZZ)/\ZZ$. Iterating this procedure generates more and more complicated contributions to the sum over geometries, e.g.:\vspace{-0.3cm}
\begin{equation}\label{eq:adding crossing}
\vcenter{\hbox{
\begin{tikzpicture}[yscale=1,xscale=0.7]
\draw[very thick, wilsonred] (0,1) -- (0,-1);
\draw[very thick, wilsonred] (0,0) ellipse (2 and 1);
\draw[very thick] (-0.07,0.2) arc (100:440:0.35 and 0.25) node[right,black] {$\gamma_2$};
\end{tikzpicture}
}} \,\,\rightarrow\,\,
\vcenter{\hbox{
\begin{tikzpicture}
    \draw[very thick] (0,0) ellipse (0.8 and 0.3) node[black] {$\gamma_2$};
    \draw[line width=4, white] ({-1+0.5*cos{-15}},{0.5*sin{-15}})  arc (-15:300:0.5);
    \draw[very thick, wilsonred] ({-1+0.5*cos{-15}},{0.5*sin{-15}})  arc (-15:320:0.5);
    \draw[line width=4, white] ({1+0.5*cos{-145}},{0.5*sin{-145}})  arc (-145:190:0.5);
    \draw[very thick, wilsonred] ({1+0.5*cos{-145}},{0.5*sin{-145}})  arc (-145:190:0.5);
    \draw[very thick,red] (-1,0.5) to[out = 90,in=90,looseness=1.5] (1,0.5);
    \draw[very thick] (0.15,1.45) arc (15:345:0.2 and 0.3) node[above right, black] {$\gamma_4$};
\end{tikzpicture}
}}
\,\,\rightarrow\,\,
\vcenter{\hbox{
\begin{tikzpicture}[yscale=1,xscale=0.75,baseline={([yshift=10ex]current bounding box.south)}
]
\draw[very thick, wilsonred] (0,1) -- (0,-1);
\draw[very thick, wilsonred] (0,0) ellipse (2 and 1);
\draw[very thick] (-0.07,0.2) arc (100:440:0.35 and 0.25) node[right,black] {$\gamma_5$};
\draw[very thick] (0,-1.75) arc (-90:90:0.2 and 0.3);
\draw[line width=4, white] (-0.5,-1.01) arc (180:360:0.5 and 0.3);
\draw[line width=4, white] (-0.9,-1) arc (180:360:0.9 and 0.6);
\draw[very thick] (-0.5,-1.01) arc (180:360:0.5 and 0.3);
\draw[very thick] (-0.9,-1) arc (180:360:0.9 and 0.6);
\draw[very thick] (-0.9,-1) arc (180:25:0.2);
\draw[very thick] (0.9,-1) arc (0:155:0.2);
\draw[line width=4, white] (0,-1.75) arc (270:90:0.2 and 0.3);
\draw[very thick] (0,-1.75) arc (270:90:0.2 and 0.3);
\node at (-1.2,-1.1) {$\gamma_2$};
\node at (0,-2) {$\gamma_4$};
\end{tikzpicture}
}} \rightarrow \;\; \cdots
\end{equation}
This mechanism applies to any VTQFT diagram and will be an essential ingredient in the formulation of the gravitational machine, to which we turn next.

\paragraph{The general gravitational machine.} Having introduced all the necessary ingredients, we come to formulate the general gravitational machine, which produces a set of manifolds $\mathcal{M}$ consistent with all properties of the boundary statistical ensemble that we imposed in section~\ref{sec:2}. The idea is to give a finite set of `moves' which inductively define a set of orientable 3-manifolds\footnote{$\mathcal{M}$ can in principle also contain manifolds with conical defects described by light Wilson lines. However, here we will restrict ourselves to the `pure' gravity case without defects.}
\be 
M \in \mathcal{M}\,.
\ee
As an initial condition, we assume that $\mathcal{M}$ contains all handlebodies of genus $g$, which is the minimal set needed for consistency with crossing. From this initial set, we define a finite number of operations that produce new manifolds, which we also have to include in the sum over topologies for consistency with the other conditions imposed in section \ref{sec:2} (including the ones discussed above). We will again think of the resulting manifolds at the level of networks of Wilson lines (i.e.~in the fixed-$P$ formalism), but they can as always be thickened to pass to the canonical ensemble.

The three moves of the gravitational machine are as follows:
\begin{enumerate}
    \item For $M_1$ and $M_2$ in $\mathcal{M}$, the disconnected manifold $M_1 \sqcup M_2$ is also allowed. \label{move:disconnected sum}
    \item For any $M \in \mathcal{M}$, we are allowed to perform crossing transformations on the Wilson line network representing the boundary data.\label{move:crossing}
    \item For $M\in \mathcal{M}$, we can apply the cylinder surgery $M \#_{P_1 \sim P_2}$ on any pair of Wilson lines, where $P_1$ and $P_2$ are identified. This surgery can either glue different connected components of $M$ or it can glue the same component to itself. \label{move:gluing}
\end{enumerate}
For the third move, we have to make some additional technical assumptions that we discuss below. Next,
for each application of one of these moves, one should check\footnote{Hyperbolic manifolds are completely specified topologically by their fundamental group $\pi_1(M)$. Since the machine generates only hyperbolic manifolds (as we will show), checking homeomorphism comes down to testing for isomorphism of the fundamental group. This can be highly non-trivial in practice, see for example the computation in appendix \ref{subapp:fundamental group}.} whether the resulting topology is already part of the set $\mathcal{M}$, and if not, add it to $\mathcal{M}$. This algorithm generates an infinite set of bulk manifolds. Then, when we compute the average of any CFT quantity $\overline{Z[\Sigma]}$ (where $\Sigma$ may be disconnected), we use the dictionary \eqref{eq:sumovertopologies} to sum over all $M\in \mathcal{M}$ with $\partial M = \Sigma$. 

This defines a minimal sum over topologies that is consistent with the statistical bootstrap defined in section \ref{sec:2}. From the point of view of the statistical description of the boundary CFT$_2$, move~\ref{move:disconnected sum} ensures that any statistical moment $\overline{XY}$ contains at least the disconnected contribution $ \overline{X}\,\overline{Y}$, which always exists. Move \ref{move:crossing} ensures crossing symmetry of each moment, as defined in \eqref{eq:crossingOPE}. Lastly, move \ref{move:gluing} ensures that all index contractions exist as demanded by typicality \eqref{eq:generaltypicality}. 

There are three more technical conditions that we have to impose on the cylinder surgery operation $M \#_{P_1 \sim P_2}$ in move \ref{move:gluing}. Namely, consider the Wilson lines $P_1$ and $P_2$ in the manifold $M$ before gluing, and let $m_1$ and $m_2$ be loops around these Wilson lines. These loops define elements of the fundamental group $\pi_1(M)$.\footnote{For $M$ disconnected, say $M=M_1 \sqcup M_2$ with $m_1$ a loop in $M_1$ and $m_2$ a loop in $M_2$, they define elements in $\pi_1(M_1)$ and $\pi_1(M_2)$ respectively.} We assume that these group elements satisfy the following conditions:
\begin{enumerate}[label=\alph*)]
        \item \label{cond:no Omega loops 2} $m_1$ and $m_2$ are not contractible. In other words, the elements they define in $\pi_1(M)$ are non-trivial. 
        \item \label{cond:no parallel rings 2} $m_1$ \emph{or} $m_2$ is primitive, i.e.~not a power of another element in $\pi_1(M)$.
        \item \label{cond:no Wilson line self gluing} For self-gluings, i.e.~where $m_1$ and $m_2$ are in the same connected component, $\langle m_1 \rangle$ does not have a non-trivial intersection with a conjugate of $\langle m_2 \rangle$ in $\pi_1(M)$. 
\end{enumerate}
The first condition \ref{cond:no Omega loops 2} means that neither $m_1$ nor $m_2$ are $\Omega$-loops, since they are not contractible in the bulk. If either $m_1$ or $m_2$ would be contractible, i.e.~bound a disk, then the Wilson line label is projected to the identity Wilson line.\footnote{For example, this happens for the middle edge of the dumbbell diagram as illustrated in \eqref{eq:contractible dumbbell}.}
However, the identity does not behave in a statistical way since it is constant throughout the ensemble. In particular, for the typicality ansatz \eqref{eq:genus2typicality} it was assumed that all operators are different from the identity, because for OPE coefficients with an identity $C_{\id 23}$ the right hand side is simply dictated by conformal symmetry. So condition \ref{cond:no Omega loops 2} is natural from the ensemble point of view. 

The second condition \ref{cond:no parallel rings 2}  appears from requiring that we do not include disconnected moments more than once. To see why this is necessary, consider a Wilson line that has already been Dehn filled, and apply a second Dehn surgery:
\be \label{cond:no parallel rings}
\tikzset{every picture/.style={line width=1.2pt}} %set default line width to 0.75pt        
\,
\ee
Using \eqref{eq:complete to MWK}, this Dehn surgery along parallel loops would lead to the enhancement of $\rho_0(P)\rho_0(\bar P)$ to the \emph{square} of $\rho_{\text{MWK}}(P,\bar P)$, which does not make sense from the CFT point of view. To see why the condition \ref{cond:no parallel rings 2} forbids the move in \eqref{cond:no parallel rings}, we note that 
on the LHS, after a single Dehn filling $p/q$, the loop around the Wilson line admits a $p$-th root and is no longer primitive as an element of $\pi_1$. The converse also holds: if a loop around a Wilson line is not primitive, then we can write the geometry as a Dehn filling of a simpler geometry where the loop is primitive. Thus primitivity of the loop around the Wilson line formalizes the notion of a Dehn filling already being present, which explains condition \ref{cond:no parallel rings 2}.\footnote{We remark that it can be non-trivial to ascertain whether a parallel Dehn-filled arc already exists, since this is in principle a complicated topological condition and the parallel arc might be knotted in non-trivial ways. The cleanest way to formulate this condition is in terms of the fundamental group, which is what we have done in condition \ref{cond:no parallel rings 2}.}

The third condition \ref{cond:no Wilson line self gluing} simply says that we are not allowed to glue the same Wilson line to itself, since that would clearly not make sense from a boundary statistical viewpoint. The phrasing in \ref{cond:no Wilson line self gluing} of this simple statement is a bit technical, due to the fact that $m_1$ and $m_2$ are only defined up to conjugation and thus we have to demand that $m_1$ is not conjugate to $m_2$. In fact, we required the slightly stronger condition that powers of $m_1$ are also not conjugate to powers of $m_2$. This says that we are also not allowed to glue a Wilson line to a power of itself, i.e.~to a loop that winds several times around the same Wilson line.

\subsection{The gravitational machine is hyperbolic}\label{subsec:machine is hyperbolic}
In this section, we prove one of the important results of this paper: 
that the gravitational machine maps hyperbolic manifolds to hyperbolic manifolds.  
This will follow from a rather nice interaction between topology and the conditions imposed on the statistical ensemble. It will turn out that the conditions \ref{cond:no Omega loops 2}, \ref{cond:no parallel rings 2} and \ref{cond:no Wilson line self gluing} on the gluing are precisely what is topologically needed to stay in the class of hyperbolic manifolds. The following section is somewhat mathematical, but central to our argument, which is why it is included in the main text. The reader that trusts us on this point can however jump to section~\ref{subsec:what 3-manifolds are generated}.

The proof proceeds by induction. Namely, suppose $M \in \mathcal{M}$ is hyperbolic. Then we will prove that after applying any of the moves \ref{move:disconnected sum}-\ref{move:gluing}, the resulting manifold $\widetilde{M}$ is also hyperbolic. Since our initial set of manifolds was the set of handlebodies (which are hyperbolic), we conclude that $\mathcal{M}$ only contains hyperbolic manifolds. 

\paragraph{Thurston's theorem.} To prove the induction step, we will examine the homotopy groups of the manifolds produced by the gravitational machine. The main workhorse in the proof is Thurston's hyperbolization theorem, which gives a topological criterion that decides whether a topological 3-manifold admits a hyperbolic structure. More precisely, it can be stated as \cite{Thurston:1982}:
\emph{
\begin{quote}
    The interior of a compact 3-manifold $M$ with nonempty boundary has a hyperbolic structure if and only if $M$ is prime, homotopically atoroidal and not homeomorphic to $(T^2\times [0,1])/\ZZ_2$. 
\end{quote}}
\noindent Let us clarify the terminology used in this theorem. Assuming that the closure of $M$ is \emph{compact} is a relatively weak finiteness condition, ensuring that the hyperbolic manifold is not too wild, for example it does not have infinitely many boundary components. 
A \emph{prime} manifold is a manifold that cannot be written as a non-trivial connected sum, meaning that $M=M_1 \# M_2$ implies that $M_1=\mathrm{S}^3$ or $M_2=\mathrm{S}^3$. Prime is implied and essentially equivalent to being \emph{irreducible}, which is the condition that the second homotopy group is trivial, $\pi_2(M) =0$.\footnote{The only manifold that is prime, but not irreducible is $\mathrm{S}^2 \times \mathrm{S}^1$, which is not hyperbolic.} Thus we will in the following replace prime by irreducible. \emph{Homotopically atoroidal} means that every map of $T^2$ to $M$ which acts injectively on the fundamental groups is homotopic to $\partial M$. This condition is equivalent to the condition that any subgroup of $\pi_1(M)$ isomorphic to $\mathbb{Z}\times \mathbb{Z}$ is conjugate to a subgroup of $\pi_1(\partial M)$.
We will in the following take atoroidal to mean homotopically atoroidal.\footnote{
In the literature, there are two similar but distinct notions of atoroidal 3-manifolds: homotopically atoroidal and geometrically atoroidal. A manifold is geometrically atoroidal if every \emph{embedded} torus with injective $\pi_1$ in $M$ is isotopic to a component of $\partial M$. In contrast, a manifold is homotopically atoroidal if every \emph{immersed} torus, including those with self-intersections, with injective $\pi_1$ is homotopic to a component of $\partial M$. The definition of homotopically atoroidal implies geometrically atoroidal, but the converse does not hold. There are irreducible geometrically atoroidal 3-manifolds $M$ that satisfy $\mathbb{Z}\times \mathbb{Z} \subset \pi_1(M)$, where the $\mathbb{Z}\times \mathbb{Z}$ factor is carried by an immersed torus that cannot be homotoped to $\partial M$. Notable examples include $\Sigma_{0,3}\times$S$^1$ (the product of a 3-punctured sphere with S$^1$), and the complement in S$^3$ of any torus knot. See e.g.~\cite{thurston1986hyperbolic} for more details.} 

We also need the notion of an \emph{incompressible surface}. These are surfaces $S\subset M$ whose fundamental group injects into the fundamental group of the 3-manifold. An incompressible surface is called \emph{boundary parallel} if it can be isotoped into a boundary component of $M$; otherwise, it is called \emph{essential}. Thurston's hyperbolization theorem implies that hyperbolic 3-manifolds do not contain essential tori. Connecting to the 3d gravity vocabulary, we note that a regular neighborhood of an essential torus is topologically a torus wormhole $T^2\times I$, showing that any bulk topology containing a torus wormhole is automatically off-shell. 

As an example of the above criteria, consider the complement in S$^3$ of the following handlebody-knot (to be discussed in section~\ref{subsec:handlebody-knots}), called $\mathbf{5}_4$ in table \ref{table:handlebodyknots}:
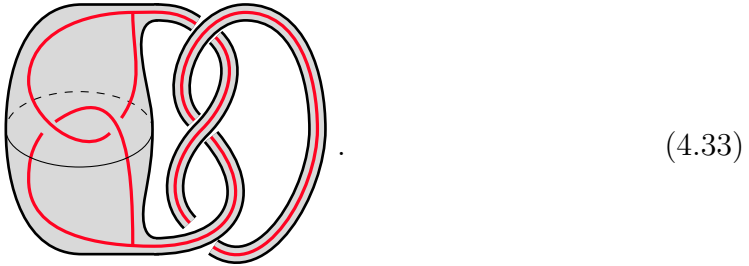
\begin{equation}
\label{eq:54}
\vcenter{\hbox{

}}.
\end{equation}
The handlebody-knot (in red) has a complement in $\mathrm{S}^3$ with a genus-2 boundary, but it is not atoroidal. In the figure above, the essential torus is highlighted in gray. Thus, by Thurston's theorem the $\mathbf{5}_4$-complement does not carry a hyperbolic metric. 

\paragraph{Proving hyperbolicity.} We now come back to the proof that the gravitational machine does not map out of the space of hyperbolic 3-manifolds. Since we only cut and paste manifolds one-by-one, the procedure does not change the compactness properties of $M$. Taking the disconnected union of manifolds (move \ref{move:disconnected sum}) as well as performing crossing transformations (move \ref{move:crossing}) do not map us out of hyperbolicity. For the latter statement, we use Mostow rigidity \cite{Sullivan} which ensures that a hyperbolic structure in the bulk either exists for \emph{all} boundary moduli or for none of them. In particular, in the canonical picture crossing transformations act on the boundary moduli, which do not affect hyperbolicity of the bulk manifold.

Moreover, the non-hyperbolic exceptional manifold $(T^2 \times [0,1])/\ZZ_2$ is not generated by the machine. The $\ZZ_2$ quotient flips the interval and corresponds to the covering transformation of $T^2$ such that $T^2/\ZZ_2$ is a Klein bottle. This exceptional manifold does have a genus-1 boundary, but it is not one of the $\mathrm{PSL}(2,\ZZ)$ family of Euclidean black holes, and these are the only manifolds the machine generates at genus 1. This follows directly from conditions \ref{cond:no parallel rings 2} and \ref{cond:no Wilson line self gluing}. Namely, at genus 1, there is a single Wilson line, so self-gluing is not permitted. The only available operation is the cylinder surgery of two BTZ black holes, $M_1$ and $M_2$. This either produces again a single BTZ black hole (for the trivial Dehn filling) or it introduces parallel rings, which are not allowed by condition \ref{cond:no parallel rings 2}.

Thus, it only remains to prove that the cylinder surgery operation, i.e.~move \ref{move:gluing}, keeps $M$ irreducible and atoroidal. For irreducibility, we will use a simple geometric argument known as the innermost/outermost curve argument. Our proof that the manifold stays atoroidal is entirely group-theoretic and requires some input from geometric group theory that we review in appendix~\ref{app:Bass-Serre theory}. By induction, it will follow that any manifold produced by the gravitational machine is hyperbolic.

\paragraph{Irreducible.} Let $M$ be the manifold before gluing and $\widetilde{M}=M \#_{P_1 \sim P_2}$ the manifold after gluing. For the irreducibility part of the proof, it doesn't matter whether we are gluing two disconnected components of $M$ or the same component to itself. As the induction hypothesis, we assume that $M$ is hyperbolic.

Recall that to define cylinder surgery, we had to excise the interiors of a pair of 2-punctured spheres in $M$. This does not change the homotopy groups of $M$. Thus, we will not distinguish between $M$ and $M$ with the balls removed. In particular, we will write $\partial M= \Sigma \cup \Sigma_{0,2} \sqcup \Sigma_{0,2}$, where $\Sigma$ is the original boundary of $M$ and $\Sigma_{0,2} \sqcup \Sigma_{0,2}$ is the disjoint union of two 2-punctured spheres. In $\widetilde{M}$, these 2-punctured spheres get identified.
We will only use the $\Omega$-loop condition \ref{cond:no Omega loops 2}, which says that neither $m_1$ nor $m_2$ is contractible in $M$. It also implies that a power $m_i^n$ for $n \in \ZZ \setminus \{0\}$ is not contractible. Otherwise, $\pi_1(M)$ would have torsion, which is impossible for hyperbolic manifolds. In other words, the inclusion $\pi_1(\Sigma_{0,2}) \hookrightarrow \pi_1(M)$ is an injection, for both of the boundary components $\Sigma_{0,2}$.

As a proof by contradiction, suppose that $\pi_2(\widetilde{M})\ne 0$. Choose a non-trivial element in $\pi_2(\widetilde{M})$, which can be represented by an embedded 2-sphere $S$.\footnote{For this one formally uses the sphere theorem of 3-manifolds, saying that any element of $\pi_2(M)$ may always be represented as a properly embedded sphere.} 
W.l.o.g., assume that $S$ intersects the 2-punctured sphere $\Sigma_{0,2}$ transversely in a number of curves $C_i$, with $i=1,\dots,n$. We choose the representative and the non-trivial element in $\pi_2(\widetilde{M})$ such that the number of connected components $n$ of the intersection is minimal. This $n$ cannot be zero since then $S$ would lie entirely in $M$. But since $\pi_2(M)=0$ by hypothesis (for any connected component of $M$), this implies that $S$ is null-homotopic in $M$ and thus  in $\widetilde{M}$, contradicting non-triviality of $S$ in $\pi_2(\widetilde{M})$.

Now choose an innermost curve of the intersection, i.e.~a $C_i$ that bounds a disk on $S$. Then $[C_i]$ cannot be a non-trivial element of $\pi_1(\Sigma_{0,2})$. For if it were, the disk would define a null-homotopy of $[C_i]$ in one of the connected components of $M$, thus contradicting the injectivity assumption of the maps $\pi_1(\Sigma_{0,2}) \xhookrightarrow{\quad} \pi_1(M)$. Thus $C_i$ has to be a contractible curve on $\Sigma_{0,2}$. 
\begin{figure}
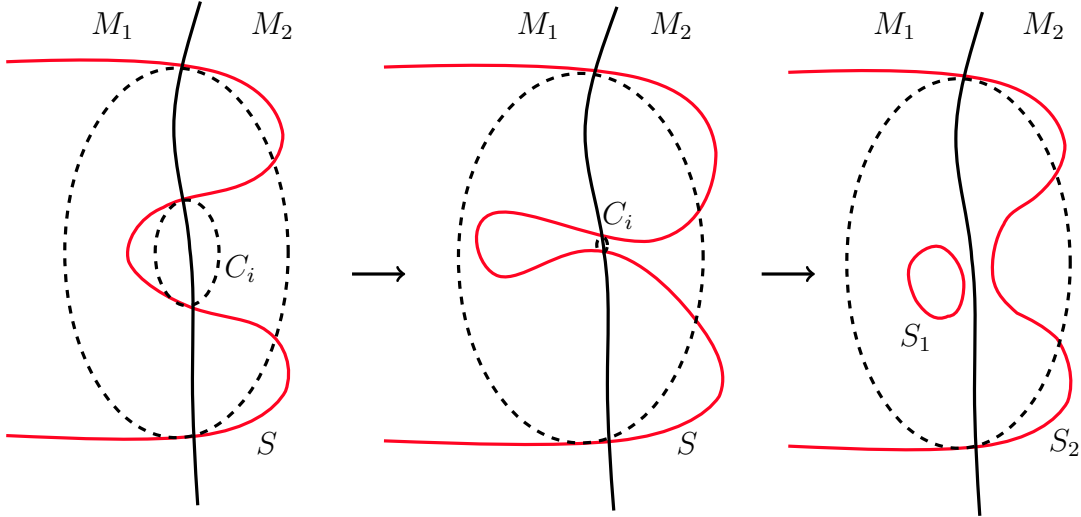

\begin{center}

\end{center}
\caption{Pinching the innermost curve leads to two surfaces with fewer intersections with the boundary $\partial M_1=\partial M_2$.} \label{fig:pinching curve}
\end{figure}
Since $C_i$ is contractible on $\Sigma_{0,2}$ and doesn't contain any other intersection curves in its interior, we can isotope it to a point. This pinches the 2-sphere $S$ along the curve $C_i$ and we obtain two 2-spheres $S_1$ and $S_2$ connected at a single point. If we separate them, at least one of the 2-spheres still defines a non-trivial element of $\pi_2(\widetilde{M})$ by the definition of $\pi_2$. See figure~\ref{fig:pinching curve} for an illustration of this procedure. Say $[S_1]$ is non-trivial. However, the number of intersections with $\Sigma_{0,2}$ of $S_1$ is at most $n-1$ since we removed the curve $C_i$. This contradicts the assumption that we chose $n$ minimal. Thus we have arrived at a contradiction and conclude that $\pi_2(\widetilde{M})=0$.

\paragraph{Computing $\pi_1$.} Next, we want to show that the gluing move \ref{move:gluing} also preserves the fact that the manifold $M$ is atoroidal. This can be checked by computing $\pi_1(\widetilde{M})$. The computation differs depending on whether we glue disconnected components or the same connected component to itself.

For gluing two disconnected components $M_1 \#_{P_1\sim P_2} M_2$, this is a standard application of the Seifert-Van-Kampen theorem. Before gluing, we have $M = M_1\sqcup M_2$, with $M_1 \cap M_2 \cong \Sigma_{0,2}$ with fundamental group $\pi_1(\Sigma_{0,2}) \cong \ZZ$. The theorem tells us that after gluing, the fundamental group has the form of an amalgamated product,
\be 
\pi_1(\widetilde{M})=\pi_1(M_1) *_\ZZ \pi_1(M_2)\,. \label{eq:amalgamated product}
\ee
Let us recall what amalgamation means. Suppose, for $i=1,2$, that $\pi_1(M_i)$ is generated by a set $S_i$ with some relations $R_i$, i.e.~has the presentation $\pi_1(M_i)=\langle S_i \, | \, R_i \rangle$. Then the free product $\pi_1(M_1)*\pi_1(M_2)$ has the presentation $\langle S_1,\, S_2 \, | \, R_1,\, R_2 \rangle$, so we are not imposing any further relations than the ones already present in the two factors. In particular, we are not imposing that generators in $S_1$ and $S_2$ commute (which would be the case in the direct product). With amalgamation, denoted by the subscript $*_H$, we are additionally identifying a joint subgroup $H$ in $\pi_1(M_i)$, in this case $\pi_1(\Sigma_{0,2})$, i.e.~we are adding the additional relations $\iota_1(g)=\iota_2(g)$ for all $g$ in the subgroup, where $\iota_i: H \hookrightarrow \pi_1(M_i)$ are the inclusions. In our case, we hence add the additional relation $m_1=m_2$.\footnote{In the above constructions, we have to choose a basepoint. We choose a basepoint for $\pi_1(M_i,x_i)$ in the respective boundary $ x_i\in \Sigma_{0,2} \subset M_i$, so that $x_1$ and $x_2$ get identified after identifying the two boundaries.} 

When writing the amalgamated product \eqref{eq:amalgamated product}, we used the fact that $\ZZ$ is naturally a subgroup of $\pi_1(M_1)$ and $\pi_1(M_2)$, i.e.~that the inclusion is injective according to \ref{cond:no Omega loops 2}. The amalgamated product can be viewed as a group-theoretic analogue of the cylinder surgery, where we `glue' two groups along a common subgroup.

In case that we are performing a self-gluing $\widetilde{M} = M\#_{ P_1\sim P_2}$, a similar presentation of the fundamental group exists. Let $M$ be the original manifold before gluing. Let $m_1$ and $m_2$ be (based) loops around the two Wilson lines that are to be identified, so that the fundamental group of $M$ has the presentation:
\be 
\pi_1(M)=\big \langle S,m_1,m_2 \, | \, R \big \rangle
\ee
for some relations $R$. Then $\pi_1(\widetilde{M})$ takes the form of a so-called HNN extension \cite{HNN_extension}. This means that $\pi_1(\widetilde{M})$ is obtained by adding a new generator $t$ together with the relation $m_2=t m_1 t^{-1}$, i.e.
\be 
\pi_1(\widetilde{M})=\big \langle S,m_1,m_2,t \, | \, R,\, m_2=t m_1 t^{-1} \big \rangle\,. \label{eq:fundamental group self-gluing}
\ee
The generator $t$ is usually called the stable letter. 
Geometrically, it corresponds to the loop through the `wormhole' $\Sigma_{0,2}\times I$, while the self-gluing identifies $m_1$ with the conjugated loop $t m_2 t^{-1}$ that also runs through the wormhole. Notice that the choice of $m_1$ and $m_2$ is only well-defined up to conjugation, but the resulting groups are isomorphic.\footnote{This can be derived using the groupoid version of the Seifert-van-Kampen theorem. For a more down-to-earth explanation, see \cite[chapter 5]{Wall1979}.} 

In writing down the HNN extension \eqref{eq:fundamental group self-gluing}, we again made use of the injectivity of the inclusions of $\pi_1(\Sigma_{0,2})$ in $\pi_1(\widetilde{M})$. More generally, the HNN extension identifies two isomorphic subgroups $H$ and $H'$ of $\pi_1(M)$, usually denoted by $\pi_1(M)*_{H\sim H'}$. Thus it provides the group-theoretic analogue of the self-gluing $M\#_{ P_1\sim P_2}$. More details on the underlying geometric group theory can be found in appendix~\ref{app:Bass-Serre theory}.

\paragraph{Malnormality.} We now establish two further properties of the amalgamated product and the HNN extension in our case, which will let us conclude that they do not have $\ZZ \times \ZZ$ subgroups. Let us write $G_i=\pi_1(M_i)$ in the disconnected case and $G_1=\pi_1(M)$ in the connected case, and let $H_i =\langle m_i \rangle  \cong \ZZ$, $i=1,\, 2$ for the joint isomorphic subgroups that we are identifying. W.l.o.g., we can assume that $m_1$ is primitive by condition \ref{cond:no parallel rings 2}.
For the case of a connected gluing, the condition \ref{cond:no Wilson line self gluing} that a Wilson line cannot be glued to itself states that
\be 
H_1 \cap g H_2 g^{-1}=\{1\}, \quad \text{for all } g \in G_1\,. \label{eq:H1 H2 intersection condition}
\ee
We next claim that $H_1 \subset G_1$ is a \emph{malnormal} subgroup in both the disconnected and the connected case. This means that 
\be 
 H_1 \cap g H_1 g^{-1}  =\{1\}, \quad \text{for all } g \in G_1 \setminus H_1 \,.\label{eq:malnormality}
\ee
To show this, assume that $g m_1^k g^{-1}=m_1^\ell$ for $k,\, \ell \in \ZZ \setminus \{0\}$. Since $m_1$ is primitive, the left-hand side is a $k$-th power of a primitive element and the right-hand side is an $\ell$-th power of a primitive element. Thus, they can only be equal if $|k|=|\ell|$. Furthermore, since $m_1$ is primitive, we only have to analyze $g m_1 g^{-1}=m_1$ and $g m_1 g^{-1}=m_1^{-1}$. In either case, $[m_1,g^2]=1$. Since $G_1$ does not have $\ZZ \times \ZZ$ subgroups by the induction hypothesis, $m_1$ and $g^2$ have to lie in the same infinite cyclic subgroup, which is necessarily generated by $m_1$, since it is primitive. Thus $g^2=m_1^n$ for some $n$. Since $m_1$ is again primitive, it follows that $n$ has to be even and since the group is torsion-free we can take the square root of the relation to conclude that $g \in H_1 =\langle m_1 \rangle$. This establishes \eqref{eq:malnormality}.

\paragraph{Atoroidal.} We can now finish the argument quickly by using geometric group theory \cite{SerreTrees}. We give some more mathematical  background on this theory in appendix~\ref{app:Bass-Serre theory}. Let us first treat the amalgamated product case, where we have the following theorem \cite{Karrass:1971}.\footnote{We will actually need a small generalization that is for example explained in \cite{mathoverflow_amalgamatedproduct}. In \cite{Karrass:1971}, it is assumed that $H$ is malnormal in $G_1$ \emph{and} $G_2$. We explain the derivation of this more general version in appendix~\ref{app:Bass-Serre theory}.} Consider an amalgamated product $G=G_1*_H G_2$. Assume that $H$ is malnormal in $G_1$. Let $g \in G$. Then the centralizer $\mathcal{C}(g)$ in $G$ is either infinite cyclic or $\mathcal{C}(g)$ is conjugate to $G_1$ or $G_2$.

We set everything up such that we can apply the theorem. Centralizers in $G_1$ and $G_2$ are all infinite cyclic groups (since $G_i$ does not have torsion or $\ZZ \times \ZZ$ subgroups). Thus, in both possibilities in the above theorem, the centralizer is an infinite cyclic group. Thus $G$ does not have $\ZZ \times \ZZ$ subgroups.

We have a similar theorem for the HNN extension. Consider an HNN extension $G=G_1 *_{H_1 \sim H_2}$. It states: if $H_1$ does not intersect any of the conjugates of $H_2$ and is malnormal in $G_1$, then the centralizer $\mathcal{C}(g)$ of $g \in G$ is either infinite cyclic or $\mathcal{C}(g)$ is conjugate to $G_1$. The assumptions of this theorem are guaranteed by \eqref{eq:H1 H2 intersection condition} and \eqref{eq:malnormality}. Since $G_1$ does not have $\ZZ \times \ZZ$ subgroups by the induction hypothesis, it follows that also $G$ does not have $\ZZ \times \ZZ$ subgroups.

Thus in both gluing types, we do not get $\ZZ \times \ZZ$ subgroups of $\pi_1(\widetilde{M})$, and hence $\widetilde{M}$ is homotopically atoroidal. Using Thurston's theorem, this concludes our proof that the operations of the machine preserve hyperbolicity.

\subsection{What 3-manifolds are generated?} \label{subsec:what 3-manifolds are generated}
At this point, we constructed a minimal set of manifolds $\mathcal{M}$ compatible with the statistical bootstrap, and we saw that this set only contains hyperbolic manifolds. This is remarkable, because it implies that in a statistical ensemble that is fully consistent with crossing symmetry and typicality (but without spectral correlations), it suffices to sum over \emph{on-shell} topologies only. This leads to two natural follow-up questions: first, does $\mathcal{M}$ contain non-handlebodies? And second, are \emph{all} hyperbolic manifolds with appropriate boundary conditions generated by the gravitational machine?  

The first question is answered affirmatively, as we have already alluded to in section \ref{subsec:gravitational machine 1}. In this section, we will prove this for the simplest application of the machine at genus 2. However, the answer to the second question is negative, and we will demonstrate that some hyperbolic manifolds are \emph{not} generated. In particular, the machine does not generate \emph{acylindrical} manifolds, as will be explained below.

\paragraph{The gravitational machine produces non-handlebodies.}

 First, it is not so obvious whether the moves of the gravitational machine that we described actually produce non-handlebodies, since 3-manifolds can often be represented in different ways. To show that the machine does in fact produce non-handlebodies, it suffices to examine the fundamental groups of some of the resulting 3-manifolds. The fundamental group of a genus two handlebody is the free group on two generators, $\langle a,b\rangle$; if the machine produces a manifold with a different fundamental group, it cannot be a handlebody. 
 
 The simplest examples of non-handlebody geometries generated by the machine are the manifolds given by three independent Dehn surgeries:
\begin{equation}
\label{manifold}
\vcenter{\hbox{
\begin{tikzpicture}[scale=1]
\draw[very thick, wilsonred] (0,1) -- (0,-1);
\draw[very thick, wilsonred] (0,0) ellipse (2 and 1);
\draw[very thick] (-0.07,0.2) arc (100:440:0.35 and 0.25);
\draw[very thick] (-2,0.2) arc (100:440:0.35 and 0.25);
\draw[very thick] (1.9,0.2) arc (100:440:0.35 and 0.25);
\node at (-0.7,1.2) {$m_1$};
\node at (0.75,1.2) {$m_2$};
\node at (0.4,0.6) {$m_3$};
\node at (-2,-0.6) {$l_1$};
\node at (-0.2,-0.6) {$l_2$};
\node at (2.1,-0.6) {$l_3$};
\end{tikzpicture}
}}
\end{equation}
where the loops labeled by $l_i$, $i=1,2,3$, have been Dehn filled with the parameters $p_i/q_i$, and $(p_i,q_i)$ are co-primes with $p_i\neq 0$. The fundamental group can be computed via the so-called Wirtinger presentation (see chapter 9 of \cite{RonaldBrown} or appendix~\ref{subapp:fundamental group}). This yields a fundamental group with six generators and the relations 
\begin{equation}
\label{fundgroup}
   \pi_1(M) =  \left\langle 
    m_i,l_i\vert \;\;
    [m_i,l_i] = 1,\; 
    l_i^{p_i} m_i^{q_i} = 1,\;
    m_1 m_2m_3 = 1
    \right\rangle.
\end{equation}
The generators $m_i$ and $l_i$ correspond to loops around the arcs depicted in \eqref{manifold}. The operation of Dehn filling a loop amounts to the relation $l^pm^q =1$, where $l,m$ denote the longitude and meridian of the filled torus. 

We can reduce the group \eqref{fundgroup} to one with only three generators by simplifying the relations associated with each Dehn filling. We focus on the abelian subgroups generated by each meridian and longitude. Since the argument applies independently to each pair of generators, we omit the subscript $i$. First, we note that because $p,q$ are relatively prime, there exist integers $u_{1,2}$ such that 
\begin{equation}
    u_1 p + u_2 q =1. 
\end{equation}
This means that the following generators also form a basis for the subgroup generated by $m$ and $l$:
\begin{equation}
    v = l^{p} m^{q},\quad w = l^{-u_2} m^{u_1}.
\end{equation}
The inverse relations are: 
\begin{equation}
    l = v^{u_1}w^{-q}, \quad m =  v^{u_2}w^p.
\end{equation}
In this new basis, the relation $l^pm^q=1$ becomes $v=1$, effectively eliminating one generator from the group. The resulting presentation is then given by: 
\begin{equation}
     \pi_1(M) =  \left< w_1,w_2,w_3\, \vert \,
w_1^{p_1} w_2^{p_2} w_3^{p_3} = 1
    \right>. \label{eq:fundamental group triple Dehn filling}
\end{equation}
We can see that when one of the Dehn fillings is trivial, e.g.~$p_1 =1$, this group reduces to the free group on two elements. This is expected as one can always perform any two Dehn fillings on the handles of the genus-2 handlebody, corresponding to two modular transformations on the two dumbbells, which are part of the mapping class group (recall \eqref{eq:genus 2 to genus0,4 reduction}). By contrast, when $p_1,p_2,p_3 \neq 1$, this group is no longer free and thus $M$ is not a handlebody.\footnote{This should be intuitively obvious, but can also be rigorously proved from Whitehead's algorithm described in appendix~\ref{subapp:fundamental group}. There is no Whitehead automorphism that reduces the length of the word $w_1^{p_1}w_2^{p_2}w_3^{p_3}$ for $p_1,p_2,p_3>1$. Thus it cannot be equivalent to, say, the relation $w_3=1$ and the group is not isomorphic to a free group.}

The above example is only one simple instance of the gravitational machine. As we keep adding crossing transformations, as in \eqref{eq:adding crossing}, and combining them with cylinder surgeries, we can generate a multitude of hyperbolic non-handlebodies.  In section \ref{sec:5} we will give a number of notable examples of more complicated non-handlebodies produced by the machine.

\paragraph{The gravitational machine only produces cylindrical manifolds.}
Although $\mathcal{M}$ contains many hyperbolic non-handlebodies, there is an important class of non-handlebody topologies that are \emph{not} produced by the machine: the so-called `acylindrical' 3-manifolds. These are manifolds without embedded essential cylinders (annuli). The formal definition of an essential cylinder is discussed in appendix \ref{subapp:boundary compressibility}, but it is similar to the notion of an essential torus described in section~\ref{subsec:machine is hyperbolic}: it should be incompressible in the bulk and not parallel to the boundary.  

 While hyperbolic manifolds are necessarily atoroidal by Thurston's theorem, they are not necessarily acylindrical. For example, a genus-2 handlebody is cylindrical---below, we have drawn a solid torus with a handle attached (thus making a genus-2 handlebody) and we have highlighted an embedded essential annulus in red, which cuts the torus like a bagel:
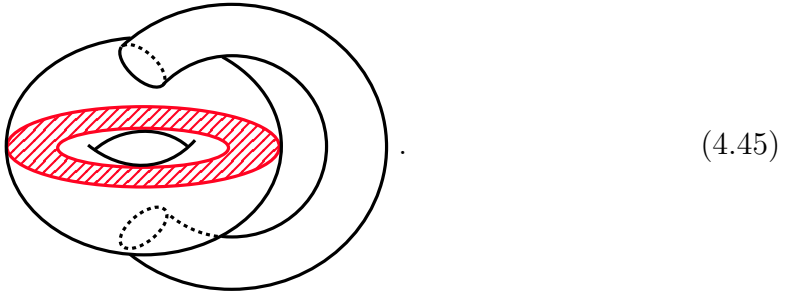
\begin{equation}
 % Pattern Info
\tikzset{
pattern size/.store in=\mcSize, 
pattern size = 5pt,
pattern thickness/.store in=\mcThickness, 
pattern thickness = 0.3pt,
pattern radius/.store in=\mcRadius, 
pattern radius = 1pt}
\makeatletter
\pgfutil@ifundefined{pgf@pattern@name@_u3ckhdq9l}{
\pgfdeclarepatternformonly[\mcThickness,\mcSize]{_u3ckhdq9l}
{\pgfqpoint{0pt}{0pt}}
{\pgfpoint{\mcSize+\mcThickness}{\mcSize+\mcThickness}}
{\pgfpoint{\mcSize}{\mcSize}}
{
\pgfsetcolor{\tikz@pattern@color}
\pgfsetlinewidth{\mcThickness}
\pgfpathmoveto{\pgfqpoint{0pt}{0pt}}
\pgfpathlineto{\pgfpoint{\mcSize+\mcThickness}{\mcSize+\mcThickness}}
\pgfusepath{stroke}
}}
\makeatother
\tikzset{every picture/.style={line width=1.2pt}} %set default line width to 0.75pt        
\ . \label{eq:bulk incompressible cylinder}
\end{equation}
Another example is the Maldacena-Maoz wormhole $\Sigma_g\times I$: take any non-contractible loop $C$ on $\Sigma_g$, then the essential cylinder is $C\times I$, as illustrated in \eqref{eq:embedded cylinder}. 

In fact, our main observation is that (with the exception of the solid torus), all the manifolds produced by the gravitational machine are cylindrical. This is simple to see. The starting point of the gravitational machine are the genus-$g$ handlebodies, which for $g\geq 2$ are cylindrical as remarked above.\footnote{The only exception is $g=1$: for a solid torus, each bulk incompressible annulus (which slices the manifold in half like a bagel) is boundary parallel.} Every time we apply the cylinder surgery (move \ref{move:gluing}), we introduce a cylinder into the geometry, namely the cylinder along which we glue. This cylinder is bulk incompressible because of condition \ref{cond:no Omega loops 2} on the gluing. It is also easily seen to be non-boundary parallel by the same criterion, and hence it is an essential cylinder.\footnote{The only case where we have to be slightly careful is the Dehn filling operation discussed in section~\ref{subsec:gravitational machine 1}. For $p>1$, the Dehn filling changes the fundamental group and thus the glued annulus is indeed non-boundary parallel (if it were boundary parallel, then cutting off the part of the manifold through which the cylinder is isotoped to the boundary is a homeomorphism and cannot change the fundamental group). The Dehn filling with $p=1$ is just a Dehn twist, which only changes the boundary moduli and does not change the fact that the manifold is cylindrical.}
Thus it follows that the gravitational machine only produces cylindrical manifolds.

However, there exist also acylindrical (i.e.~not cylindrical) hyperbolic manifolds. An example is Thurston's `tripus' manifold \cite{thurston_book}, which is homeomorphic to the $\mathbf{5}_3$ handlebody-knot. This example will be discussed in detail in section~\ref{subsec:handlebody-knots}. 

There is a nice characterization of acylindrical hyperbolic manifolds, again due to Thurston \cite{Thurston1986HyperbolicI}, which (roughly) says that 
a hyperbolic 3-manifold $M$ with boundary admits a hyperbolic structure \emph{with totally geodesic boundary} if and only if $M$ is acylindrical. 
See \cite{yoshida2013minimal} for self-contained exposition to this theorem, and see below for the more precise statement of the theorem.
Recall that in section \ref{subsec:machine is hyperbolic}, we used the most general version of Thurston's hyperbolization theorem, which guarantees the existence of a hyperbolic metric regardless of whether this metric is of finite or infinite volume and regardless of whether the boundary of the convex core is geodesic or a pleated surface. By contrast, for geodesic boundaries (i.e.~with vanishing extrinsic curvature) the relevant version of Thurston's theorem reads:
\emph{
\begin{quote}
Let $M$ be a compact, orientable, irreducible 3-manifold. Then $M$ admits a finite-volume hyperbolic structure where $\partial M$ is a totally geodesic boundary if and only if:
\begin{enumerate}
    \item $M$ is atoroidal.  
    \item $M$ is $\partial$-irreducible. 
    \item $M$ is acylindrical.
\end{enumerate}
\end{quote}
}
In appendix \ref{subapp:boundary compressibility}, we define what $\partial$-irreducible means, and explain the simple direction of the implication (why the existence of essential cylinders obstructs the presence of a hyperbolic metric with totally geodesic boundary). 

\begin{figure}
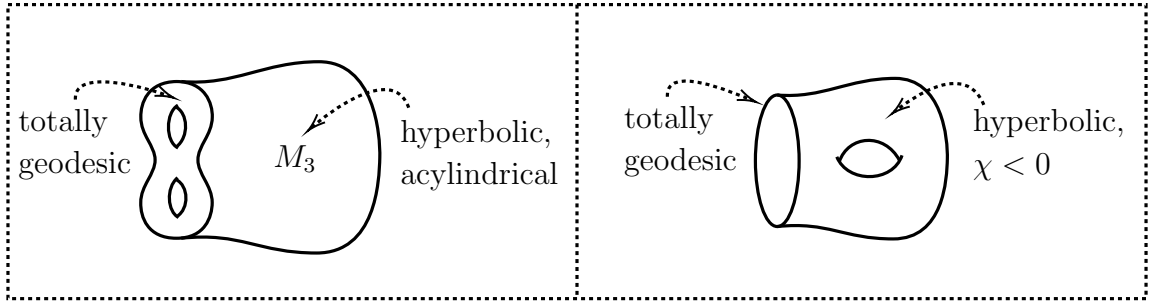

    \centering
    \tikzset{every picture/.style={line width=1.2pt}} %set default line width to 0.75pt        

    \caption{Cartoon comparing 3d gravity to 2d JT gravity: in 3d, the hyperbolic manifolds with totally geodesic boundary are acylindrical, and in 2d the hyperbolic manifolds with geodesic boundaries have negative Euler characteristic.}
    \label{fig:acylindrical}
\end{figure}

\paragraph{Geodesic boundaries and comparison to JT gravity.}
The above theorem suggests an analogy to JT gravity \cite{Saad:2019lba}, illustrated in figure \ref{fig:acylindrical}. Namely, in JT gravity (with negative cosmological constant), one evaluates the path integral on a Riemann surface $\Sigma_{g,n}$ with Euler characteristic $\chi(\Sigma_{g,n})<0$ by gluing trumpets to a bordered Riemann surface with geodesic boundaries. The only manifolds for which this cannot be done are the disk and the cylinder (double trumpet): if we remove the trumpets, we are not left with a 2-manifold. 
Similarly, for 3-manifolds we can also ask if it is possible to amputate the 3-dimensional trumpets, i.e.~the asymptotic regions of the manifold.  A formal way to do so is to take the \emph{convex core} of the 3-manifold, explained for example in \cite{Schlenker:2022dyo}. The question now is whether the convex core admits a finite-volume hyperbolic metric with a geodesic boundary surface, i.e.~with extrinsic curvature $K=0$.

Here, it is important to distinguish between geodesic boundaries and pleated boundaries. As explained in section \ref{sec:fixedP} and \cite{Hartman:2025ula}, cylindrical manifolds may admit a finite-volume hyperbolic metric if we allow for a pleated boundary, i.e.~a surface that is geodesic except for `corners’ along a finite collection of geodesics. By Thurston's theorem quoted above, none of the manifolds produced by the machine carry a hyperbolic metric with totally geodesic boundary. Thus, they are in a sense analogous to the disk or the double trumpet of JT-gravity, and the gravitational machine does not generate all other geometries. In the \hyperref[sec:discussion]{discussion}, we comment on the implications of this rather striking outcome. 

%%%%%%%%%%%%%%%%%%%%%%%%%%%%%%%%%%%%
\section{The gravitational machine at work}
\label{sec:5}
In the previous section, we have seen that the topologies generated by the action of the gravitational machine on handlebodies are not always themselves handlebodies, and so we learn that consistency of the boundary statistical ensemble with crossing symmetry and typicality requires topologies that go beyond the handlebody sum.
In this section, we discuss in detail a number of examples of non-handlebodies generated by the gravitational machine, directly in quantum gravity using Virasoro TQFT, and explore their consequences for the boundary ensemble.

First, we show that the machine generates the Maldacena-Maoz wormhole, i.e.\ the two-boundary Euclidean wormhole with topology $\Sigma_{g,n}\times I$ \cite{Maldacena:2004rf}, which we construct using the cylinder surgery move \ref{move:gluing}.  Then we show that the prototypical single-boundary non-handlebody in 3d gravity, namely the genus-2 `twisted $I$-bundle' \cite{Yin:2007at,Collier:2024mgv}, is generated by the machine as a triple application of Dehn surgery on a genus-2 handlebody. In the boundary ensemble this topology contributes to $\overline{\rho_1\rho_2\rho_3C_{123}C^*_{123}}$ and is interpreted as one of the terms in the $\PSL(2,\mathbb{Z})$ sum that defines the average density of states, weighted by the $C_0$ formula for the leading contribution to the variance $\overline{C_{123}C^*_{123}}$. 

There are also non-handlebodies produced by the machine that give genuine non-perturbative corrections to the statistics of the structure constants themselves, which go beyond what is required by identity dominance and modular invariance. To demonstrate this, we discuss another class of hyperbolic topologies generated by the machine known as `handlebody-knots,' which are complements of knotted trivalent graphs in $\mathrm{S}^3$. The non-trivial irreducible genus-2 handlebody-knots with up to seven crossings are fully classified \cite{ISHII:2012,bellettini2025tablegenushandlebodyknotsseven}, and we work through all examples with fewer than six crossings in detail. Each of these examples showcases one of the results obtained in section \ref{sec:minimal completion}. These have been summarized in table \ref{table:handlebodyknots}.

\subsection{Euclidean wormholes} \label{subsec:Euclidean wormhole}
The most standard example of a hyperbolic non-handlebody geometry is the Euclidean Maldacena-Maoz wormhole \cite{Maldacena:2004rf}. Its boundary is a pair of Riemann surfaces of genus $g$ and $n$ punctures (we assume that $2g-2+n>0$). In the case that the moduli of both boundaries agree, it is simple to write down the hyperbolic metric:
\be 
\d s^2_{\Sigma_{g,n}}=\d \rho^2+\cosh^2 \rho \, \d s_{\Sigma_{g,n}}^2\ ,
\ee
where $\d s_{\Sigma_{g,n}}^2$ denotes the hyperbolic metric of $\Sigma_{g,n}$. If the boundary moduli are not the same, we have a quasi-Fuchsian wormhole which still admits a hyperbolic metric.

In the bulk-boundary dictionary \eqref{eq:sumovertopologies}, the above wormhole contributes to the $2(2g-2+n)$-th moment of structure constants. This wormhole topology is very simply generated by the machine as follows. Choose a pair of pants decomposition of the surface $\Sigma_{g,n}$. The cycles $C_i$ in the decomposition uplift to cylinders $C_i\times I$ in the 3-dimensional manifold, which we can picture for the case $g=0$, $n=4$ as follows:
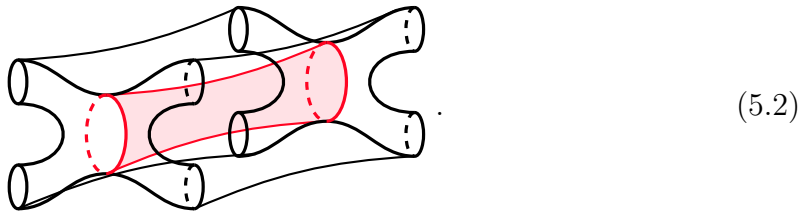
\be \label{eq:embedded cylinder}
\ .
\ee
Thus we can in general glue wormholes with $g=0$ and $n=3$ along those cylinders with the gluing operation \ref{move:gluing} to obtain the general Euclidean wormhole from the basic one with $\Sigma_{0,3}$. This makes sense from the point of view of the statistical ensemble, since the wormhole just describes the quadratic contractions of all the structure constants involved \cite{Chandra:2022bqq}. This quadratic moment comes from $\Sigma_{0,3}$. However $\Sigma_{0,3}\times I$ \emph{is} actually just a genus-2 handlebody, which becomes manifest when we draw it as in \eqref{eq:theta graph MM wormhole}. Alternatively, its fundamental group is free on two generators, showing that the manifold is topologically indeed a handlebody. Thus we learn that, not surprisingly, the Euclidean wormhole is indeed generated by the gravitational machine.

More generally, also multi-boundary wormholes, such as those studied in \cite{Collier:2024mgv,deBoer:2024mqg}, are straightforwardly constructed using cylinder surgery: given a product of structure constants $X_1\, X_2\cdots X_n$, where each $X_i$ represents the OPE decomposition of a genus $g_i$ boundary Riemann surface (e.g.~$X_1 = C_{123}C_{123}^*$), the connected contractions between the indices of the $X_i$'s are geometrized by the surgery of the corresponding pairs of Wilson lines. In this way, it is straightforward to reconstruct the multi-boundary wormhole computations of \cite{Chandra:2022bqq,deBoer:2024mqg,Collier:2024mgv} using the gravitational machine. 

\subsection{Twisted \texorpdfstring{$I$}{I}-bundles} \label{subsec:twisted I bundles}
Another class of well-studied hyperbolic 3-manifolds are so-called twisted $I$-bundles, which were studied in \cite{Yin:2007at, Maxfield:2016mwh} as an example of non-handlebody saddlepoints of the gravitational path integral in 3d gravity. Let us discuss the simplest of those. Consider the Maldacena-Maoz wormhole $\Sigma_2 \times I$. We choose the moduli of the genus-2 surface $\Sigma_2$, such that it admits an orientation reversing fixed-point free diffeomorphism $\Phi: \Sigma_2 \to \Sigma_2$ with the quotient being the non-orientable surface of Euler characteristic $-1$ (also known as the van Dyck surface and topologically equal to a sphere with three crosscaps). In other words, $\Phi$ is the deck transformation of the orientation double-cover of $\Sigma_2/\Phi$. Then $\Phi$ extends to the 3-manifold $\Sigma_2 \times I$ by reflecting the interval $x \to 1-x$. Since $x \to 1-x$ is also orientation reversing, $\Phi$ \emph{preserves} the orientation of $\Sigma_2 \times I$ and the quotient $M=(\Sigma_2 \times I)/\Phi$ is an orientable hyperbolic 3-manifold with a single genus-2 boundary. 

We will now demonstrate that $M$ is generated by the machine in a simple way. Namely, we claim that the twisted $I$-bundle corresponds to the following Dehn filling of a genus-2 handlebody:
\begin{equation}
M\cong\vcenter{\hbox{
\begin{tikzpicture}
    \draw[line width=7pt, black] (0,-1.5) -- (0,1.5);
    \draw[line width=7pt, black] (0,0) ellipse (2 and 1.5);
    \draw[line width=5pt, black!15] (0,0) ellipse (2 and 1.5);
    \draw[line width=5pt, black!15] (0,-1.5) -- (0,1.5);
    \draw[line width=3pt,white] (-0.18,0.2) arc (120:420:0.35 and 0.25);
    \draw[line width=3pt,white] (-2.15,0.2) arc (120:420:0.35 and 0.25);
    \draw[line width=3pt,white] (1.8,0.2) arc (120:420:0.35 and 0.25);
    \draw[very thick,black] (-0.18,0.2) arc (120:420:0.35 and 0.25);
    \draw[very thick,black] (-2.15,0.2) arc (120:420:0.35 and 0.25);
    \draw[very thick,black] (1.8,0.2) arc (120:420:0.35 and 0.25);
    \node at (-2.2,-0.5) {$2$};
    \node at (0.3,-0.5) {$2$};
    \node at (2.2,-0.5) {$2$};
\end{tikzpicture}
}}, \label{eq:twisted I bundle identification}
\end{equation}
where the label $2$ means that we perform Dehn surgery along the black curves with slope $p/q = 2$ as in section~\ref{subsec:gravitational machine 1}.

Let us first verify this equivalence topologically by computing the fundamental group. For hyperbolic 3-manifolds, this is enough to show their equivalence. From the presentation as a twisted $I$-bundle, it is clear that the manifold is homotopic to the van Dyck surface $\Sigma_2/\Phi$, since we can retract the interval to its midpoint. The fundamental group of this non-orientable surface is well-studied and admits the presentation:
\be 
\pi_1(M)=\langle w_1, \, w_2,\, w_3\, | \, w_1^2w_2^2w_3^2=1 \rangle\ , \label{eq:fundamental group twisted I bundle}
\ee
which can be read off from the fundamental polygon for the surface. This presentation of $\pi_1(M)$ is precisely of the form \eqref{eq:fundamental group triple Dehn filling} that we derived for the triple Dehn filling, which shows the claim \eqref{eq:twisted I bundle identification}. In particular, this also shows that $M$ is a non-handlebody, because the fundamental group is not the free group on two elements.

As an interesting aside, we note that the twisted $I$-bundle is one of the simplest examples that is $\partial$-irreducible, i.e.~the map $\pi_1(\partial M) \longrightarrow \pi_1(M)$ is injective. This property was important in section~\ref{subsec:what 3-manifolds are generated}. This is simple to see, since the map corresponds to the homomorphism induced by a covering map $\pi_1(\Sigma_2) \to \pi_1(\Sigma_2/\Phi)$ and is thus injective. In fact, the image is an index-2 subgroup.

Let us now also check that the Virasoro TQFT partition functions agree in both presentations. The computation of the Virasoro TQFT partition function from the point of view of the twisted $I$-bundle was discussed in \cite{Collier:2024mgv} and is simply given by the Liouville partition function on the van Dyck surface. Let us pick a convenient conformal block channel to expand this partition function:
\begin{multline} 
Z_\text{Vir}(M)=\int_0^\infty \!\d^3P\,\Gamma(P_1)\Gamma(P_2)\Gamma(P_3)\, C_0(P_1,P_2,P_3)\, \mathcal{F}_{g=2}\!\left[\!\!\vcenter{\hbox{\begin{tikzpicture}[scale=.6]
    \draw[very thick] (0,1) to[out=180,in=180,looseness=2] (0,-1) to[out=0,in=180] (1,-0.75) to[out=0,in=180] (2,-1) to[out=0,in=0,looseness=2] (2,1) to[out=180, in =0] (1,.75) to[out=180,in=0] (0,1);
    \draw[very thick, out=180, in=180, looseness=2] (0, 1) to (0,-1);
    \draw[very thick, out=0, in = 180] (0,1) to (1,0.75) to (2, 1);
    \draw[very thick, out = 0, in=180] (0,-1) to (1,-0.75) to (2,-1);
    \draw[very thick, out=0, in=0,looseness=2] (2,1) to (2,-1);
    \draw[very thick, bend right=30] (-1/2,0+.06) to (1/2, 0+.06);
    \draw[very thick, bend right=30] (2-1/2,0+.06) to (2+1/2, 0+.06);
    \draw[very thick, bend left = 30] (-1/3,-.08+.06) to (+1/3,-.08+.06);
    \draw[very thick, bend left = 30] (2-1/3,-.08+.06) to (2+1/3,-.08+.06);
    \draw[very thick, wilsonred, out=180, in=180, looseness=6] (1,-.5) to (1,.5);
    \draw[very thick, wilsonred] (1,-.5) to node[right, black, shift={(-.1,0)}] {\footnotesize 3} (1,.5);
    \draw[very thick, wilsonred, out=0, in=0, looseness=6] (1,-.5) to (1,.5);
    \node at (0,.7) {\footnotesize 1};
    \node at (2,.7) {\footnotesize  2};
    \end{tikzpicture}}}\!\!\right] ,
\end{multline}
where the picture denotes as usual schematically the relevant conformal block, and
\be 
\Gamma(P)=\big(\mathbb{T}^{\frac{1}{2}}\mathbb{S} \mathbb{T}^2 \mathbb{S} \mathbb{T}^{\frac{1}{2}}\big)_{\id,P}=4 \cosh(\pi b P) \cosh(\pi b^{-1} P) \label{eq:cross cap normalization}
\ee
is the cross cap normalization. This has exactly the right form to come from our gravitational machine. Since $\Gamma(P)$ is essentially the modular matrix $\mathbb{S}\mathbb{T}^2\mathbb{S}$, this is precisely the effect of performing the $2/1$ Dehn surgery. The additional terms of $\mathbb{T}^{1/2}$ appearing in the cross cap normalization \eqref{eq:cross cap normalization} give a non-trivial identification of the boundary moduli as described in the picture as produced by the gravitational machine \eqref{eq:twisted I bundle identification} and as a twisted $I$-bundle by an additional half Dehn twist.

\subsection{Handlebody-knots}
\label{subsec:handlebody-knots}

As a third class of examples of non-handlebodies generated by the machine, we study \emph{handlebody-knots}.
A genus-$g$ handlebody-knot is the 3-manifold obtained by removing a genus-$g$ handlebody from $\mathrm{S}^3$. These manifolds can be represented using knotted trivalent graphs; they are generalizations of regular knots, now including trivalent vertices. A handlebody-knot is said to be trivial if the manifold, that is the complement of the knot in S$^3$, is itself a handlebody. The simplest example of a non-trivial genus-2 handlebody-knot has four crossings:
\begin{equation}
\label{eq:41}
\hspace{-0.5cm}
\vcenter{\hbox{
}}\,. 
\end{equation}
Handlebody-knots up to seven crossings have been classified and tables for these knots can be found in the literature \cite{ISHII:2012,bellettini2025tablegenushandlebodyknotsseven}. The knot depicted in \eqref{eq:41} is referred to as the $\mathbf{4}_1$ handlebody-knot, and has been already introduced in section \ref{subsec:gravitational machine 1}. As with regular knots, the 4 in the name corresponds to the minimum number of crossings required to draw such a knot, the 1 means that this is the first knot that appears in the table of \cite{ISHII:2012}. Note that the handlebody-knot $\mathbf{4}_1$ has the same pattern of contractions as the dumbbell channel, thus it contributes to $\overline{\rho_1\rho_2\rho_3C_{112}C_{233}}$. 

Two handlebody-knot diagrams correspond to the same handlebody-knot if and only if they are related by a sequence of Reidemeister moves together with two additional moves, which correspond to the $\mathbb{F}$ and the $\mathbb{B}$ transforms. For example, the sunset and dumbbell diagrams are equivalent since they are related by a single $\mathbb{F}$ transform. A more complicated example is given in the following sequence of diagrams:
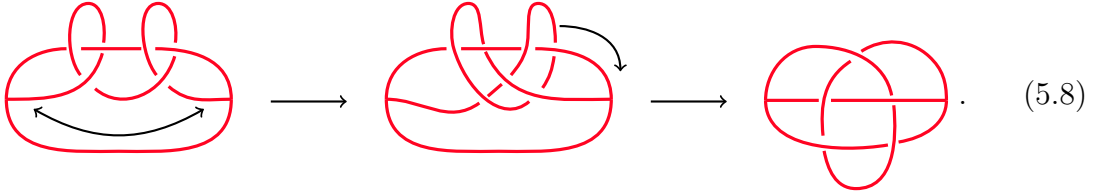
\begin{equation}
\label{eq:52}
 \vcenter{\hbox{

}}.
\end{equation}
In the first step, we applied a sequence of $\mathbb{F}$, $\mathbb{B}$ and $\mathbb{F}$ transforms to exchange the end points of the red line. Then we applied a sequence of Reidemeister moves and braiding to simplify the diagram; the resulting graph corresponds to the $\mathbf{5}_2$ handlebody-knot (see table \ref{table:handlebodyknots}). 

We will in the following discuss these handlebody-knots up to five crossings and how they are (or are not) generated by the machine. They nicely illustrate the different moves of the machine. It turns out that $\mathbf{4}_1$ and $\mathbf{5}_1$ are generated by the machine via the self-gluing move, $\mathbf{5}_2$ is also generated via Dehn fillings, $\mathbf{5}_3$ is hyperbolic but acylindrical and thus \emph{not} generated and $\mathbf{5}_4$ is not even hyperbolic and thus also not generated. We summarize the situation in table \ref{table:handlebodyknots}.
\begin{table}\label{tab:handlebody-knots}
\centering
\begin{tabular}{|c|c|c|}
    \hline
    Handlebody-knot 
    & Hyperbolic & Generated by the machine \\
    \hline\hline
    \begin{tikzpicture}
        \begin{scope}[scale=0.5]
            \draw[very thick, wilsonred]
            (-1.8,0.2) to[out=0, in ={30-180}]
            (-0.37,0.4) to[out = 30, in =-60 ]
            (-0.3,1.3);
            \draw[very thick, wilsonred]
            (-0.46,1.53) to[out = 135, in=90]
            (-1.8,0.2) to[out=270, in=205]
            (0.1,-1.1);
            \draw[very thick, wilsonred]
            (0.3,-1) to[out=25, in =270]
            (0.63,-0.5) to[out = 90, in =-45]
            (-0.24,0.3);
            \draw[very thick, wilsonred]
            (-0.5,0.45) to[out=130, in = 180, looseness =1.3]
            (0.27,1.6) to[out=0, in = 90]
            (1.67,0.2) to[out = 270, in = 0]
            (0.69, -1.2) to[out = 180, in =-135, looseness=1.3]
            (0.14, -0.05);
            \draw[very thick, wilsonred]
            (0.38,0.1) to[out = 35, in =180] (1.67,0.2);
            \node[left] at (-1.8,0.2) {$\mathbf{4}_1$};
        \end{scope}
    \end{tikzpicture}
    & Yes & Yes (self-gluing of handlebody)\\
\hline
\end{tabular}
\caption{The irreducible non-trivial handlebody-knots with fewer than six crossings (omitting their mirror images). Among these examples, the gravitational machine generates the hyperbolic and cylindrical handlebody-knots, either through a self-gluing or a Dehn surgery on a handlebody. It does not generate the acylindrical or non-hyperbolic handlebody-knots.}
\label{table:handlebodyknots}
\end{table}

\paragraph{The $\mathbf{4}_1$ handlebody-knot.} Starting with the handlebody-knot $\mathbf{4}_1$ drawn in \eqref{eq:41}, the VTQFT partition function associated to this diagram is given by:
\begin{equation}
\label{eq:vir41}
Z_{\text{Vir}}\left[
\vcenter{\hbox{
\begin{tikzpicture}
\begin{scope}[xscale=0.65, yscale=0.65]
    \draw[very thick, wilsonred]
    (-1.8,0.2) to[out=0, in ={30-180}]
    (-0.37,0.4) to[out = 30, in =-60 ]
    (-0.3,1.3);
    \draw[very thick, wilsonred]
    (-0.46,1.53) to[out = 135, in=90]
    (-1.8,0.2) to[out=270, in=205]
    (0.1,-1.1);
    \draw[very thick, wilsonred]
    (0.3,-1) to[out=25, in =270]
    (0.63,-0.5) to[out = 90, in =-45]
    (-0.24,0.3);
    \draw[very thick, wilsonred]
    (-0.5,0.45) to[out=130, in = 180, looseness =1.3]
    (0.27,1.6) to[out=0, in = 90]
    (1.67,0.2) to[out = 270, in = 0]
    (0.69, -1.2) to[out = 180, in =-135, looseness=1.3]
    (0.14, -0.05);
    \draw[very thick, wilsonred]
    (0.38,0.1) to[out = 35, in =180] (1.67,0.2);
    \node at (-1.6,1.6) {$1$};
    \node at (1.4,-1.4) {$3$};
    \node at (1.2,1.7) {$2$};
\end{scope}
\end{tikzpicture}}} 
\,\right] = 
\frac{1}{\sqrt{\mathsf{C}_{112}\mathsf{C}_{233}}}\,\skerhat{P_2}{P_3}{P_2}\skerhat{P_2}{P_1}{P_2}^*
\end{equation}
where recall that $\mathsf{C}_{ijk} = C_0(P_i,P_j,P_k)$ and the complex conjugation just gives a phase $\skerhat{P_1}{P_2}{P_2}^* = \e^{-\pi i h_2}\skerhat{P_1}{P_2}{P_2}$.
This amplitude was computed using the Verlinde loop operators on the Wilson lines with momenta $P_1$ and $P_3$, see appendix \ref{app:technology}. To evaluate the correction to the moment $\overline{\rho_1\rho_2\rho_3C_{112}C_{233}}$ associated to this diagram, we first need to multiply by the normalization factor $\rho_{g=2}(\boldsymbol{P})$, and then take the product of left and right movers. The first step can be interpreted as replacing the trivalent vertices of the diagram with asymptotic spherical boundaries as in (\ref{eq:vertex replacement rule}).

There is also a nontrivial handlebody-knot that differs from the $\mathbf{4}_1$ example above by changing the pattern of over- and under-crossings on one of the two loops (changing the crossings on both loops gives the same amplitude since (\ref{eq:vir41}) is real). This gives a configuration whose VTQFT partition function differs from (\ref{eq:vir41}) by an overall phase:
\begin{equation}
\label{eq:vir41-2}
Z_{\text{Vir}}\left[
\vcenter{\hbox{
\begin{tikzpicture}
\begin{scope}[xscale=0.65, yscale=0.65]
    \draw[very thick, wilsonred]
    (-1.8,0.2) to[out=0, in ={30-180}]
    (-0.47,0.3);
    \draw[very thick, wilsonred] (-0.27,0.5) to[out=30,in=-60] (-0.35,1.4) to[out=135,in=90] (-1.8,0.2) to[out=270, in=205]
    (0.1,-1.1);
    \draw[very thick, wilsonred] (0.3,-1) to[out=25, in =270]
    (0.63,-0.5) to[out = 90, in =-45]
    (-0.37,0.4) to[out=135, in=-90] (-0.70,0.9) to[out=90, in=225] (-.47,1.3);
    \draw[very thick, wilsonred] (-.27,1.5) to[out=25,in=180,looseness=1.3] (0.27,1.6) to[out=0, in = 90]
    (1.67,0.2) to[out = 270, in = 0]
    (0.69, -1.2) to[out = 180, in =-135, looseness=1.3]
    (0.14, -0.05);
    \draw[very thick, wilsonred]
    (0.38,0.1) to[out = 35, in =180] (1.67,0.2);
    \node at (-1.6,1.6) {$1$};
    \node at (1.4,-1.4) {$3$};
    \node at (1.2,1.7) {$2$};
\end{scope}
\end{tikzpicture}}} 
\,\right] = 
\frac{1}{\sqrt{\mathsf{C}_{112}\mathsf{C}_{233}}}\,\skerhat{P_2}{P_3}{P_2}\skerhat{P_2}{P_1}{P_2}\ .
\end{equation}
Summing up the amplitudes in (\ref{eq:vir41}) and (\ref{eq:vir41-2}), and changing variables to conformal weights $h_i$, the resulting contribution to the second moment of the structure constants in the dumbbell contraction is given by:
\begin{equation}
\overline{\rho_1\rho_2\rho_3C_{112}C_{233}} \supset 
(1+(-1)^{J_2}) \left|\rho_0(h_1)\rho_0(h_2)\rho_0(h_3)
\sqrt{\mathsf{C}_{112}\mathsf{C}_{233}}\,
\skerhat{h_2}{h_3}{h_2}\skerhat{h_2}{h_1}{h_2}^*
\right|^2. 
\end{equation}
We notice that there is a projection to even spin $J_2$. The reason for this is the reality properties of the structure constants \eqref{eq:OPEpermutation}, see also the discussion below \eqref{eq:genus2typicality}. Comparing to the handlebody contribution of the linked dumbbell \eqref{eq:cherryOPE}, the $\mathbf{4}_1$ manifold has one extra factor of $\widehat{\mathbb{S}}$.

Let us now explain how the $\mathbf{4}_1$ handlebody-knot is generated by the gravitational machine.
Consider again the linked dumbbell diagram \eqref{eq:cherry}, which we recall is a handlebody; i.e.~it can be obtained by applying a sequence of crossing transformations on the dumbbell diagram.  Now apply the self-gluing surgery operation of the machine, which amounts to a self-contraction of the indices $2$ and $3$:
\begin{equation}
    M \#_{P_2\sim P_3} =
\end{equation}
The conditions \ref{cond:no Omega loops 2}, \ref{cond:no parallel rings 2} and \ref{cond:no Wilson line self gluing} are clearly satisfied.
We claim that the resulting topology is homeomorphic to the $\mathbf{4}_1$ handlebody-knot in \eqref{eq:41}, if we perform an additional S-transform on one of the bells. To show this, let us represent the self-gluing operation using the $\Omega$ loop, and then use \eqref{eq:modular_on_graphs} to exchange the $\Omega$ loop for an S-transform:
\begin{equation}
    M \#_{P_2\sim P_3} \cong 
    %
\label{eq:41 from omega loop}
\end{equation}
The prime indicates that the $P_3$ bell is in the S-dual channel. We can also explicitly confirm that the VTQFT partition functions agree: 
\begin{align}
Z_{\text{Vir}}\left[
\vcenter{\hbox{
%
}} 
\right]
 &= \int_0^\infty \d P \,\sker{P_3}{P}{P_2}
\frac{\skerhat{P_1}{P_2}{P_2} \skerhat{P}{P_2}{P_2}^*}{\sqrt{\mathsf{C}_{112}\mathsf{C}_{2PP}}} \\
&= \frac{\delta(P_2-P_3)}{\rho_0(P_2)}\frac{\skerhat{P_1}{P_2}{P_2}}{\sqrt{\mathsf{C}_{112}\mathsf{C}_{233}}}\ ,
\end{align}
where we used the idempotency of the one-point S-kernel (\ref{eq:idempotency}). (Note that the first S-kernel is not hatted). The right-hand side is indeed equal to the partition function of the linked dumbbell diagram \eqref{eq:cherry} times the self-contraction $\delta_{23}$.

One can also see the homeomorphism \eqref{eq:41 from omega loop} at the level of the fundamental group, which can be written as (using the technology described in appendix~\ref{subapp:fundamental group}):
\be 
\pi_1(\mathbf{4}_1)=\big\langle a,\,b,\,c\, | \, c a c^{-1}=abab^{-1} \big \rangle\ . \label{eq:4_1 fundamental group}
\ee
Note that $\pi_1$ indeed has the form of an HNN extension of the free group generated by $a$ and $b$, with $c$ playing the role of the stable letter. This is expected by the general observation in section \ref{subsec:machine is hyperbolic} that cylinder surgery turns the fundamental group into an HNN extension; in this case the handlebody before self-gluing has fundamental group $\langle a, b\rangle$ and self-gluing turns it into \eqref{eq:4_1 fundamental group}. 

As remarked in section \ref{subsec:gravitational machine 1}, if we perform yet a further modular $S$ transformation on the bell labeled by the $P_1$ Wilson line of the $\mathbf{4}_1$ handlebody-knot, this generates a contribution to the variance of the structure constants with all indices equal, corresponding to the double contraction $\delta_{12}\delta_{23}$. This corresponds to a double self-gluing of the sunset handlebody, recall equations (\ref{eq:41prime self-gluing}) and (\ref{eq:4_1 from surgery}).

\paragraph{The $\mathbf{5}_1$ handlebody-knot.}
We now discuss the $\mathbf{5}_1$ handlebody-knot. Its VTQFT amplitude is straightforwardly computed to be:
\begin{equation}\label{eq:ZVir 51}
    Z_{\text{Vir}}\!\!\left[\!\!\!\!
    \vcenter{\hbox{
    \begin{tikzpicture}
       \draw[very thick, wilsonred] (-1,0) to[out=0,in=225] (-0.5+0.25,0.25) to[out=45,in=-45] (-0.45+0.25,0.45);
        \draw[very thick, wilsonred] (-0.55+0.25,0.55) to[out=135,in=225] (-0.5+0.25,0.75) to[out=45,in=90] (1,0);
        \draw[very thick, wilsonred] (-0.55+0.25,0.8) to[out=135,in=90] (-1,0);
        \draw[very thick, wilsonred] (-0.45+0.25,0.7) to[out=-45,in=45] (-0.5+0.25,0.5) to[out=225,in=135] (-0.55+0.25,0.3);
        \draw[very thick, wilsonred] (-0.45+0.25,0.2) to[out=-45,in=135] (0.5-0.25,-0.25) to[out=-45,in=45] (0.55-0.25,-0.7);
        \draw[very thick, wilsonred] (0.45-0.25,-0.8) to[out=225,in=-90] (-1,0);
        \draw[very thick, wilsonred] (1,0) to[out=180,in=45] (0.55-0.25,-0.2);
        \draw[very thick, wilsonred] (0.45-0.25,-0.3) to[out=225,in=135] (0.5-0.25,-0.75) to[out=-45,in=-90] (1,0);
        \node[left] at (-.85,0.5) {$3$};
        \node[right] at (.82,0.5) {$2$};
        \node[above] at (0.6,-0.05) {$1$};
    \end{tikzpicture}
    }}
    \!\!\!\!\right] = \frac{\widehat{\mathbb{S}}_{P_1P_3}[P_2]}{\sqrt{\mathsf{C}_{112}\mathsf{C}_{233}}}\int_0^\infty\d P\, \rho_0(P)\, \e^{3\pi i (h_P-h_2-h_3)}
    \!\begin{Bmatrix}
        P_2 & P_3 & P \\ P_2 & P_3 & P_3
    \end{Bmatrix}\, .
\end{equation}
This is computed by unhooking the $P_1$ and $P_3$ Wilson lines using the rule (\ref{eq:hookedlines}), then inserting an identity line between the $P_2$ and $P_3$ Wilson lines and unbraiding them using $\mathbb{B}$. 

Similarly to the case of the $\mathbf{4}_1$ handlebody-knot studied above, there is another configuration that amounts to changing the orientation of the linking of the $P_1$ and $P_3$ Wilson lines. Adding up its contribution to the statistics of the structure constants $\overline{\rho_1\rho_2\rho_3C_{112}C_{233}}$ with \eqref{eq:ZVir 51} gives an overall factor $(1+(-1)^{J_2})$, which accounts for the projection to even spin $J_2$.

We now explain how this topology is generated by the machine. The mechanism is very similar to that of the $\mathbf{4}_1$ handlebody-knot discussed above so we will be brief. Like the $\mathbf{4}_1$ handlebody-knot, the $\mathbf{5}_1$ handlebody-knot may be obtained by self-gluing of a handlebody. Indeed, consider the following Wilson line configuration:
\begin{equation}
    \text{``twisted cherry''} \equiv
    \vcenter{\hbox{
    \begin{tikzpicture}
        \draw[very thick, wilsonred] (-1,0) to[out=0,in=225] (-0.5+0.5,0.25) to[out=45,in=-45] (-0.45+0.5,0.45);
        \draw[very thick, wilsonred] (-0.55+0.5,0.55) to[out=135,in=225] (-0.5+0.5,0.75) to[out=45,in=90] (1,0);
        \draw[very thick, wilsonred] (-0.55+0.5,0.8) to[out=135,in=90] (-1,0);
        \draw[very thick, wilsonred] (-0.45+0.5,0.7) to[out=-45,in=45] (-0.5+0.5,0.5) to[out=225,in=135] (-0.55+0.5,0.3);
        \draw[very thick, wilsonred] (-0.45+0.5,0.2) to[out=-45,in=180] (1,0);
        \draw[very thick, wilsonred] (-1,0) to[out=-90,in=180] (0,-1) to[out=0,in=-90] (1,0);
        \node[left] at (-1,0.3) {$3$};
        \node[right] at (1,0.3) {$2$};
        \node[above] at (0,-1) {$1$};
    \end{tikzpicture}
    }}\, .
\end{equation}
While its name may suggest otherwise, this is a handlebody in the sunset contraction. It is simply related to the linked dumbbell diagram (\ref{eq:cherry}) by a sequence of $\mathbb{F}$, $\mathbb{B}$, and $\mathbb{F}$ transformations. Its VTQFT amplitude is given by:
\begin{equation}
    Z_{\text{Vir}}(\text{twisted cherry}) = \frac{1}{\mathsf{C}_{123}}\int_0^\infty \d P\, \rho_0(P) \, \e^{3\pi i (h_P-h_2-h_3)}
    \begin{Bmatrix}
        P_2 & P_3 & P_1 \\
        P_2 & P_3 & P
    \end{Bmatrix}\, .
\end{equation}
We claim that the modular $S$-transform of the $\mathbf{5}_1$ handlebody-knot along the $P_1$ bell is equivalent to self-gluing the twisted cherry handlebody along the $P_1$ and $P_3$ Wilson lines. While this is transparent from the visual depiction of the knots, one can also confirm that their VTQFT amplitudes indeed agree:
\begin{multline}
    \int_0^\infty \d P\, \mathbb{S}^*_{P_1 P}[P_2]\,Z_{\text{Vir}}(\mathbf{5}_1;P,P_2,P_3) = \underbrace{\frac{\delta(P_1-P_3)}{\rho_0(P_1)}}_{\text{self-gluing factor}}\\
    \times 
    \underbrace{\frac{1}{\mathsf{C}_{233}}\int_0^\infty\d P'\, \rho_0(P')\,\e^{3\pi i (h_{P'}-h_2-h_3)}
    \begin{Bmatrix}
        P_2 & P_3 & P_3 \\ P_2 & P_3 & P'
    \end{Bmatrix}}_{Z_{\text{Vir}}(\text{twisted cherry};P_1=P_3)}\, . 
\end{multline}
One can also see the similarity of the $\mathbf{5}_1$ handlebody-knot with the $\mathbf{4}_1$ handlebody-knot at the level of the fundamental group. The fundamental group reads:
\be 
\pi_1(\mathbf{5}_1)=\big\langle a,\,b,\,c\, | \, c a c^{-1}=ab^2ab^{-1} \big \rangle\ ,
\ee
which again has the form of an HNN extension similar to \eqref{eq:4_1 fundamental group}.

\paragraph{The $\mathbf{5}_2$ handlebody-knot.} We next discuss the $\mathbf{5}_2$ handlebody-knot. Its VTQFT amplitude can be similarly computed and takes the slightly more complicated form:
\begin{multline}
\label{eq:52amp}
Z_{\text{Vir}}\smash[b]{\left[
\vcenter{\hbox{
\begin{tikzpicture}
\begin{scope}[xscale=0.65, yscale=0.65]
    \draw[dashed, thick] (0,0.9) -- (0,1.7);
    \draw[dashed, thick] (-1.15,-1.15) -- (-0.5,-0.5);
    \draw[very thick, wilsonred]
    (-0.67, -0.2) to[out = -80, in =180]
    (0.76,-1.2) to[out=0, in=-90]
    (1.7, -0.1) to[out = 90, in = 10]
    (-0.6,0.81);
    \draw[very thick, wilsonred]
    (-0.8,0.75) to[out = {180+10}, in = 90]
    (-1.70, -0.1) to[out= 270, in = 135]
    (-1.37, -1) to[out=-45, in = 215]
    (-0.1, -1.1);
    \draw[very thick, wilsonred]
    (0.1,-0.9) to[out=35, in = -85]
    (0.62,0.62);
    \draw[very thick, wilsonred]
    (0.6,0.93) to[out = 100, in=0]
    (0, 1.7) to[out = 180, in = 95]
    (-0.71, 0);
    \draw[very thick, wilsonred]
    (-1.7, -0.1) -- (0.47, -0.1);
    \draw[very thick, wilsonred]
    (0.7, -0.1) --
    (1.7, -0.1);
    \node at (-1.4,1.2) {$1$};
    \node at (1.2,-0.4) {$3$};
    \node at (0.8,1.7) {$2$};
\end{scope}
\end{tikzpicture}}} 
\right]} =
\frac{1}{\mathsf{C}_{123}}\int_0^\infty \d P_s \d P_t\, \rho_0(P_s)\rho_0(P_t)\\ \vspace*{-1cm} \times \e^{\pi i(4h_1-2h_2-4h_s+3h_t)}\sixj{P_1}{P_2}{P_3}{P_t}{P_s}{P_2}^2 \ .
\end{multline}
This amplitude is computed by inserting two identity Wilson lines, indicated by the black dashed lines, followed by an $\mathbb{F}$ transform acting on them. Equation \eqref{eq:52amp} leads to a non-perturbative correction to the second moment in the sunset contraction:
\begin{equation}
    \overline{C_{123}C^*_{123}}\supset \left|\mathsf{C}_{123} \int_0^\infty \d P_s \d P_t\, \rho_0(P_s)\rho_0(P_t)\e^{\pi i(4h_1-2h_2-4h_s+3h_t)}\sixj{P_1}{P_2}{P_3}{P_t}{P_s}{P_2}^2\right|^2\, .
\end{equation}
This should be compared to the leading-order contribution to this quadratic moment $\overline{C_{123}C_{123}^*}$, which is given by the $C_0$ formula itself as computed by \eqref{eq:C0 contribution}.

To see how $\mathbf{5}_2$ is generated by the machine, we examine the fundamental group. Using the Wirtinger presentation of $\pi_1$, combined with the Whitehead algorithm, we explain in detail in appendix~\ref{subapp:fundamental group} that the fundamental group can be presented as:
\be 
\pi_1(\mathbf{5}_2)=\langle a,\, b,\, c \, | \, a^3b^2c^2=1 \rangle\ .
\ee
Similarly to the twisted $I$-bundle \eqref{eq:twisted I bundle identification}, this has the correct form to be generated by 3 Dehn surgeries, as in \eqref{eq:twisted I bundle identification}, except that one of the Dehn filling parameters is 3 instead of 2. This is rather difficult to see on the level of the partition function \eqref{eq:52amp}, which should be equivalent to: 
\be 
Z_{\text{Vir}}\left[
\vcenter{\hbox{
\begin{tikzpicture}
\begin{scope}[baseline=(current bounding box.center),xscale=0.65, yscale=0.65]
    \draw[very thick, wilsonred]
    (-0.67, -0.2) to[out = -80, in =180]
    (0.76,-1.2) to[out=0, in=-90]
    (1.7, -0.1) to[out = 90, in = 10]
    (-0.6,0.81);
    \draw[very thick, wilsonred]
    (-0.8,0.75) to[out = {180+10}, in = 90]
    (-1.70, -0.1) to[out= 270, in = 135]
    (-1.37, -1) to[out=-45, in = 215]
    (-0.1, -1.1);
    \draw[very thick, wilsonred]
    (0.1,-0.9) to[out=35, in = -85]
    (0.62,0.62);
    \draw[very thick, wilsonred]
    (0.6,0.93) to[out = 100, in=0]
    (0, 1.7) to[out = 180, in = 95]
    (-0.71, 0);
    \draw[very thick, wilsonred]
    (-1.7, -0.1) -- (0.47, -0.1);
    \draw[very thick, wilsonred]
    (0.7, -0.1) --
    (1.7, -0.1);
    \node at (-1.4,1.2) {$1$};
    \node at (1.2,-0.4) {$3$};
    \node at (0.8,1.7) {$2$};
\end{scope}
\end{tikzpicture}}} 
\right]\sim \frac{1}{\mathsf{C}_{123}}\, (\mathbb{S}\mathbb{T}^3\mathbb{S})_{\id P_1}(\mathbb{S}\mathbb{T}^2\mathbb{S})_{\id P_2}(\mathbb{S}\mathbb{T}^2\mathbb{S})_{\id P_3}
\ee
after a suitable crossing transformation $\mathbb{K}^\gamma_{\boldsymbol{P}\boldsymbol{P}'}$. The relevant transformation is in principle contained in the map \eqref{eq:5_2 pi_1 automorphism}, but it is rather difficult to translate this in practice to crossing kernels.

\paragraph{The $\mathbf{5}_3$ handlebody-knot.} The next handlebody-knot in the table \ref{table:handlebodyknots} is quite famous and is known under many names, such as Thurston's knotted Y-manifold, tripus or tripos manifold \cite{thurston_book}. It will be our primary example of a hyperbolic manifold that is \emph{not} generated by the machine because it is acylindrical. It has the following equivalent forms, related by Reidemeister moves:
\begin{equation}
\vcenter{\hbox{
}}\ 
\end{equation}
The right-hand side is also known as Kinoshita's theta curve \cite{kinoshita1972elementary}. It is also equivalent (via Reidemeister moves) to the presentation of the handlebody-knot $\mathbf{5}_3$ in table 1 of \cite{tablehandlebodyknots}.
Its complement in $\mathrm{S}^3$ is a non-trivial handlebody-knot \cite{ozawa2019two} with genus-2 boundary. Crucially, it admits a hyperbolic metric with finite volume and totally geodesic boundary \cite{thurston_book}. As discussed in section~\ref{subsec:what 3-manifolds are generated}, this means that it is acylindrical and \emph{not} generated by the gravitational machine. 

This can also be seen on the level of the fundamental group, which is written as:
\be 
\pi_1(\mathbf{5}_3)=\langle a,\, b,\, c \, | \, aba^{-1}b^{-1}cbc^{-1}a^{-1}c=1 \rangle\ ,
\ee
where the relation has been chosen to have minimal word length.
The fact that this manifold is not produced by the machine can already be guessed from this presentation (although the proof requires the knowledge that the manifold is acylindrical). It does not (manifestly) have the form of an HNN extension, since every generator appears three times in the relation and thus cannot serve a stable letter. It also cannot be written as a Dehn filling of a simpler expression since this would lead to powers of elements in the defining relation. 

Let us for completeness evaluate the VTQFT amplitude of this counter-example. First, we use the S-kernel to unlink the hooked lines (see also \eqref{eq:hookedlines}) and then apply two braiding transformations:
\begin{align}
    \hbox{
    }\ .
\end{align}
Next, we insert an identity line and perform a fusion transformation, followed by three braiding moves to put the graph in planar form:
\begin{equation}
    \hbox{%
}\ .
\end{equation}
Finally, we recognize two Virasoro $6j$ symbols (or `Wilson triangles'; see \eqref{eq:Wilson triangle}), which allows us to evaluate the planar graph. Putting everything together, we have thus found the following VTQFT partition function for the tripus manifold:
\begin{multline}
Z_\text{Vir}(\text{Tripus};P_1,P_2,P_3)= \\[1em] \frac{\e^{-\pi i(h_1+h_2)}}{\mathsf{C}_{123}}\int_0^\infty \d \mu(P)\d \mu(P')\, \widehat{\mathbb{S}}_{12}[P] \,\e^{3\pi i(h_{P'}-h_P)}%
\,.
\end{multline}
In simplifying, we used that $\widehat{\mathbb{S}}^*_{12}[P] = \e^{-\pi i h_P}\widehat{\mathbb{S}}_{12}[P]$ and we defined $\d\mu(P) \equiv \d P\rho_0(P)$.

\paragraph{The $\mathbf{5}_4$ handlebody-knot.} Finally, the last handlebody-knot with 5 crossings is not hyperbolic, which we already remarked in \eqref{eq:54}, where we highlighted the essential torus. Moreover, one can check that the VTQFT partition function diverges. We can see the non-hyperbolicity also directly at the level of the fundamental group, which takes the form:
\be 
\pi_1(\mathbf{5}_4)=\langle a,\, b,\,c\, | \, a^2 [b,c]a^{-1}[b,c]=1 \rangle\ . 
\ee
Let $r=[b,c]$. Then the relation reads $a^2r a^{-1}=r^{-1}$. Using the relation twice gives:
\be 
a^3r a^{-3}=ar^{-1}a^{-2}=(a^2r a^{-1})^{-1}=r\ .
\ee
Thus $a^3$ and $[b,c]$ generate a $\ZZ \times \ZZ$ subgroup, showing that the manifold cannot be hyperbolic. Hence it is not generated by the machine. 

%%%%%%%%%%%%%%%%%%%%%%%%%%
\section{Discussion} \label{sec:discussion}

In this paper we formulated a set of constraints that any consistent ensemble of 2d CFT data --- defined by a collection of statistical moments of the structure constants and density of primary states --- must obey. These requirements combine the rigid CFT bootstrap constraints on the CFT data, together with internal consistency of the statistical description, including typicality at high energies. Applied to the dual description of pure AdS$_3$ quantum gravity, these constraints give predictions for the set of topologies that must appear in the gravitational path integral and allows one to `discover' the sum over topologies from the statistical ensemble. 

In our analysis, a crucial role is played by requiring the moments of structure constants to exhibit consistent index contractions, similar to typicality in ETH \cite{Foini:2018sdb}. This led us to conclude that in addition to the handlebodies (required by the existence of the identity operator together with crossing symmetry and modular invariance), the gravitational path integral must also include an infinite set of non-handlebody topologies. In the Virasoro TQFT description of 3d gravity, using networks of Wilson lines, this led us to formulate a list of modifications on the bulk side that reflect the statistical properties of the boundary moments. In particular, we found that one has to allow bulk topologies obtained by a generalized connected sum that involves gluing along 2-punctured spheres in order for the boundary theory to have a statistical interpretation. This collective set of operations defines a ``gravitational machine'' that generates an infinite set of manifolds (given a seed manifold) such that the resulting sum over topologies satisfies our ensemble properties. 

Importantly, we proved that this set contains only on-shell saddles of the gravitational path integral. Moreover, this set contains many non-handlebodies, which we showcased through a number of examples, including the Euclidean wormhole, the twisted $I$-bundle, and nontrivial handlebody-knots with fewer than six crossings. We also showed that not all hyperbolic 3-manifolds are generated by the machine. Hence, surprisingly the ensemble dual of 3d gravity is not uniquely singled out by our general requirements. 

We conclude with some open questions and comment on prospects for future~work. 

\paragraph{Non-minimality of pure 3d gravity and analogy with JT gravity.}
One of the most striking consequences of this work is the conclusion that pure AdS$_3$ quantum gravity, defined by a gravitational path integral that sums over \emph{all} hyperbolic 3-manifolds, represents a highly non-minimal solution to the statistical bootstrap. Indeed, the minimal completion (in the sense that it includes the smallest set of bulk manifolds) of the sum over handlebody geometries is the set $\mathcal{M}$ generated by the machine. As discussed in section \ref{subsec:what 3-manifolds are generated}, this set excludes infinitely many hyperbolic 3-manifolds, namely at least all acylindrical hyperbolic 3-manifolds. 

In principle, we could build other consistent solutions of the ensemble properties of section \ref{sec:2} by applying the moves of the gravitational machine to two seeds: one being a handlebody $M$ and the other any of the hyperbolic 3-manifolds $M'$ not produced by the machine (such as an acylindrical manifold). We could then add the full `orbit' of $M'$ under the action of the machine to the gravitational path integral with any positive coefficient, while preserving a consistent statistical interpretation of the boundary ensemble. This suggests the existence of infinitely many consistent solutions to the statistical bootstrap equations as articulated in this paper, all of which are characterized by the maximal gap in the spectrum of twists of primary operators.

Interestingly, one may regard JT gravity, whose gravitational path integral involves a sum over all constant negative curvature surfaces, as a similarly non-minimal theory of quantum gravity, in the following sense. One may consider the restriction of the JT gravity path integral to genus-zero surfaces, which is a consistent truncation of the topological recursion of the boundary matrix integral.\footnote{Together with a suitable rescaling of the boundaries, one may think of this as the strict $\e^{S_0}\to\infty$ limit of JT gravity.} This restriction of the JT gravity path integral may be loosely regarded as analogous to the minimal completion of the handlebody sum described in this paper.\footnote{Also recall the comparison made in section \ref{subsec:what 3-manifolds are generated} between the acylindrical 3-manifolds and higher-genus surfaces in JT gravity: both are characterized by the existence of a hyperbolic metric with totally geodesic boundaries.} However, it is not a perfect analogy: in JT gravity all higher-genus contributions are entirely fixed by genus-zero data through topological recursion \cite{Mirzakhani:2006fta,Eynard:2007kz,Saad:2019lba}, whereas in our ensemble dual we lack such an organizing principle that uniquely fixes the full sum over hyperbolic 3-manifolds from the minimal completion of the handlebody sum. Moreover, in the JT limit described above, the disk is the unique one boundary contribution to the density of states. In our case, we have an infinite number of contributions already in the minimal completion of the handlebody sum.

\paragraph{Missing conditions.}
The fact that pure AdS$_3$ quantum gravity represents such a non-minimal solution to the constraints imposed by the statistical bootstrap immediately raises a question: what further assumptions must be incorporated such that the \emph{full} sum over all hyperbolic 3-manifolds represents the minimal solution to the resulting constraints? Similarly, by what operations must the gravitational machine be augmented in order to generate all hyperbolic 3-manifolds with a given boundary?

Although we do not have an answer to these questions, let us note that from the point of view of the boundary ensemble we already know that we have not incorporated all of the constraints that the moments of a fully consistent statistical ensemble of 2d CFTs should obey.  
For example, we have not enforced unitarity, in particular we have not imposed positivity of the average density of states over the full range of twists: the Maloney-Witten density of states famously suffers from violations of unitarity near extremality \cite{Maloney:2007ud,Keller:2014xba,Benjamin:2019stq}, and ameliorating this is expected to require non-hyperbolic 3-manifolds \cite{Maxfield:2020ale,Benjamin:2020mfz} (more on that below). 

Perhaps more importantly, we have also not considered consistency conditions associated with the moment problem, which descend from the existence of an underlying positive probability distribution over the space of CFT data. For the Hamburger moment problem, this is characterized by the positivity of the Hankel matrix. Here, we can imagine a more sophisticated version, which in any case implies an infinite set of positivity constraints on the various moments. It is possible that fully satisfying these constraints (or saturating them) imposes the sum over all hyperbolic 3-manifolds with the 3d gravity path integral as the only solution. 

Here, we should mention the connection to the matrix/tensor model of \cite{Belin:2023efa,Jafferis:2025vyp} (and its BCFT incarnation \cite{Jafferis:2025yxt}), which was proposed as a non-perturbative definition of the 3d gravity path integral including the sum over topologies. This tensor model, which has a constraint-squared potential, generates as triple-line `Feynman diagrams' many 3-manifolds beyond those generated by our gravitational machine; including potentially off-shell diagrams (like the $\mathbf{5}_4$ handlebody-knot in \eqref{eq:54}, whose VTQFT partition function diverges). However, the potential of the tensor model is not unique, e.g.~higher powers of the constraint can be introduced, which in turn affect the Feynman diagrammatics. It would be interesting to see if on general grounds, having any tensor model whatsoever (i.e.~a probability distribution of OPE coefficients) necessarily imposes the existence of Feynman diagrams corresponding to geometries beyond those generated by the machine.

\paragraph{Typicality and tensor models.} This brings us to another point. In this paper, we assumed the existence of an abstract measure $\mu[\{h_i,\bar h_i, C_{ijk}\}]$ satisfying general properties, without committing to one specific model. However, one natural tensor model that satisfies the properties of section \ref{sec:2} is the `maximum ignorance' model briefly described in the discussion of \cite{deBoer:2023vsm}. That is, as input we give all genus-$g$ partition functions and we demand that the measure $\mu$ reproduces these input data on average; then we maximize the Shannon entropy of $\mu$ subject to these constraints. This model 
satisfies crossing symmetry on average, by construction (because the partition functions that we input are crossing symmetric), and, moreover, it satisfies the typicality property because the potential is single-trace and the envelope functions in the potential are smooth. 
However, in practice this is a difficult statistical theory to work with because the potential is of arbitrarily high degree. A similar comparison between a single-trace maximum ignorance model and a double-trace constraint-squared model can be found in ETH/JT gravity with matter \cite{Jafferis:2022wez,Jafferis:2022uhu}.

\paragraph{Off-shell topologies: unitarity and spectral statistics.} One of the main results of our analysis is that the gravitational machine only produces on-shell topologies. 
 It was in particular non-trivial to see that the gluing operation---which is a bulk manifestation of the index contractions of the CFT moments---preserves hyperbolicity of the manifold to which it is applied.

However, from the perspective of the boundary ensemble there are some clear shortcomings of a sum over on-shell saddles only. As we have  discussed in section \ref{sec:spectral_statistics}, the average density of states $\bar\rho = \rho_{\text{MWK}}$ becomes negative for twists near extremality. Moreover, in this paper we have treated the density of states entirely classically in the sense that all higher moments are taken to factorize into products of the average and the correlation between the density of states and structure constants vanishes. The bulk topologies generated by the machine hence do not capture nontrivial spectral statistics that are expected to characterize the spectra of chaotic conformal field theories \cite{Cotler:2020ugk,DiUbaldo:2023qli,Haehl:2023tkr,Haehl:2023mhf,Haehl:2023xys, Boruch:2025ilr}. 

The solutions to both of these problems are believed to require incorporating off-shell topologies in the gravitational path integral, for which new technical machinery must be developed. For example, it has been argued (through dimensional reduction to JT gravity) that the inclusion of a class of non-hyperbolic Seifert manifolds to the torus partition function can render the average density of states unitary \cite{Maxfield:2020ale}. The leading corrections to the density of states are expected to come from single-boundary Seifert manifolds that are circle fibrations over an $(n+1)$-punctured sphere, which admit a description in terms of Dehn fillings of an $n$-punctured disk times a circle (see \cite{deBoer:2025rct}). The effect of these Seifert manifolds on the density of states may be incorporated by relaxing the condition 3\ref{cond:no parallel rings 2} in section~\ref{subsec:general gravitational machine} in the action of the gravitational machine on internal Wilson lines, allowing for parallel Dehn fillings as follows: 
\begin{equation}
    \vcenter{\hbox{
    \begin{tikzpicture}
        \draw[very thick, wilsonred] (-1.4,0) to (1.4,0);
        \fill[white] (-.95,0) circle (.07);
        \fill[white] (-.45,0) circle (.07);
        \fill[white] (.55,0) circle (.07);
        \draw[very thick] (-3/4,0) [partial ellipse=10:350:.2 and .5];
        \draw[very thick] (-1/4,0) [partial ellipse=10:350:.2 and .5];
        \node[above] at (.25,.25) {$\ldots$};
        \draw[very thick] (3/4,0) [partial ellipse=10:350:.2 and .5];
        \node[right] at (1.4,0) {$P$};
        \node[below] at (-3/4,-1/2) {$\tfrac{p_1}{q_1}$};
        \node[below] at (-1/4,-1/2) {$\tfrac{p_2}{q_2}$};
        \node[below] at (3/4,-1/2) {$\tfrac{p_n}{q_n}$};
    \end{tikzpicture}
    }}\ .
\end{equation}
Such configurations may be realized by cutting networks of Wilson lines along 2-punctured spheres and gluing in the Seifert manifolds discussed above rather than torus handlebodies as discussed in section \ref{subsec:general gravitational machine}.
The resulting 3-manifolds are however non-hyperbolic, since this operation typically introduces incompressible tori.
Similarly to how the sum over a single Dehn filling of internal Wilson lines promoted the average density of states from the Cardy formula to the MWK spectrum associated with the $\PSL(2,\ZZ)$ black holes, the sum over multiple Dehn fillings of internal Wilson lines would incorporate the contributions of Seifert manifolds to the average density of states. 
However, how the above topological operation is to be computed in gravity is an open problem; the outstanding question is how to quotient by the nontrivial bulk mapping class group of the Seifert manifolds \cite{deBoer:2025rct}.

The second problem, the absence of connected spectral statistics $\overline{\rho^n}_{\text{conn}}$, may also be cured by adding off-shell contributions, such as multi-boundary torus wormholes $\Sigma_{g,n}\times S^1$, to the gravitational path integral.\footnote{In fact, they are closely related to the Seifert manifolds discussed above: the $n$-punctured disk times a circle may be thought of as an $(n+1)$-boundary torus wormhole if one puffs up the Wilson lines rather than Dehn filling along them as in the construction of the Seifert manifolds.} In \cite{deBoer:2025rct} it was explained how to construct a class of off-shell topologies that capture the two-body spectral statistics $\overline{\rho\rho}_{\text{conn}}$ via RMT surgery, which involves cutting along embedded tori and gluing in the two-boundary torus wormhole. In order to capture the higher moments $\overline{\rho^n}_{\text{conn}}$, one may consider a generalization of RMT surgery that involves cutting along multiple embedded tori and gluing in the corresponding torus wormhole. However, with the possible exception of the two-boundary torus wormhole \cite{Cotler:2020ugk}, the gravitational path integral on torus wormholes has resisted direct computations, due to the fact that Virasoro TQFT does not capture the associated bulk mapping class group. 

\paragraph{Not all cylindrical manifolds are generated by the machine.} Coming back to the more specific technical results of this paper, in section \ref{subsec:what 3-manifolds are generated} we saw that the gravitational machine generates only cylindrical hyperbolic 3-manifolds. However it turns out that not all cylindrical hyperbolic 3-manifolds are generated by the machine. For example, the self-gluing (move \ref{move:gluing} of the gravitational machine) of an acylindrical hyperbolic manifold (such as the $\mathbf{5}_3$ handlebody-knot discussed in section \ref{subsec:handlebody-knots}) generates a cylindrical hyperbolic manifold that will never be generated by the machine. So the notion of cylindrical vs.~acylindrical is not a perfect characterization of the set $\mathcal{M}$ generated by the machine. We could study $\mathcal{M}$ by defining an abstract graph whose vertices are the manifolds $M\in \mathcal{M}$ and whose edges represent the moves of the machine; it may then be interesting to mathematically characterize the properties of this graph (i.e.~is it a tree, and is it connected?).

\paragraph{More complicated gluings.}
The main ingredient in the gravitational machine was the `cylinder surgery' operation of gluing networks of Wilson lines along 2-punctured spheres.
In terms of the fundamental groups of the involved 3-manifolds, both the amalgamated product and the HNN extension constructions allow for gluing along more complicated sub-surfaces. 
The simplest example would be to glue along a 3-punctured sphere $\Sigma_{0,3}$: one could imagine excising a 3-ball containing a trivalent vertex from a network of Wilson lines, and gluing back in a manifold with a more complicated configuration of Wilson lines along the resulting 3-punctured sphere boundary. 
While the generalized connected sum as implemented by cylinder surgery was essential to satisfy typicality, the statistical impetus for allowing these more general gluings is less clear.
The main technical difficulty in this construction is to identify the correct conditions under which the group used for the gluing is malnormal in either the single component of the HNN extension or in one of the components in the amalgamated product of fundamental groups.

\paragraph{Simplified gravity theories.} There is a simplified version of 3d gravity known as $\U(1)$-gravity \cite{Maloney:2020nni, Afkhami-Jeddi:2020ezh,Collier:2021rsn, Datta:2021ftn, Benjamin:2021wzr}, which is dual to an ensemble average of Narain CFTs in a precise sense. Similarly, there are also simplified settings where one creates a toy model for 3d gravity from a TQFT with a finite-dimensional Hilbert space, such as those involving ensembles of rational CFTs (e.g. ``code CFTs,'' or modular invariants at minimal model values of the central charge, $c=1-\frac{6}{m(m+1)}$) \cite{Castro:2011zq, Jian:2019ubz, Meruliya:2021lul, Meruliya:2021utr, Dong:2021wot,Aharony:2023zit, Dymarsky:2024frx,Dymarsky:2025agh,Barbar:2025vvf,Angelinos:2025zek,Romaidis:2023zpx}. In these examples, some of the formulas for summing over topologies can be made more precise. In particular, the handlebody sum is finite. It has been recently argued that a simple mechanism to extract the correct bulk topologies in these frameworks exist. Essentially, the duality trivializes when summing over handlebodies of very high genus and lower genus boundary surfaces can be obtained by degenerating higher genus surfaces \cite{Dymarsky:2024frx}. Such a strategy does not seem to work for actual 3d gravity since a boundary cycle that is non-contractible in the bulk does not support an identity contribution, so the relevant degeneration limits are ill-defined in Virasoro TQFT. Our gravitational machine is in fact inherently a ``bottom-up'' machine and builds more complicated manifolds by gluing manifolds of lower boundary genera, while genus reduction reduces the consistency conditions on a higher genus surface to a lower genus surface by degenerating the boundary surface and is thus a ``top-down'' procedure.

\paragraph{Supergravity and other generalizations.} One can repeat much of what we analyzed for $\mathcal{N}=1$ 3d supergravity. In that case, one also has to account for spin structures, which are permuted by the mapping class group. However, every orientable 3-manifold is spin, see e.g.\ \cite{Kirby1989}. The corresponding bulk theory is expected to be the super Virasoro TQFT and has partially been developed \cite{Aghaei:2020otq, Bhattacharyya:2024vnw}. Operators can belong to the NS- or the R-sector and there are correspondingly NS- and R-type Wilson lines in the bulk. There are NS-NS-NS and NS-R-R structure constants. With the obvious modifications, we expect that everything that we said in this paper will continue to hold.

The extension of the theory to the setting of boundary CFT's is also possible and has been considered in \cite{Hung:2024gma, Wang:2025bcx, Jafferis:2025yxt, Hung:2025vgs, Geng:2025efs, Jafferis:2025jle}. 

Lastly, one can also drop the orientability assumption, in which case the boundary ensemble has new independent data in form of the cross-cap normalization $\Gamma_i$, which non-trivially interacts with the structure constants. From the bulk point of view, a boundary crosscap allows a bulk Wilson line to end there. This means that in the non-orientable theory, the moment $\overline{\Gamma_i C_{ijj}}$ can be non-trivial, see e.g. \cite{Tsiares:2020ewp}. It would be interesting to work out the details.

\section*{Acknowledgments}
It is a pleasure to thank Jan de Boer,  Jeevan Chandra, Tolya Dymarsky, Laura Foini, Thomas Hartman, Luca Iliesiu, Daniel Jafferis, Juan Maldacena, Alex Maloney, Eric Perlmutter, Julian Sonner and Gabriele di Ubaldo for useful discussions. We also thank Diandian Wang for explaining the installation of the software Orb to us. BP is supported by the ERC Consolidator Grant GeoChaos No 101169611. DL is supported by the SNF starting grant “The Fundamental Description of the Expanding Universe”. DL is also supported by the Fonds National Suisse de la Recherche Scientifique (Schweizerischer Nationalfonds zur Förderung der wissenschaftlichen Forschung) through the Project Grant 200021\_215300 and  NCCR51NF40-141869 The Mathematics of Physics (SwissMAP). LE is supported by the European Research Council (ERC) under the European Union’s Horizon 2020 research and innovation programme (grant agreement No 101115511). While at MIT, SC was supported by the U.S. Department of Energy, Office of Science, Office of High Energy Physics of U.S. Department of Energy under grant Contract Number DE-SC0012567 (High Energy Theory research), DOE Early Career Award  DE-SC0021886 and the Packard Foundation Award in Quantum Black Holes and Quantum Computation.

\appendix

\section{3-manifold technology} \label{app:3-manifolds}
In this appendix, we explain a few concepts in the theory of hyperbolic 3-manifolds that are relevant to our computations in 3d gravity. We refer to \cite{Collier:2023fwi} for many other relevant tools and concepts of hyperbolic 3-manifolds, as well as appendix A of \cite{Eberhardt:2021jvj}.

\subsection{(In)compressible surfaces in 3-manifolds} \label{subapp:boundary compressibility}

\paragraph{Bulk incompressible surfaces.}  Let $S$ be a properly embedded orientable surface in a 3-manifold $M$. If $S$ has a boundary, we also assume that $\partial S \subset \partial M$. A compressing disk $D\subset M$ for $S$ is a disk with $\partial D = D\cap S$ such that $\partial D$ does not bound a disk in $S$. The action of `pinching' along the disk $D$ is called a compression, and it transforms $S$ into a new simpler surface, e.g.~compressing along a meridian turns the torus into a (marked) 2-sphere. A properly embedded connected surface $S$ with Euler characteristic $\chi(S) \leq 0$ is called \emph{compressible} if it has a compressing disk, and \emph{incompressible} otherwise. In an incompressible surface, every disk $D$ with $\partial D = D\cap S$ bounds a ball in $M$, as depicted in the picture below:
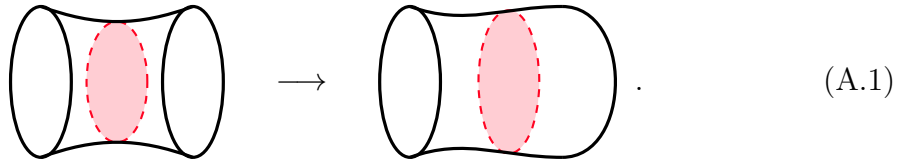
\begin{equation}
    \vcenter{\hbox{
    \begin{tikzpicture}
        \draw[very thick, black] (-1,0) ellipse (0.4 and 1);
        \draw[very thick, black] (1,0) ellipse (0.4 and 1);
        \draw[thick,wilsonred, dashed,fill=wilsonred!20] (0,0) ellipse (0.4 and 0.8);
        \draw[very thick, black] (-1,1) to[out =-20, in = 200] (1,1);
        \draw[very thick, black] (-1,-1) to[out =20, in = 160] (1,-1);
    \end{tikzpicture}
    }}\quad \longrightarrow \quad
    \vcenter{\hbox{
    \begin{tikzpicture}
        \draw[very thick, black] (-1,0) ellipse (0.4 and 1);
        \draw[thick,wilsonred, dashed,fill=wilsonred!20] (0.3,0) ellipse (0.4 and 0.95);
        \draw[very thick, black] (-1,1) to[out =-20, in = 180] (1,1);
        \draw[very thick, black] (-1,-1) to[out =20, in = 180] (1,-1);
        \draw[very thick, black] (1,1) to[out =0, in = 0, looseness = 1.2] (1,-1);
    \end{tikzpicture}}}.
\end{equation}
A basic characterization of incompressible surfaces is given by the following result: if $S$ is a $\pi_{1}$-injective, orientable, connected and properly embedded surface with $\chi(S) \leq 0$, then $S$ is incompressible. We also refer to such surfaces as \emph{bulk}-incompressible, to distinguish them from the $\partial$-incompressible surfaces that we discuss below.

A surface is called \emph{boundary-parallel} if the embedding map $\iota:(D,\partial D) \to (M,\partial M)$ can be isotoped into $\partial M$. 
A surface $S$ that is both bulk-incompressible and non-boundary parallel is called \emph{essential}.

There is a useful way to diagnose if a bulk-incompressible surface is boundary parallel. Cutting $M$ along a boundary-parallel incompressible surface cuts $M$ into two pieces. Moreover, one of the pieces is homotopic to $S$ itself, while the other is homotopic to the original manifold $M$. Thus if cutting $M$ along a surface $S$ yields a connected manifold or a disconnected manifold where both pieces have fundamental groups different from $S$ itself, then $S$ is not boundary parallel.

\paragraph{Boundary incompressible surfaces.}
 It is also useful to discuss the notion of a boundary incompressible (or $\partial$-incompressible) surface. Roughly speaking, boundary compressible surfaces are allowed to be compressed on the boundary, rather than in the bulk.
More precisely, a $\partial$-compressing disc for $S$ is a disk whose boundary lies on both $S$ and $\partial M$ such that $\partial D$ does not bound another disk $D'$ with components in $S$ and $\partial M$, see the picture below:
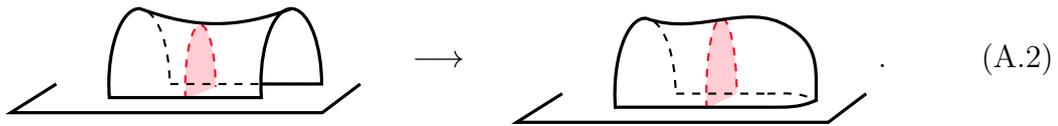
\begin{equation}
    \vcenter{\hbox{
    \begin{tikzpicture}
        \draw[very thick, black] (-1.4,-0.2) arc (191.5:88:0.4 and 1);
        \draw[very thick, black] (0.6,-0.2) arc (191.5:0:0.4 and 1);
       \draw[thick, black, dashed] (-1.4,-0.2) arc (191.5:0:0.4 and 1);
        \fill[white] (-1,0) -- (0.45,0) -- (0.45,-0.2) -- (-1,-0.2);
        \fill[white] (1,0) -- (1.5,0) -- (1.5,-0.2) -- (1,-0.2);
        \draw[very thick, black] (0.6,0) -- (1.413,0);
        \draw[thick,wilsonred, dashed,fill=wilsonred!20] (-0.4,-0.2) arc (191.5:0:0.2 and 0.82);
        \draw[very thick, black] (-1,1) to[out =-20, in = 200] (1,1);
        \draw[very thick, black] (-1.4,-0.18) -- (0.6,-0.18);
        \draw[thick, dashed, black] (-0.6,0) -- (1.4,0);
        \draw[very thick] ({-0.6-1.5},{0}) -- ({-1.4-1.3},{-0.18-0.2}) -- (1.4,-0.38) -- ({0.6+1.3},{0});
    \end{tikzpicture}
    }}\quad \longrightarrow \quad
    \vcenter{\hbox{
    \begin{tikzpicture}
        \draw[very thick, black] (-1.4,-0.2) arc (191.5:88:0.4 and 1);
       \draw[thick, black, dashed] (-1.4,-0.2) arc (191.5:0:0.4 and 1);
        \fill[white] (-1,0) -- (0.45,0) -- (0.45,-0.2) -- (-1,-0.2);
        \fill[white] (1,0) -- (1.5,0) -- (1.5,-0.2) -- (1,-0.2);
        
        \draw[thick,wilsonred, dashed,fill=wilsonred!20] (-0.2,-0.2) arc (191.5:0:0.2 and 1);
        \draw[very thick, black] (-1,1) to[out =-20, in = 140] (1,0.8) to[out =-40, in = 90] (1.24,-0.1);
        \draw[very thick, black] (-1.4,-0.18) -- (0.6,-0.18);
        \draw[thick, dashed, black] (-0.6,0) -- (0.6,0);
        \draw[thick, dashed] (0.6,0) to[out = 0, in = 150] (1.25,-0.1);
        \draw[very thick, black] (1.26,-0.08) to[out = 205, in = 0] (0.6,-0.18);
        \draw[very thick] ({-0.6-1.5},{0}) -- ({-1.4-1.3},{-0.18-0.2}) -- (1.4,-0.38) -- ({0.6+1.3},{0});
    \end{tikzpicture}
    }}.
\end{equation}
A surface $S$ with no $\partial$-compressing disks is called $\partial$-incompressible. The two notions of incompressibility are independent: a surface may be bulk incompressible but $\partial$-compressible, or $\partial$-incompressible but bulk compressible.

\paragraph{Relation to hyperbolization theorem with geodesic boundaries.} 
Bulk incompressible cylinders (or equivalently, annuli) play an important role in Thurston’s hyperbolization theorems quoted in section \ref{subsec:what 3-manifolds are generated}. These annuli may be either $\partial$-compressible or $\partial$-incompressible. An example of a bulk incompressible annulus inside the genus-2 handlebody is displayed in \eqref{eq:bulk incompressible cylinder}. 
This annulus is $\partial$-compressible. Indeed, a known result about handlebodies is that they do not contain surfaces that are both bulk-incompressible and $\partial$-incompressible, see e.g.~\cite{martelli2016introduction}. For a discussion and classification of incompressible annuli in the genus-$g$ handlebody, see \cite{rubinstein1999genus,doi:10.1142/S0218216509006914}.

We also mention here the notion of $\partial$-irreducibility of a manifold $M$, which features in the variant of the hyperbolization theorem with geodesic boundaries. This notion implies that all essential annuli are $\partial$-incompressible. More precisely,
\begin{quote}
Let $M$ be irreducible and $\partial M$ incompressible, i.e.~$\partial M$ is $\pi_1$-injective. Then an essential annulus $S$ in $M$ is  also $\partial$-incompressible.
\end{quote}
To show this, assume that $S$ is $\partial$-compressible, i.e.~that there exist a $\partial$-compressing disk $D$. Compressing along $D$ turns the annulus $S$ into a disk $D' \subset M$. The boundary of $D'$ is a circle in $\partial M$ which is homotopically trivial in $M$. Since $\partial M$ is incompressible, it follows that $\partial D'$ bounds a disk in $\partial M$. The union of the boundary disk and $D'$ defines a sphere inside $M$, but since $M$ is prime (i.e.~$\pi_2(M)$ is trivial), this sphere must bound a ball. It follows that $D'$ is boundary parallel; however, this means that the original annulus $S$ is also boundary parallel. Thus, $S$ cannot be essential and we have reached a contradiction.

Let us now show how the presence of an essential cylinder is an obstruction to the existence of a hyperbolic metric with a totally geodesic boundary. Let $M$ be a manifold with boundary $\partial M$, and suppose $M$ contains an essential cylinder. We can consider the doubled manifold $M \cup \overline{M}$ glued along the common boundary. Now suppose that $M$ admits a finite-volume hyperbolic metric with geodesic boundary, then the doubled manifold $M \cup \overline{M}$ also admits a finite-volume hyperbolic metric. However, an essential annulus in $M$ uplifts to an essential torus in $M\cup \overline{M}$ (incompressibility follows for one cycle from incompressibility of the annulus and for the other cycle by non-boundary parallelness). Since $M \cup \overline{M}$ has an incompressible torus, it cannot carry a finite-volume hyperbolic metric so we reached a contradiction. Hence $M$ cannot carry a finite-volume metric with a geodesic boundary.

There is a second obstruction to the existence of a finite-volume hyperbolic metric. If the boundary $\partial M$ is an incompressible surface, i.e.~if the inclusion $\pi_1(\partial M) \to \pi_1(M)$ is injective, the manifold is said to be $\partial$-irreducible (or boundary-irreducible). Manifolds that are not $\partial$-irreducible cannot carry a finite-volume hyperbolic metric with geodesic boundary for the folloing argument. For such a manifold, there is a curve $\gamma$ on the boundary that is null-homotopic in the bulk, but not null-homotopic on the boundary, i.e.~bounds a disk $D$ in the bulk. We can then run the same argument as above. In the doubled manifold $M \cup \overline{M}$, $D$ uplifts to a sphere in $M \cup \overline{M}$, i.e.~the manifold is not irreducible and thus not hyperbolic. 

Although the machine does not produce acylindrical manifolds, it \emph{does} produce manifolds which are $\partial$-irreducible, an example being the twisted $I$-bundles discussed in section~\ref{subsec:twisted I bundles}.

\subsection{Wirtinger presentation of fundamental groups} \label{subapp:fundamental group}
We often make use of the fundamental group of 3-manifolds. There is a systematic algorithm to compute these for complements of Wilson line networks known as the Wirtinger presentation. We refer for further background to \cite[chapter 9]{RonaldBrown}.

Suppose we consider the complement of a network of Wilson lines in $\mathrm{S}^3$. Alternatively, we could also thicken the Wilson lines and consider a knotted genus-$g$ boundary, but we will continue to think of network of Wilson lines. Choose a projection to the plane and endow all arcs with orientations. For concreteness, we will explain this for the handlebody-knot $\mathbf{5}_2$ appearing in \eqref{eq:52}. We use the presentation
\be 
\mathbf{5}_2= \vcenter{\hbox{
\begin{tikzpicture}
\begin{scope}[xscale=1, yscale=1]
    \draw[very thick, wilsonred,->]
    (-0.67, -0.2) to[out = -80, in =180]
    (0.76,-1.2) node[below, black] {$a$};
    \draw[very thick, wilsonred] (0.76,-1.2) to[out=0, in=-90]
    (1.7, -0.1);
    \draw[very thick, wilsonred,->]
    (1.7, -0.1) to[out = 90, in = 10] node[above, black, shift={(.5,-.2)}] {$b$}
    (-0.6,0.81);
    \draw[very thick, wilsonred,->]
    (-0.8,0.75) to[out = {180+10}, in=40] (-1.4,.5) node[above, black] {$e$};
    \draw[very thick, wilsonred] (-1.4,.5) to[out={40-180}, in = 90]
    (-1.70, -0.1);
    \draw[very thick, wilsonred,->]
    (-1.70, -0.1) to[out= 270, in = 135]
    (-1.37, -1) node[left, black, shift={(0,-.1)}] {$d$};
    \draw[very thick, wilsonred] (-1.37, -1) to[out=-45, in = 215]
    (-0.1, -1.1);
    \draw[very thick, wilsonred,->]
    (0.1,-0.9) to[out=35, in = -85]
    (0.62,0.62) node[left, black, shift={(0.1,-.3)}] {$c$};
    \draw[very thick, wilsonred,->]
    (0.6,0.93) to[out = 100, in=0]
    (0, 1.7) node[below, black] {$h$};
    \draw[very thick, wilsonred] (0,1.7) to[out = 180, in = 95]
    (-0.71, 0);
    \draw[very thick, wilsonred,->]
    (-1.7, -0.1) to node[above, shift={(.3,-.1)}, black] {$f$} (0.47, -0.1);
    \draw[very thick, wilsonred,->]
    (0.7, -0.1) to node[above, black, shift={(0.2,-.05)}] {$g$}
    (1.2, -0.1);
    \draw[very thick, wilsonred] (1.2,-.1) to (1.7,-.1);
\end{scope}
\end{tikzpicture}
}} \label{eq:5_2 handlebody-knot generators}
\ee
Every arc will lead to one generator in the fundamental group, and we labeled the arcs with $a,b,c,d,e,f,g,h$ in the picture. We choose a base point for the fundamental group outside the screen closer to the reader. Each generator associated to an arc corresponds to the loop starting at the base point, going around the arc according the right-hand rule (so that the thumb goes along the direction of the chosen orientation and the remaining fingers indicate the orientation of the loop) and returning to the base point.

\paragraph{Relations.} Clearly, reversing the direction of the chosen orientation corresponds to choosing the inverse generator. For every junction and every over- and under-crossing, we get one relation. For a junction with only outgoing arrows, the counterclockwise product of all the generators is unity, since the concatenation of the paths is contractible behind the figure (drawn for concreteness for a cubic junction):
\be 
\vcenter{\hbox{

}}:\quad x_1x_2x_3=1\ .
\ee
For a crossing, we also produce a relation in which the new generator after the crossing gets conjugated by the generator of the strand that overcrosses it. Concretely, if we orient the arrows as in the figure, then the blue path corresponding to $x_3$ is equivalent to the path $x_2 x_1 x_2^{-1}$,
\be 
\vcenter{\hbox{
%
}}: \quad x_3=x_2x_1x_2^{-1}
\ .
\ee
In any given network of lines, we can always omit one of the relations coming from a vertex or a crossing. The relations can all be understood as the statement that the loop behind the vertex or the crossing is contractible. Any such given loop may be rewritten as a concatenation of loops around all the other vertices and its triviality is thus implied by all the other relations.

\paragraph{The $\mathbf{5}_2$ handlebody-knot.} If we apply this to the generators as in \eqref{eq:5_2 handlebody-knot generators}, then we obtain the following presentation for the fundamental group:
\begin{align}
    \pi_1(\mathbf{5}_2)\cong \left\langle \begin{array}{c} a,\,b,\,c,\,d\,,\\
    e,\,f,\,g,\,h \end{array} \, \left| \, \begin{array}{c} df e^{-1}=1 , \,  a^{-1}b g^{-1}=1 , \, e=h b h^{-1},\, h=b c b^{-1},\\
    g= c f c^{-1},\, a =f h f^{-1},\,  c=a^{-1}da \end{array}\right.\right\rangle\ .
\end{align}
The first two relations are the vertex relations, while the other relations are the crossing relations. We can clearly simplify this presentation dramatically. Let us intially do this by hand. We can easily eliminate the generators $ d,\, e,\, f,\,g,\,h$ by solving relations $1$, $2$, $3$, $4$ and $7$ and expressing them in terms of the other generators $a$, $b$ and $c$,
\be 
g=a^{-1}b\,,\ d=ac a^{-1}\,,\ e=b c b c^{-1} b^{-1}\,,\ f=ac^{-1}a^{-1}b c b c^{-1} b^{-1}\,,\   h=b c b^{-1}\ .
\ee
This leaves only the relations 5 and 6, which can easily be seen to be equivalent, in line with what we explained above that one of the relations may be dropped since it is implied by all the others.

This simplifies the group to
\be 
\pi_1(\mathbf{5}_2)\cong\langle a,\, b,\, c \, | \, cac^{-1}a^{-1}bcbc^{-1}b^{-1}c^{-1}b^{-1}a =1 \rangle\ . \label{eq:5_2 fundamental group reduced}
\ee
Thus there is only one relation $R(a,b,c)=1$, with
\be 
R(a,b,c)=cac^{-1}a^{-1}bcbc^{-1}b^{-1}c^{-1}b^{-1}a \label{eq:5_2 relation}
\ee
in the fundamental group. All fundamental groups of interest to us in this paper are of this form.

\paragraph{Whitehead algorithm.} The fundamental group in the form \eqref{eq:5_2 fundamental group reduced} is simplified, but it can be difficult to see whether it is isomorphic to another group of the form
\be 
\mathrm{G}=\langle a,b,c \, | \, R'(a,b,c) = 1 \rangle
\ee
where $R'$ is a different relation. This is the case if there is an (possibly outer) automorphism $\phi: F_3 \to F_3$ with $R(a,b,c)=R'(\phi(a),\phi(b),\phi(c))$, where $F_3$ is the free group on three generators. For the free group, the Whitehead algorithm constructs $\phi$ (or rules out its existence).

Let us describe the algorithm for a free group $F_n$ with generating set $X=\{x_1,\dots,x_n\}$. We find $\phi$ as a composition of basic automorphisms known as the Whitehead automorphisms. There are two types:
\begin{enumerate}
    \item For $\sigma \in S_n$ and $s \in \{\pm\}^n$, define 
    \be 
        \phi_{\sigma,s}^{(1)}(x_i)=x_{\sigma(i)}^{s_i}
    \ee
    \item For $x \in X \cup X^{-1}$ and $A \subseteq (X \cup X^{-1}) \setminus \{x,x^{-1}\}$, define
    \be 
        \phi_{x,A}^{(2)}(x_i)=\begin{cases}
            x_ix\,,\quad &x_i \in A \text{ and }x_i \not \in A^{-1} \cup \{x,x^{-1}\}\,, \\
            x^{-1}x_i\,,\quad &x_i \in A^{-1} \text{ and }x_i \not \in A \cup \{x,x^{-1}\}\,, \\
            x^{-1}x_i x\,,\quad &x_i \in A \text{ and }x_i \in A^{-1}\,,\\
            x_i\,,\quad&\text{otherwise}\,.
        \end{cases} 
    \ee
\end{enumerate}
Every automorphism of $F_n$ can be written as a composition of such basic Whitehead automorphisms. 

Let now $w \in F_n$ be a reduced word in $F_n$ (meaning that occurrences of $x_i x_i^{-1}$ have been canceled). Let $|w|$ be its length, i.e.~the number of letters. By definition, $|x_i^n|=|n|$. The Whitehead algorithm aims to find an automorphism $\phi$, such that the length $|\phi(w)|$ is minimized. The crucial fact about Whitehead automorphisms is the following: If $w$ is not of minimal length (i.e.~if $\phi$ exists such that $|\phi(w)|<|w|$), then we can choose $\phi$ to be a Whitehead automorphism. Thus, the algorithm tests if $|\phi(w)|<|w|$ for any Whitehead automorphism $\phi$ (for which there are finitely many). If yes, we pick $\phi$ for which the length decreased and repeat the procedure. If no (i.e.~if $|\phi(w)|=|w|$ for all Whitehead automorphisms $\phi$), then $w$ has minimal length. We then find all words with minimal length equivalent to $w$ (of which there are finitely many) by repeatedly applying Whitehead algorithms until we do not produce any new words of minimal length. When applying this algorithm to two different words, we can conclusively check whether the sets of equivalent minimal-length words agree. If yes, then the resulting group obtained from the relation $w=1$ is isomorphic, otherwise they are non-isomorphic.

We exemplify this produce at the example of $w_1=R(a,b,c)$ given in \eqref{eq:5_2 relation}, which has length 12. Then
\be 
\phi^{(2)}_{c,\{a^{-1},b^{-1}\}}(w_1)=w_2=ac^{-1}a^{-1}b^2c^{-1}b^{-2}a\ ,
\ee
which has length 9. It then finds
\be 
\phi^{(2)}_{b,\{a,c,c^{-1}\}}(w_2)=w_3=ac^{-1}a^{-1}b c^{-1}b^{-1}ab\ ,
\ee
which has length 8. Finally
\be 
\phi^{(2)}_{a^{-1},\{b^{-1}\}}(w_3)=w_4=ac^{-1}bcb^{-1}ab\ ,
\ee
which has length 7. At this point, no Whitehead automorphism exists that reduces the length further. We then apply all Whitehead automorphisms to $w_4$ and keep all words of length 7. After iterating this procedure 5 times, we have found all 6720 words of length 7 that are equivalent to $w_4$. We can pick out a canonical representative by taking the word that appears first in lexicographic order given by $a^3b^2c^2$. It is obtained from the composition
\be 
a^3b^2c^2=\phi^{(2)}_{b^{-1},\{c,c^{-1}\}}\circ \phi^{(2)}_{a^{-1},\{b^{-1},c,c^{-1}\}}\circ \phi^{(2)}_{a^{-1},\{b^{-1},c\}}\circ \phi^{(1)}_{(12),\{+,+,-\}}\circ \phi^{(2)}_{b^{-1},\{a,a^{-1}\}}(w_4)\ .
\ee
The final automorphism that satisfies $R(\phi(a),\phi(b),\phi(c))=a^3b^2c^2$ is the rather non-trivial automorphism
\begin{subequations}
\begin{align}
    \phi(a)&=b^{-1}a^{-2}bcb^{-1}aba^3ba^2b\ , \\
    \phi(b)&=b^{-1}a^{-2}bcb^{-1}aba^3b\ , \\
    \phi(c)&=b^{-1}a^{-1}bc^{-1}b^{-1}a^2b\ .
\end{align} \label{eq:5_2 pi_1 automorphism}
\end{subequations}
In particular, this algorithm showed that the group $\pi_1(\mathbf{5}_2)$ is not free (since otherwise $R(a,b,c)=1$ would be equivalent to the relation $a=1$) and thus the handlebody-knot $\mathbf{5}_2$ is not a handlebody.

\subsection{Tricolorability and handlebodyness} \label{subapp:tricolorability}
Computing the fundamental group and applying the Whitehead algorithm is a systematic way to check whether two handlebody-knots are equivalent, as long as the fundamental group only has one relation. There is another simpler test to distinguish handlebody-knots.
In the main text, we used the fundamental group as an invariant as discussed in appendix~\ref{subapp:fundamental group} for the handlebody-knot $\mathbf{5}_2$.
The goal of this section is to present a much simpler criterion for detecting non-trivial handlebody-knots. This test is a generalization of tricolorability for classical knots.

\paragraph{Tricolorability for knots.} Let us first review tricolorability for regular knots. A knot is called tricolorable if each segment of the knot when projected onto the plane can be colored using the following set of rules:
\begin{enumerate}
    \item All three colors must be used.
    \item At every crossing all three colors are used, or only one color is used. 
\end{enumerate}
If any two knots are related to each other via Reidemeister moves, then either both knots are tricolorable, or neither is. In particular, the unknot is not tricolorable. The classic example of a tricolorable knot is the trefoil,
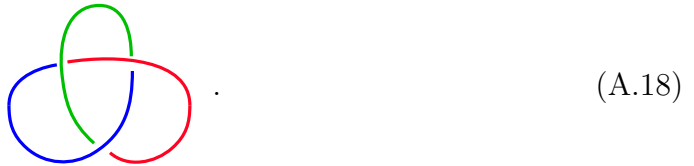
\begin{equation}
\vcenter{\hbox{
\begin{tikzpicture}
\begin{scope}[xscale = 0.7, yscale={0.7*0.9}]
    \draw[very thick, wilsonred]
    (1.7, -0.1) to[out = 90, in = 10]
    (-0.6,0.81);
    \draw[very thick,  blue]
    (-0.8,0.75) to[out = {180+10}, in = 90]
    (-1.70, -0.1);
    \draw[very thick, blue]
    (-1.70, -0.1) to[out= 270, in = 135]
    (-1.37, -1) to[out=-45, in = 215]
    (-0.1, -1.1) to[out=35, in = -90]
    (0.62,0.62);
    \draw[very thick, green!75!black]
    (0.6,0.93) to[out = 90, in=0]
    (0, 2) to[out = 180, in = 95]
    (-0.71,0.62) to[out = -85, in =135]
    (-0.1, -0.9);
    \draw[very thick, wilsonred]
    (0.2, -1.1) to[out = -45, in = 225]
    (1.37, -1) to[out=45, in=-90]
    (1.7, -0.1);
\end{scope}
\end{tikzpicture}
}}\ .
\end{equation}
\paragraph{Tricolorability for handlebody-knots.} For handlebody-knots, a very similar test applies with a few modifications. First, we need to pick a link component of the diagram. We call this component the \emph{enhanced constituent link} of the handlebody-knot. Examples of enhanced constituent links are given below in \eqref{eq:examples}; they correspond to the edges with double lines in the diagrams. Note that the enhanced constituent link may have more than one component. We say that an edge is \emph{thick}, if it is part of an enhanced constituent link; we refer to the rest of the edges as \emph{thin}.

A handlebody-knot diagram with an enhanced constituent link is tricolorable if we can color it following the rules:
\begin{enumerate}
    \item All three colors must be used.
    \item At every trivalent vertex, the colors of each edge must be the same.
    \item At every crossing:
    \begin{enumerate}
        \item[a)] If the over-edge is thick, then either all colors are used, or only one color is used.
        \item[b)] If the over-edge is thin, then the under-edges do not change color. 
    \end{enumerate}
\end{enumerate}
Below are a few examples of tricolorable handlebody-knots:
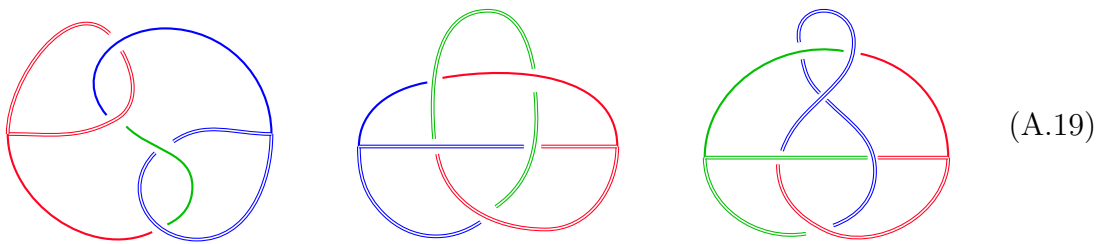
\begin{equation}
\label{eq:examples}
\vcenter{\hbox{

}}
\end{equation}
Note that depending on the choice of enhanced constituent link, the diagram may or may not be colorable. For a handlebody-knot to be tricolorable, we simply require that an enhanced constituent link exists such that the diagram can be colored.

Importantly, the trivial handlebody-knot is not tricolorable, as it can be seen from the diagram below; regardless of the choice of enhanced constituent link, only the trivial coloring is possible:
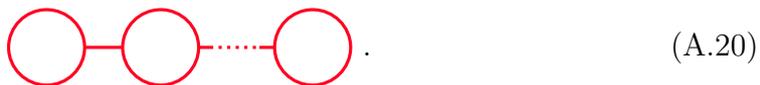
\begin{equation}
\vcenter{\hbox{
%
}}.
\end{equation} 

The tricolorability criterion was introduced in \cite{Ishii:2008}, where it was demonstrated that if two handlebody-knots are related to each other by a sequence of moves which include the  $\mathbb{F}$ and $\mathbb{B}$ transforms, and all the Reidemeister moves, then they are both tricolorable or neither is. 
This shows that no tricolorable handlebody-knot, in particular those in \eqref{eq:examples}, can be a handlebody.

\section{The fundamental group of the gravitational machine} \label{app:Bass-Serre theory}

In this appendix, we further discuss the group theoretic description of the fundamental groups produced by the machine. In particular, we prove the theorems used in section~\ref{subsec:machine is hyperbolic} to show that all fundamental groups of the 3-manifolds produced by the machine do not have $\ZZ \times \ZZ$ subgroups. This requires us to introduce some elements of Bass-Serre theory, which turns out to be a very useful tool. Here, we give a brief introduction to the subject, with \cite{SerreTrees} being the standard reference.  We begin in \ref{app:motivationB-S} with examples that motivate and explain the main ideas without giving any formal definitions. Then, we move to the definitions and main constructions in \ref{app:definitionsBS} and \ref{app:FTBS}. These will allow us to prove the relevant theorems in \ref{app:applicationBS}.

\subsection{What is Bass-Serre theory?}
\label{app:motivationB-S}

The idea behind Bass-Serre theory is to characterize a group $G$ through its action on a tree (a connected graph without loops). By studying this action, one can describe the group as built from simpler groups, glued together using amalgamated products and HNN extensions.
The structure of the group and its tree is encoded by a \emph{graph of groups}, which consists of a finite graph, not necessarily a tree,  together with a group assigned to each vertex and each edge and so-called boundary maps. To construct the graph of groups, one considers the quotient of the tree by the group action. The groups assigned to the vertices and edges of the graph correspond to stabilizer subgroups. These are the subgroups that fix a given vertex or edge of the tree under the group action. Bass-Serre theory can be seen as the dictionary that relates graphs of groups with groups acting on trees.

A very simple example of a tree with a group action is the integer line
\be 
\vcenter{\hbox{\begin{tikzpicture}[
  scale=.6,
  every node/.style={circle,fill,inner sep=1.5pt},
  line cap=round
]
  \tikzset{arrowedge/.style={->, very thick, shorten >=2pt, shorten <=2pt}}
  \node (a0) at (0,0) {};
  \foreach \i [evaluate=\i as \j using \i-1] in {1,2,3,4,5} {
    \node (a\i) at (1.5*\i,0) {};
    \draw[arrowedge] (a\j) -- (a\i);
    }
  \end{tikzpicture}
}}\ , \label{eq:integers tree}
\ee
continued to infinity on both the left and the right. The acting group is the group of integers $\mathbb{Z}$, which acts by discrete translation. The graph of groups is the quotient of the tree under this action, 
\begin{equation}
\label{eq:HNNintegers}
\vcenter{\hbox{\begin{tikzpicture}[
  scale=.6,
  line cap=round
]
  \tikzset{arrowedge/.style={->, very thick, shorten >=2pt, shorten <=2pt}}
  \tikzset{circlenode/.style={circle,fill,inner sep=1.5pt}}
  \node[circlenode, label=below:{$1$}] (a0) at (0,0) {};
  \draw[arrowedge, out=45, in=135, looseness=20] (a0) to node[above] {$1$} (a0);
  \end{tikzpicture}
}},
\end{equation}
where $1$ denotes the trivial group. The action has no fixed vertices or edges, so we attach the trivial group to the vertex and edge in the graph.
The graph of groups allows one to reconstruct the original group $G$ acting on the tree from the invariant subgroups attached to it. We explain this reconstruction in detail in the next section. For the case of a single loop, the group $G$ turns out to be an HNN extension:\footnote{
The HNN extension $G*_{A\sim B}$ of a group $G = \langle S\, \lvert\, R\rangle$ defined via some generators $S$ and relations $R$, with $\tau: A \rightarrow B$ an isomorphism between two subgroups $A,B\subset G$, is given by 
\begin{equation}
    G*_{A\sim B} = \langle S, t\, \lvert\, R, t a t^{-1} = \tau(a), \forall a\in A \rangle.  
\end{equation}
It is also sometimes written as $G*_{\tau}$, with $\tau$ the isomorphism between $A$ and $B$. For an introduction to HNN extensions, including an HNN extended version of the Seifert–van Kampen theorem see 
\cite{monster:2024}, page 1685. 
} 
\begin{equation}
\vcenter{\hbox{\begin{tikzpicture}[
  scale=.6,
  line cap=round
]
  \tikzset{arrowedge/.style={->, very thick, shorten >=2pt, shorten <=2pt}}
  \tikzset{circlenode/.style={circle,fill,inner sep=1.5pt}}
  \node[circlenode, label=below:{$G_1$}] (a0) at (0,0) {};
  \draw[arrowedge, out=45, in=135, looseness=20] (a0) to node[above] {$H$} (a0);
  \end{tikzpicture}
}} \quad \longrightarrow \qquad  G = G_1*_{H}.
\label{eq:HNNEX}
\end{equation}
For the group of the integers, \eqref{eq:HNNintegers} corresponds to the presentation:
\begin{equation}
   \ZZ \cong 1*_{1} = \langle e\rangle,
\end{equation} 
where $\langle e\rangle$ is the infinite cyclic group on a single generator $e$.

A different example is to consider the action of the infinite dihedral group D$_\infty \cong \ZZ_2 * \ZZ_2$ acting on the same tree (which one can consider as an $\infty$-gon).
Now there are two kinds of vertices: those serving as targets and those serving as sources. We color code them as follows
\be 
\vcenter{\hbox{
}}\ , \label{eq:Z2 Z2 tree}
\ee
again continued indefinitely to the left and the right. The action of D$_\infty$ is defined as follows. We start by choosing a pair of adjacent blue and red vertices. We let one of the $\ZZ_2$ factors act by a half-turn around the blue vertex; the other $\ZZ_2$ factor acts by a half-turn around the red vertex. The graph of groups is given by
\begin{equation}
\vcenter{\hbox{%
}}.
\end{equation}
This graph makes the structure of D$_\infty$ as an amalgamated product of two components very explicit. More generally, the group reconstructed from a graph of groups which is given by a single edge and two vertices is given by 
\begin{equation}
\vcenter{\hbox{%
}} \quad \longrightarrow \qquad G= G_1*_{H}G_2.
\label{eq:amalgamatedP}
\end{equation}

A more complicated example is the action of the modular group, which we present as $\PSL(2,\ZZ)=\langle S,\, T\, | \, S^2=1=(ST)^3 \rangle$, acting on the Farey tree, which is a tree with two kinds of vertices, namely 2-valent vertices and 3-valent vertices. Edges connect the 2-valent vertices with 3-valent vertices and we can take the orientation to be towards the 3-valent vertices. Here is a portion of the Farey tree with a similar color coding,
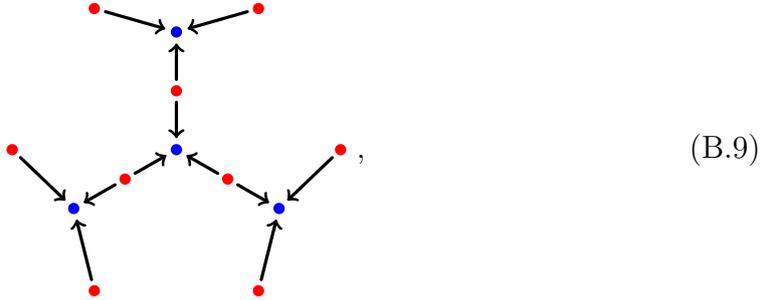
\be 
\vcenter{\hbox{%
}} \ ,\label{eq:Farey tree}
\ee
again continued indefinitely.
The $S$ and $ST$ elements rotate adjacent 2-valent and 3-valent vertices, respectively. As in  D$_\infty$, the fundamental domain of this tree is a single edge with two vertices; however, the stabilizer groups of the vertices are different:
\begin{equation}
\vcenter{\hbox{%
}}.
\end{equation}
The idea is that the graph of groups makes the relation $\PSL(2,\ZZ) \cong \ZZ_2* \ZZ_3$ explicit.

As a last example, consider the similar-looking group $\SL(2,\ZZ)=\langle S,\, T \, | \, S^4=(ST)^6=1,\, S^2=(ST)^3 \rangle$. We can also let it act on the Farey tree \eqref{eq:Farey tree}. In this case, the element $S^2=(ST)^3$ has a trivial action, leaving the graph invariant. The graph of groups now reads, 
\begin{equation}
\vcenter{\hbox{%
}}.
\end{equation}
This time the edge has a non-trivial stabilizer group. From the graph of groups, we can read off the group isomorphism:
\begin{equation}
    \SL(2,\ZZ) \cong \ZZ_4 *_{\ZZ_2} \ZZ_6.
\end{equation}

\subsection{Groups acting on trees}
\label{app:definitionsBS}

The motivation for studying Bass-Serre theory is that the fundamental groups generated by the gravitational machine are described using amalgamated products and HNN extensions. 
In this setting, the machine provides a graph of groups, and our goal is to reconstruct the associated group from this graph. We also want to build a tree on which this group acts.
The passage from a graph of groups to a tree with a group action is the content of the Fundamental Theorem of Bass-Serre theory.
This theorem is explained in the next section; here, we provide the formal definitions needed in the construction. 

\paragraph{Trees.} An oriented tree $T = (\tilde{\mathcal{V}},\tilde{\mathcal{E}})$\footnote{We use the tilde notation to refer to the vertices and edges of the tree. The notation without a tilde refers to the vertices and edges of the graph of groups.} is a connected graph with a collection of vertices $\tilde{v} \in \tilde{\mathcal{V}}$ and oriented edges $\tilde{e} \in \tilde{\mathcal{E}}$ such that there are no loops. That is, every pair of vertices $\tilde{v},\tilde{w}\in \tilde{\mathcal{V}}$ is connected by exactly one path. In particular, a tree has no multi edges. For each edge $\tilde{e}$ we denote its target (the vertex the edge points to) by $\tilde{e}^+$ and its source by $\tilde{e}^-$. The trees considered below are typically infinite, and vertices may have infinitely many adjacent edges.

In a tree, there is a unique path (without backtracking) that connects any two vertices. This path is called the geodesic between the vertices, and its length is the distance between them. The distance from $\tilde v$ to $\tilde w$ is denoted by $d(\tilde v,\tilde w)$. We have $d(\tilde v,\tilde v)=0$, and $d(\tilde v,\tilde w)=1$ if $\tilde v$ and $\tilde w$ are adjacent.

Let $G$ be a group acting on $T$. We assume that the action of $G$ on $T$ preserves edge orientations, so edge inversions are not allowed. Let $\text{Fix}(g) =\{\tilde x\in T \,\lvert\, g \tilde x = \tilde x \}$, then every $g \in G$ comes in two forms:
\begin{itemize}
    \item Elliptic: $\text{Fix}(g)$ is not the empty set, i.e.~a number of vertices is stabilized.
    \item Hyperbolic: $g$ stabilizes a unique geodesic set-wise; i.e.~$g$ translates along
    an infinite line through the tree.
\end{itemize}
There are no other options. For example, translations by an integer on the real line correspond to hyperbolic elements.  In the case of D$_\infty$, the two $\ZZ_2$ components as well as their conjugates are elliptic elements. For the Farey tree, $S$ and $ST$ and their conjugates are elliptic, while the hyperbolic elements are those matrices in $\PSL(2,\ZZ)$ whose trace has absolute value larger or equal to two.

\paragraph{Graph of groups.} For a group $G$ acting on a tree $T$, we can form the quotient $T/G$, where we identify all edges in a group orbit and all vertices in a group orbit. The resulting space is still a graph, but not necessarily a tree. For the four examples of $\ZZ$, D$_\infty$, $\PSL(2,\ZZ)$ and $\SL(2,\ZZ)$, the quotient graphs look like
\be
\vcenter{\hbox{\begin{tikzpicture}[
  scale=.6,
  every node/.style={circle,fill,inner sep=1.5pt},
  line cap=round
]
\useasboundingbox (-.6,0) rectangle (.6,.6);
  \tikzset{arrowedge/.style={->, very thick, shorten >=2pt, shorten <=2pt}}
  \node (a0) at (0,0) {};
  \draw[arrowedge, out=45, in=135, looseness=20] (a0) to (a0);
  \end{tikzpicture}
}}, \quad 
\vcenter{\hbox{\begin{tikzpicture}[
  scale=.6,
  every node/.style={circle,fill,inner sep=1.5pt},
  line cap=round
]
  \tikzset{arrowedge/.style={->, very thick, shorten >=2pt, shorten <=2pt}}
  \node[red] (b0) at (0,0) {};
  \node[blue] (a0) at (1.5,0) {};
  \draw[arrowedge] (b0) -- (a0);
  \end{tikzpicture}
}}\ , \quad
\vcenter{\hbox{\begin{tikzpicture}[
  scale=.6,
  every node/.style={circle,fill,inner sep=1.5pt},
  line cap=round
]
  \tikzset{arrowedge/.style={->, very thick, shorten >=2pt, shorten <=2pt}}
  \node[red] (b0) at (0,0) {};
  \node[blue] (a0) at (1.5,0) {};
  \draw[arrowedge] (b0) -- (a0);
  \end{tikzpicture}
}}\ , \quad 
\vcenter{\hbox{\begin{tikzpicture}[
  scale=.6,
  every node/.style={circle,fill,inner sep=1.5pt},
  line cap=round
]
  \tikzset{arrowedge/.style={->, very thick, shorten >=2pt, shorten <=2pt}}
  \node[red] (b0) at (0,0) {};
  \node[blue] (a0) at (1.5,0) {};
  \draw[arrowedge] (b0) -- (a0);
  \end{tikzpicture}
}}\ .
\ee
The first example shows that the quotient need not be a tree. The other three quotient graphs look identical. Clearly, taking the quotient forgets a lot of structure. The main idea of Bass-Serre theory is to upgrade this quotient of trees to a graph of groups. For a graph of groups, we retain the additional information regarding the stabilizers of the relevant vertices and edges. Let us give the formal definition.\\[-5pt]

\noindent A graph of groups $\Gamma = (\mathcal{V},\mathcal{E},G_{\mathcal{V}},G_{\mathcal{E}})$ consists of the following objects:
\begin{itemize}
    \item An underlying connected oriented graph with a set of edge $\mathcal{E}$ and a set of vertices $\mathcal{V}$. Multi-edges and loops are allowed. For every edge $e \in \mathcal{E}$, we denote the corresponding target vertex again by $e^+$ and the source vertex by $e^-$.
    \item A group $G_e$ to every edge $e \in \mathcal{E}$ and a group $G_v$ to every vertex $v \in \mathcal{V}$. Every edge group injects naturally in both the source and the target vertex group, $G_e \xhookrightarrow{} G_{e^\pm}$. 
    \item Boundary maps, that is group homomorphisms describing the inclusions $\partial_e^+:G_e \to G_{e^+}$ and $\partial_e^-:G_e \to G_{e^-}$. 
\end{itemize}
The quotient $T/G$ described above naturally carries the structure of a graph of groups. Indeed, we can associate to every vertex $v \in T/G$ and edge $e \in T/G$ the stabilizer groups  $\Stab(\tilde{v}) = \{g\in G \, \lvert\, g \tilde{v} = \tilde{v}\}$  and $\Stab(\tilde{e})$, respectively. Here, $\tilde{v}$ and $\tilde{e}$ are elements of the tree $T$ which have been lifted from the quotient $T/G$. For a stabilized edge $\tilde{e}$, also the source and target vertices $\tilde{e}^-$ and $\tilde{e}^+$ are stabilized, which gives natural inclusions $\Stab(e) \hookrightarrow \Stab(e^\pm)$ yielding the monomorphisms $\partial^\pm_e$. These constructions depend in principle on the chosen lifts, but this only conjugates the monomorphisms. In particular the resulting graph of groups for different choices are isomorphic. 
 
\subsection{Fundamental theorem of Bass-Serre theory}
\label{app:FTBS}

The main observation of Bass-Serre theory is that one can recover both the tree $T$ as well as the original group $G$ acting on it from the corresponding graph of groups $\Gamma$. Thus, the data of a graph of groups is equivalent to the tree with a group action. 

We already saw how to construct the graph of groups of $\Gamma$ from $G$ acting on $T$. We will in the following explain how to go in the other direction and reconstruct $G$ and the tree $T$ from $\Gamma$. In the literature, the group $G$ is commonly referred to as the \emph{fundamental group of the graph of groups}, and it is denoted by $G = \pi_1(\Gamma)$. The name and notation stems from a close connection to topology that we explain below. The tree $T$ constructed from a graph of groups is called the \emph{Bass-Serre covering tree}.

\paragraph{The fundamental group of a graph of groups.}
This group is defined as the fundamental group of an auxiliary topological space $X$, which is obtained by gluing a number of pathconnected spaces $X_v$, $v \in \mathcal{V}$, along edge spaces $X_e$, $e\in\mathcal{E}$, defined as follows.
For each edge $e \in \mathcal{E}$ in the graph of groups, $X_{e_+}$ and $X_{e^-}$ overlap in a common subspace $X_e=X_{e^+} \cap X_{e^-}$, which is identified in the gluing.
Ignoring the subtleties resulting from the choice of basepoint, one chooses the spaces $X_v$ and $X_e$ such that $\pi_1(X_v) \cong G_v$ and $\pi_1(X_e) \cong G_e$. The boundary monomorphisms $\partial_e^\pm$ in the graph of groups are given naturally by the inclusions $\pi_1(X_e) \subset \pi_1(X_{e^\pm})$; which we assume to be injective.
The fundamental group of the graph of groups $\Gamma$ is then defined as $\pi_1(\Gamma) \equiv \pi_1(X)$. This group can be computed using the Seifert-van Kampen theorem.\footnote{Note that there is a direct analogy between the space $X$ and the hyperbolic manifold $M$ generated by the machine. The vertex spaces $X_v$ correspond to the hyperbolic manifolds $M_i$, and the gluing along 2-punctured spheres are the edge spaces $X_e$.}

Equivalently, the fundamental group can be algebraically defined as follows. Consider first the group 
\begin{equation}
\label{eq:freeG}
\Big(\mathlarger{\mathlarger{\mathlarger{\mathlarger{*}}}}{}_{v \in \mathcal{V}} \; G_v\Big) * F_{|\mathcal{E}|}\ ,
\end{equation}
where $F_{|\mathcal{E}|}$ is the free group on $|\mathcal{E}|$ generators. That is, \eqref{eq:freeG} is the free group generated by all the vertex groups as well as additional generators, one for each edge in the graph of groups. In what follows, we simply denote these additional generators by $e$, for $e \in \mathcal{E}$.  We then impose the following relations:
\begin{enumerate}
    \item For all $e \in \mathcal{E}$ and $g \in G_e$, set $e \partial_e^+(g) e^{-1}=\partial_e^-(g)$.
    \item Choose a \emph{tree} inside the graph of groups that includes all its vertices, $(\mathcal{V},\mathcal{E}_0) \subset (\mathcal{V},\mathcal{E})$. For all $e \in \mathcal{E}_0$, set $e=1$.
\end{enumerate}
The first relation describes the amalgamation in the Seifert-van Kampen theorem. The second relation is perhaps more subtle. It requires us to choose a spanning tree in $(\mathcal{V},\mathcal{E})$, i.e.~a subgraph of $(\mathcal{V},\mathcal{E})$ that is a tree as defined in section \ref{app:definitionsBS} whose vertex set coincides with the original vertex set. One can show that the fundamental group for different choices of spanning of trees is isomorphic and this choice is analogous to the choice of base point in the topological fundamental group.\footnote{There is also an equivalent definition of the fundamental group, where we instead choose an arbitrary vertex and the second relation is replaced by another condition saying loosely that we only consider in a certain sense loops based at this vertex.} The second relation now formally realizes the intuition that only every loop in the graph should provide a new generator for the fundamental group, whereas edges in a tree do not provide new generators since the tree is contractible. 

With this definition, one can recover the groups $\ZZ$, D$_\infty$, $\PSL(2,\ZZ)$ and $\SL(2,\ZZ)$ from the graphs of groups given as examples in section \ref{app:motivationB-S}. For the group of integers, the spanning tree in the graph of groups is the single vertex without the edge. For the other groups, the spanning tree is the whole graph since there are no loops. More generally, the fundamental group of the graphs \eqref{eq:HNNEX} and \eqref{eq:amalgamatedP}
are the HNN extension $G_1 *_H$ and the amalgamated product $G_1 *_H G_2$, respectively. 

\paragraph{The Bass-Serre covering tree.} Let us now construct the covering tree $T = (\tilde{\mathcal{V}},\tilde{\mathcal{E}})$ on which the fundamental group $G= \pi_1(\Gamma)$ acts. 
Identify the vertices $\tilde{\mathcal{V}}$ and edges $\tilde{\mathcal{E}}$ of the tree $T$ with  the left cosets of the groups $G_v$ and $G_e$, 
\begin{equation}
    \tilde{\mathcal{V}} =  \{G/G_v\}_{v \in \mathcal{V}}\ , \quad \text{and} \quad 
    \tilde{\mathcal{E}} = \{G/G_e\}_{e \in \mathcal{E}}\ .
\end{equation}
Namely, we label vertices and edges as $gG_v$ and $g G_e$, respectively.\footnote{Notice that we set up the fundamental group such that $G_v$ is a subgroup of it, but the edge groups are only naturally included via the boundary monomorphisms, so the coset $G/G_e$ should be understood as, say, $G/\partial_e^-(G_e)$.}  The edges are attached to the vertices as follows. For $g G_e \in \tilde{\mathcal{E}}$, set
\be 
(g G_e)^+=g e G_{e^+}\quad \text{and}\quad (g G_e)^-= g G_{e^-}\ . \label{eq:covering tree edge relation}
\ee
Pictorially, we have: 
\begin{equation}
\vcenter{\hbox{\begin{tikzpicture}[
  scale=.6,
  line cap=round
]
  \tikzset{arrowedge/.style={->, very thick, shorten >=2pt, shorten <=2pt}}
  \tikzset{circlenode/.style={circle,fill,inner sep=1.5pt}}
  \node[circlenode, wilsonred, label=below:{$gG_{e^-}$}] (b0) at (0,0) {};
  \node[circlenode, blue, label=below:{$g e G_{e^+}$}] (a0) at (4,0) {};
   \draw[arrowedge] (b0) to node[above] {$g G_{e}$} (a0);
  \end{tikzpicture}
}}.
\end{equation}
The group $G$ acts on the tree $T$ by left multiplication: for $g'\in G$, $gG_v$ maps to $g'gG_v$ and $gG_e$ to $g'gG_e$. One can easily check that the group action preserves the edge relation. One can also show that this construction defines a tree, but this is perhaps the most difficult part of the construction.

To build some intuition, we can consider the case where the graph of groups corresponds to a single amalgamated product with two vertices $e^-$ and $e^+$ and one edge $e$, 
\begin{equation}
\vcenter{\hbox{\begin{tikzpicture}[
  scale=.6,
  line cap=round
]
  \tikzset{arrowedge/.style={->, very thick, shorten >=2pt, shorten <=2pt}}
  \tikzset{circlenode/.style={circle,fill,inner sep=1.5pt}}
  \node[circlenode, wilsonred, label=below:{$G_{e^-}$}] (b0) at (0,0) {};
  \node[circlenode, blue, label=below:{$ G_{e^+}$}] (a0) at (4,0) {};
   \draw[arrowedge] (b0) to node[above] {$ G_{e}$} (a0);
  \end{tikzpicture}
}}.
\end{equation}
Since the graph is itself a tree, the generators $e$ is set to one, so the vertices $gG_{e^-}$ and $gG_{e^+}$ are joined through the edge $gG_e$. The Bass-Serre covering tree $T$ of this graph has two basic properties. First, it contains two types of vertices: the sources $gG_{e^-}$ and the targets $gG_{e^+}$ (think of the 2-valent and 3-valent vertices in the Farey tree). 
Second, the number of edges attached to the vertices $gG_{e^-}$ and $gG_{e^+}$ are $|G_{e^-}|/|G_e|$ and $|G_{e^+}|/|G_e|$, respectively.  This follows because the left coset of a vertex $gG_v$ is unchanged under $g \rightarrow g g_v$ with $g_v \in G_v$, so the possible edges attached to $gG_v$ are of the form $g g_v G_e$. Thus, portions of the tree look like: 
\be 
\vcenter{\hbox{\begin{tikzpicture}[
  scale=.6,
  line cap=round
]
  \tikzset{arrowedge/.style={->, very thick, shorten >=2pt, shorten <=2pt}}
  \tikzset{circlenode/.style={circle,fill,inner sep=1.5pt}}
    \node[circlenode, blue, label=below:{$gG_{e^+}$}] (a0) at (-5.5,0) {};
  \node[circlenode, wilsonred, label=below:{$gG_{e^-}=g g_{e^{-}} G_{e^-}$}] (b0) at (0,0) {};
  \node[circlenode, blue, label=below:{$g g_{e^{-}} G_{e^+}$}] (c0) at (5.5,0) {};
  \draw[arrowedge] (b0) to node[above] {$gG_e$} (a0);
  \draw[arrowedge] (b0) to node[above] {$g g_{e^{-}} G_{e}$} (c0);
  \end{tikzpicture}
}}
\ee
where $g_{e^{-}} \in G_{e^{-}}$. 
However, since $G_e \subset G_v$, these edges are not all distinct, and any choice with $g_v \in G_e$ does not produce a new edge. Therefore, edges are in one to one correspondence with the left cosets $G_v/G_e$.
From these two simple observations one can recover, for example, the trees for D$_{\infty}$, $\PSL(2,\ZZ)$ and $\SL(2,\ZZ)$ from their graph of groups.

For the HNN extension $G=G_1 *_H$ depicted in \eqref{eq:HNNEX}, we can take $\partial_e^-$ to be the identity, i.e.~we naturally view $H \subset G_1 \subset G$. We denote by $\partial=\partial^+$. Then $(g H)^+=g e G_1$ and $(gH)^-=g G_1$. There are $2\times |G_1|/|H|$ vertices that one can attach to the vertex $gG_1$. 
These come from the distinct choices of $g g' H$ with $g' \in G_1$, that connect $g G_1$ to $gg' e G_1$, and from the edges $g g' e^{-1} H$ that connect $g g' e^{-1} G_1$ to $g G_1$.

\subsection{Application to the gravitational machine}
\label{app:applicationBS}

The fundamental groups of the gravitational machine are changed by gluing operations along 2-punctured spheres (or cylinders). By repeatedly applying this gluing, the fundamental group of the resulting manifolds can be precisely described using graph of groups where all the vertex groups are free groups (since these correspond to handlebodies) and all the edge groups are $\ZZ$ (since we are gluing along 2-punctured spheres). By the condition \ref{cond:no Omega loops 2}, all boundary monomorphisms $\partial_e^\pm$ are injective, so the graphs of groups are well defined. The main purpose of the theory introduced above is now that we can construct the Bass-Serre covering tree and study the action of the fundamental group acting on it. We use this technology to prove two results:
\begin{itemize}
    \item[] \emph{Result 1:} Given the amalgamated product $G = G_1 *_{H}G_2$, if $H$ is malnormal in $G_1$, then the centralizer $\mathcal{C}(g) = \{h\in G \ \lvert\  gh = hg\}$ is either infinite cyclic or conjugate to a subgroup of $G_1$ or $G_2$. Namely, $\mathcal{C}(g)\cong \ZZ$ or there is an injective homomorphism from $\mathcal{C}(g)$ to $G_1$ or $G_2$. 
    \item[] \emph{Result 2:} Given the HNN extension $G =  G_1*_{H\sim K}$, if $H$ is malnormal in $G_1$, and $H \cap g_1 K g_1^{-1} = \{1\}$ for all $g_1\in G_1$, then $\mathcal{C}(g)$ is either infinite cyclic or conjugate to a subgroup of $G_1$. 
 \end{itemize}
These two results will follow from the following observation: for the group $G$ satisfying the assumptions above, the action of $G$ on its Bass-Serre tree $T$ is 2-acylindrical, i.e.~$\text{diam}(\text{Fix}(g)) \leq 2$ for every $g\in G$, $g\neq 1$. The diameter defined here is the maximum distance between any two vertices $\tilde{v}, \tilde{w}$ which are fixed by the action of $g\in G$ on $T$. Intuitively, this means that when $G$ acts on $T$, it only stabilizes a small portion of the tree. 

We present the proof of these results in three parts. First, we show that $G$ acting on $T$ is 2-acylindrical. Recall that for a given $g\in G$, its action on the tree is either hyperbolic or elliptic.
In the second and third parts we show that, in either case, $\mathcal{C}(g)$ is either infinite cyclic or conjugate to $G_1$ or $G_2$. The first result is not new and appears in \cite{mathoverflow_amalgamatedproduct}. To the best of our knowledge, the second result has not been stated in this form.

\paragraph{Part I: Fixed-point sets.} For $G$ a group acting on a tree $T$, let $\text{Fix}(g)$ the set of vertices fixed by $g$.  On $T$, there is a natural metric given by the number of edges between two vertices. $\text{Fix}(g)$ is a subtree since for any two fixed vertices, the group action necessarily has to fix the geodesic connecting the two vertices.
Let
\be 
\text{diam}(\text{Fix}(g))=\sup \{ d(\tilde{x},\tilde{y}) \, | \, \tilde{x},\, \tilde{y} \in \text{Fix}(g)\}
\ee
the diameter of the fixed point set. The action of a group on a tree $T$ is called $k$-acylindrical if for all $g \ne 1$, $\text{diam}(\text{Fix}(g)) \le k$. Note that the centralizer  $\mathcal{C}(g)$ maps $\text{Fix}(g)$ to $\text{Fix}(g)$. Indeed, for $\tilde{v} \in \text{Fix}(g)$ and $c \in \mathcal{C}(g)$, $g \, c \tilde{v} = c \, g \tilde{v} = c\tilde{v}$ and so $c\tilde{v} \in\text{Fix}(g)$. 

Consider now the case of the amalgamated product $G=G_1 *_H G_2$ with $H$ malnormal in $G_1$. We show that the group action on the associated Bass-Serre tree $T$ is 2-acylindrical. Suppose a portion of the tree of the form
\be 
\vcenter{\hbox{\begin{tikzpicture}[
  scale=.6,
  line cap=round
]
  \tikzset{arrowedge/.style={->, very thick, shorten >=2pt, shorten <=2pt}}
  \tikzset{circlenode/.style={circle,fill,inner sep=1.5pt}}
    \node[circlenode, blue, label=below:{$gG_2$}] (a0) at (-4,0) {};
  \node[circlenode, wilsonred, label=below:{$gG_1=g'G_1$}] (b0) at (0,0) {};
  \node[circlenode, blue, label=below:{$g'G_2$}] (c0) at (4,0) {};
  \draw[arrowedge] (b0) to node[above] {$gH$} (a0);
  \draw[arrowedge] (b0) to node[above] {$g'H$} (c0);
  \end{tikzpicture}
}}
\ee
is fixed by some group element. For such a portion of the graph, we have $g^{-1}g' \in G_1 \setminus H$. The stabilizers of the edges are $\Stab(gH)=g H g^{-1}$, $\Stab(g'H)=g' H (g')^{-1}$. Thus the group element has to live in the intersection of both edge stabilizers, but this is impossible due to the malnormality of $H$ in $G_1$. Indeed, if $\tilde{g} \in \Stab(gH) \cap \Stab(g'H)$, there are $h,h'\in H$ such that 
\begin{equation}
g h g^{-1}  = \tilde{g} =  g' h' (g')^{-1}, \quad \text{thus}\quad 
h  = (g^{-1}g') h' (g^{-1}g')^{-1},
\end{equation}
which by malnormality means that $h = 1 = h'$ (since $g^{-1}g' \in G_1 \setminus H$).
Thus the fixed-point set cannot contain any interior red vertex. Note that the opposite orientation, with an interior blue vertex, could in principle be fixed by $g$, since we do not assume $H$ is malnormal in $G_2$. However, since edges only connect red and blue vertices, we conclude that $\text{diam}(\text{Fix}(g)) \le 2$ for any $g \in G$.

We can make a similar observation for the HNN case. There are three non-equivalent configurations to consider. By the same argument as above, the fixed-point set of $g$ cannot contain the corresponding portion
\be 
\vcenter{\hbox{\begin{tikzpicture}[
  scale=.6,
  line cap=round
]
  \tikzset{arrowedge/.style={->, very thick, shorten >=2pt, shorten <=2pt}}
  \tikzset{circlenode/.style={circle,fill,inner sep=1.5pt}}
    \node[circlenode, label=below:{$gtG_1$}] (a0) at (-4,0) {};
  \node[circlenode, label=below:{$gG_1=g'G_1$}] (b0) at (0,0) {};
  \node[circlenode, label=below:{$g'tG_1$}] (c0) at (4,0) {};
  \draw[arrowedge] (b0) to node[above] {$gH$} (a0);
  \draw[arrowedge] (b0) to node[above] {$g'H$} (c0);
  \end{tikzpicture}
}}
\ee
since $H$ is malnormal in $G_1$.

Consider now a portion of the form
\be 
\vcenter{\hbox{\begin{tikzpicture}[
  scale=.6,
  line cap=round
]
  \tikzset{arrowedge/.style={->, very thick, shorten >=2pt, shorten <=2pt}}
  \tikzset{circlenode/.style={circle,fill,inner sep=1.5pt}}
    \node[circlenode, label=below:{$gt^{-1}G_1$}] (a0) at (-4,0) {};
  \node[circlenode, label=below:{$gG_1=g'G_1$}] (b0) at (0,0) {};
  \node[circlenode, label=below:{$g'tG_1$}] (c0) at (4,0) {};
  \draw[arrowedge] (a0) to node[above] {$g t^{-1}H$} (b0);
  \draw[arrowedge] (b0) to node[above] {$g'H$} (c0);
  \end{tikzpicture}
}}\ ,
\ee
where $t$ is the stable letter.
Elements $\tilde{g}$ that stabilize this portion of the tree must be in the stabilizers $(gt^{-1})H (tg^{-1})$ and $g' H (g')^{-1}$, which means that there exist $h,h'$ such that 
\begin{equation}
    gt^{-1} \, h \, tg^{-1} = \tilde{g}
    = g' h' (g')^{-1}, \quad \text{hence} \quad t^{-1}ht  = (g^{-1} g') h  (g^{-1} g')^{-1}.
\end{equation}
Recall that $t^{-1}ht \in  K=\partial^+(H)$ and $g^{-1} g'\in G_1$. By the assumption that $H \cap g_1 K g_1^{-1}=\{1\}$ for all $g_1 \in G_1$, it follows that $h = 1 = h'$.\footnote{
Here, we are also using the fact that the homomorphism $G_1\to G$ induced by the inclusion of $G_1$ in its HNN extension is injective.  Thus $H\cap g_1 K g_1^{-1} = \{1\}$ with $g_1\in G_1$  being trivial in $G_1$ means the intersection is also trivial in $G$.} Thus, such a portion can also not be part of the fixed-point set. The largest possible fixed-point set is then of the form
\be 
\vcenter{\hbox{\begin{tikzpicture}[
  scale=.6,
  line cap=round
]
  \tikzset{arrowedge/.style={->, very thick, shorten >=2pt, shorten <=2pt}}
  \tikzset{circlenode/.style={circle,fill,inner sep=1.5pt}}
    \node[circlenode, label=below:{$gt^{-1}G_1$}] (a0) at (-4,0) {};
  \node[circlenode, label=below:{$gG_1=g'G_1$}] (b0) at (0,0) {};
  \node[circlenode, label=below:{$gt^{-1}G_1$}] (c0) at (4,0) {};
  \draw[arrowedge] (a0) to node[above] {$gt^{-1}H$} (b0);
  \draw[arrowedge] (c0) to node[above] {$g't^{-1}H$} (b0);
  \end{tikzpicture}
}}
\ee
and thus also $\text{diam}(\text{Fix}(g)) \le 2$ and the action on $T$ is 2-acylindrical.

\paragraph{Part II: Centralizer of hyperbolic elements.} 
We start with the case where $g$ is a hyperbolic element of a group $G$ whose action on the tree $T$ is $k$-acylindrical. We prove that $\mathcal{C}(g)$ is infinite cyclic.

By definition, $g$ translates along a unique geodesic $L \subset T$, that is, $g L = L$. Let $c \in \mathcal{C}(g)$, then the geodesic $c L$ is also fixed by $g$, since $g \, cL =  c \, gL = c L$. Since the geodesic fixed by $g$ is unique, it follows that $cL =  L$. Since the group $G$ acts on the tree  as an isometry (i.e.~the action preserves distances) we have the following homomorphism between the centralizer and the group of isometries of the integer line 
\be 
\varphi: \mathcal{C}(g) \longrightarrow \text{Isom}(L)
\ee
where $c$ gets mapped to the isometry induced by the action $c L$. The kernel of this map corresponds to the elements of $c$ which fix the geodesic $L$ point-wise. Thus, if $c\in \text{ker}\varphi$ then $c$ fixes infinitely many points, so $\text{diam}(\text{Fix}(c))=\infty$. This contradicts the $k$-acylindricity of the action of $G$. Therefore $\varphi$ is injective.
 
The group $\text{Isom}(L)$ consists of translations along $L$ and possibly reflections around vertices (depending on the orientation of the edge arrows). Suppose $c \in \mathcal{C}(g)$ corresponds to a reflection. But $g$ translates along the geodesic and $c g c^{-1}$ translates along the geodesic in the other direction. Since $g=cgc^{-1}$ this is impossible. Thus, the image of the homomorphism $\varphi$ has to lie in the translation subgroup along $L$, which is isomorphic to the integers. We conclude that $\mathcal{C}(g)$ is an infinite cyclic group.

\paragraph{Part III: Centralizer of elliptic elements.} Finally, we consider the case where $g$ is an elliptic element, and thus fixes a vertex $\tilde v$. 
For this part we will need the following fact about \emph{finite} trees, we include a derivation of this result at the end of the section. Let $T'$ be a tree  with finite diameter, i.e.~
\be 
\text{diam}(T')=\sup_{\tilde u,\tilde w \in T'} d(\tilde u,\tilde w)<\infty\ ,
\ee
and $G'$ a group acting on $T'$. Then there is one vertex $\tilde u_* \in T'$ that is fixed by every element $g'\in G'$. In particular all elements of $G'$ are elliptic.

Returning to the proof, we assume that $g$ is an elliptic element of a group $G$ whose action on the tree $T$ is $k$-acylindrical. Thus, $\text{Fix}(g)$ at least contains one vertex $\tilde{v}$ and is a non-empty subtree $T' \subset T$. Since the action is $k$-acylindrical, $\text{Fix}(g)$ has finite diameter.
Hence, the group $\mathcal{C}(g)$ acts on the subtree of finite diameter $\text{Fix}(g)$. It follows that there is a vertex $\tilde{v}_*$ in the subtree that is fixed by the action of $\mathcal{C}(g)$. Thus,
\begin{equation}
 \mathcal{C}(g) \subseteq \Stab(\tilde{v}_*).
\end{equation} 
If $T$ is the Bass-Serre tree of $G_1 *_{H} G_2$ or $G_1*_{H}$, then $\tilde{v}_* =  g G_i$ for some $g\in G$ and $i\in \{1,2\}$. It follows that 
$\mathcal{C}(g) \subseteq \Stab(v_*) = g G_i g^{-1}$; namely, $\mathcal{C}(g)$ is conjugate to a subgroup of one of the vertex groups. This proves the theorems stated at the beginning of the section.

\paragraph{Finite trees.} For completeness, we include two proofs of the fact that any tree $T$ with a finite diameter and a group $G$ acting on $T$ has a vertex $\tilde{v}_* \in T$ that is fixed by the group action. We suppress the tilde notation in this part of the section and simply write $v_*$ instead of  $\tilde{v}_*$. 

The first proof follows from the observation that a nontrivial tree with finite diameter must contain vertices of different degrees. The degree of a vertex is the number of edges attached to it. In particular, the vertices at the end of the tree must be of degree one. The only trees where all vertices have the same degree are the single vertex with no edges, and two vertices joined by one edge.

The action of any element $g\in G$ must preserve the degree of the vertex, so $G$ permutes between vertices of the same degree. Let $T'$ be the subtree obtained by removing the vertices of degree one. It follows that $G$ preserves $T'$, and $\text{diam}(T')<\text{diam}(T)$. Since the diameter decreases at each step, this process can occur only finitely many times. The final tree is one of the trivial cases: either a single vertex or a single edge with two vertices. In either case, since $G$ preserves edge orientations, $G$ either fixes the single vertex or the two vertices joined by an edge. 

The second proof we present identifies the fixed vertex with the  the central vertex of the graph. To explain the construction, let
\be 
e(v)=\sup_{w \in T} d(v,w)
\ee
be the eccentricity of a vertex. By definition, $e(v) \le \text{diam}(T)<\infty$. We also let $\rho=\min_{v \in T} e(v)$ the minimal eccentricity. We define the center of the graph as the set of vertices with minimal eccentricity,
\be 
C(T)=\{v \in T \, | \, e(v)=\rho\}\ .
\ee
Since the group action preserves distances, it has to preserve the set $C(T)$.
We claim that $C(T)$ consists of either one of two vertices connected with a single edge. In both cases (since edge inversions are not allowed), $G$ fixes $C(T)$ elementwise and thus we may choose $v_* \in C(T)$.

To see why $C(T)$ consists of at most two vertices, let $\gamma$ be a path of maximal length $\text{diam}(T)$ through $T$ realizing $\max_{v,\, w \in T} d(v,w)$ (without backtracking). Denote the two endpoints by $\gamma^-$ and $\gamma^+$. Let $v \in T$ be any vertex. We can find a unique closest vertex on $\gamma$ which we denote by $\pi(v)$. This is also called the projection of $v$ on $\gamma$. The geodesic path from $v$ to $\gamma^\pm$ then has to pass through $\pi(v)$. Thus
\begin{align} 
e(v) &\ge \max(d(v,\gamma^+),d(v,\gamma^-))\\
&=d(v,\pi(v))+\max(d(\pi(v),\gamma^+),d(\pi(v),\gamma^-)) \\
&\ge d(v,\pi(v))+ \lceil \tfrac{1}{2} \, \text{diam}(T) \rceil\ . 
\end{align}
In the last step, we used that $d(\pi(v),\gamma^+)+d(\pi(v),\gamma^-)=\text{diam}(T)$. Thus we see that $\rho=\lceil \tfrac{1}{2} \, \text{diam}(T) \rceil$. It is precisely attained by the central or the two central vertices of the path $\gamma$. For a vertex outside of $\gamma$, $d(v,\pi(v))>0$ and thus they are not part of $C(T)$. The claim follows.

%%%%%%%%%%%%%%%%%%%%%%%%
\section{VTQFT technology}
\label{app:technology}

\subsection{Virasoro TQFT identities}

Here we collect some Virasoro TQFT manipulations and identities that are useful for the computations that appear in the main text. All equalities should be understood to hold at the level of the VTQFT partition functions, and may be applied as local manipulations within a bigger network of Wilson lines.

\paragraph{Braiding.}
Braiding of Wilson lines produces a phase:
\begin{equation}\label{eq:appbraid}
\vcenter{\hbox{

   }}\, .
\end{equation}
The opposite braiding involves the opposite phase. Configurations with opposite patterns of over- and under-crossings are related by complex conjugation.

\paragraph{Fictitious identity line.}
We will frequently simplify Wilson line configurations via the insertion of a fictitious identity line followed by a fusion move:
\begin{equation}
   \vcenter{\hbox{
      %
   }} \label{eq:fictious identity}
\end{equation}
 This has the effect of adding two trivalent vertices to the Wilson line configuration.

\paragraph{Wilson bubble.}
The VTQFT Hilbert space on the twice-punctured sphere is one-dimensional, and is only nontrivial when the two Wilson lines carry identical Liouville momenta. Thus any cut along a 2-punctured sphere must produce an amplitude proportional to a delta function. In the case of the Wilson bubble pictured below, we have \cite{Collier:2023fwi}:
\begin{equation}\label{eq:Wilson bubble}
   \vcenter{\hbox{
   %
   }} \, .
\end{equation}

\paragraph{Fusion kernel and the $6j$ symbol.}
The Virasoro fusion kernel $\mathbb{F}$ implements the fusion transformation on Wilson line configurations:
\begin{equation}\label{eq:fusion transformation}
   \vcenter{\hbox{
      %
   }}\, .
\end{equation}
The fusion kernel is simply related to the Virasoro $6j$ symbol via the relation:
\begin{equation}\label{eq: relation btw fusion and modular kernels}
   %
\, .
\end{equation}
Here we have introduced the convenient shorthand:
\begin{equation}
   \mathsf{C}_{ijk} \equiv C_0(P_i,P_j,P_k)\, .
\end{equation}
In cases where there is little room for confusion, we will occasionally represent the $6j$ symbol using the following abbreviated notation:
\begin{equation}
    %
\ .
\end{equation}
The Virasoro $6j$ symbol enjoys a tetrahedral symmetry that manifests as an invariance under swapping the rows or columns of any $2\times 2$ sub-block. For example, from this we learn that the fusion kernel is straightforwardly related to its inverse as:
\begin{equation}\label{eq:fusioninverse}
   \frac{\mathbb{F}_{P_t P_s}%
}{\rho_0(P_t)C_0(P_1,P_4,P_t)C_0(P_2,P_3,P_t)} \, .
\end{equation}
Written in terms of the $6j$ symbol, the fusion transformation (\ref{eq:fusion transformation}) takes the form:
\begin{equation}
    \vcenter{\hbox{
      %
   }}\, .
\end{equation}

\paragraph{$u$-channel crossing.}
The Virasoro fusion kernel also describes the crossing transformations involving $u$-channel blocks. The crossing relations involve phases due to the braiding of Wilson lines:
\begin{align}
    %
\,.
    \end{align}

\paragraph{Wilson triangle.}
The VTQFT Hilbert space on the 3-punctured sphere is one-dimensional, so cutting a configuration of Wilson lines along a sphere punctured by three Wilson lines prepares a state proportional to that of the corresponding trivalent vertex. A simple example is the Wilson triangle shown below, which is related to the trivalent vertex by a factor proportional to the Virasoro $6j$ symbol \cite{Collier:2023fwi}:
\begin{equation}\label{eq:Wilson triangle}
   \vcenter{\hbox{
   %
   }}\ .
\end{equation}

\paragraph{Wilson square.}
With the Wilson triangle identity one can decompose more complicated Wilson line configurations into integrals over $6j$ symbols. For example, the ``Wilson square'' shown below prepares a state in the Hilbert space of the 4-punctured sphere, which admits the following representation:
\begin{align}
   \!\!\!\!\!\!\vcenter{\hbox{
   %
   }}\!\!\! .\label{eq:Wilson square}
\end{align}
The above relation resembles a crossing equation for a boundary 4-point function on the disk in BCFT, and indeed with a particular choice of operator normalization it is equivalent to the disk boundary 4-point function in Liouville CFT with FZZT boundary conditions \cite{Fateev:2000ik,Teschner:2000md,Ponsot:2001ng,Hosomichi:2001xc}, where the Liouville momenta of the $a,b,c,d$ Wilson lines correspond to the FZZT parameters (which are straightforwardly related to the boundary cosmological constants) of the boundary conditions. Similarly we can interpret the Wilson triangle above in terms of the FZZT boundary 3-point function \cite{Ponsot:2001ng}. The connection between simple observables in Virasoro TQFT and Liouville BCFT with FZZT boundary conditions will be explored in more detail elsewhere.

\paragraph{The modular crossing kernel and relation to the fusion kernel.}
The Virasoro modular kernel $\mathbb{S}$ locally implements the modular $S$ transformation on the once-punctured torus. It can be recast as an integral over the fusion kernel:
\begin{equation}
\label{SwithF}
   \frac{\mathbb{S}_{P_1P_2}[P_0]}{\rho_0(P_2)C_0(P_2,P_2,P_0)} =  \int_0^\infty \d P\, \e^{\pi i(2h_P + h_0 - 2h_1 -2h_2)} \frac{\mathbb{F}_{P_0 P}\begin{bmatrix} P_1 & P_2 \\ P_1 & P_2\end{bmatrix}}{C_0(P_1,P_2,P)}\, .
\end{equation}
This relation can be derived by computing the Virasoro TQFT amplitude on the Hopf link with an extra Wilson line embedded in $\mathrm{S}^3$, as discussed in \cite{Collier:2024mgv}:
\begin{equation}
   \vcenter{\hbox{
      \begin{tikzpicture}[scale=0.85]
         \draw[very thick,wilsonred] (-1/2,0) [partial ellipse=67:412:1 and 1];
         \draw[very thick,wilsonred] (1/2,0) [partial ellipse = 233:-112:1 and 1];
         \draw[very thick,wilsonred] (-1/2,0) to (1/2,0);
         \node[right] at (3/2,0) {$2$};
         \node[left] at (-3/2,0) {$1$};
         \node[above] at (0,0) {$0$};
      \end{tikzpicture}
   }}\, .
\end{equation}
This representation makes it manifest that the modular kernel satisfies the following identity that swaps the lower indices:
\begin{equation}
   \frac{\mathbb{S}_{P_1P_2}[P_0]}{\rho_0(P_2)C_0(P_2,P_2,P_0)} = \frac{\mathbb{S}_{P_2P_1}[P_0]}{\rho_0(P_1)C_0(P_1,P_1,P_0)}\, .
\end{equation}
It is sometimes cleaner to express VTQFT amplitudes in terms of the modular kernel in the `Racah-Wigner' normalization \cite{Eberhardt:2023mrq}, which following \cite{Post:2024itb} we denote with a hat:
\begin{equation}\label{eq:Shat_def}
    \widehat{\mathbb{S}}_{P_1P_2}[P_0] = \frac{1}{\rho_0(P_2)}\sqrt{\frac{C_0(P_1,P_1,P_0)}{C_0(P_2,P_2,P_0)}}\mathbb{S}_{P_1P_2}[P_0].
\end{equation}
In this normalization the modular kernel is invariant under swapping its lower indices:
\begin{equation}
    \widehat{\mathbb{S}}_{P_1P_2}[P_0] = \widehat{\mathbb{S}}_{P_2P_1}[P_0]\, .
\end{equation}
The modular kernel normalized this way can also be expressed as the following integral over the Virasoro $6j$ symbol:
\begin{align}
    \skerhat{P_1}{P_2}{P_0} &= \int_0^\infty \d P\, \rho_0(P) \e^{\pi i(2h_P+h_0-2h_1-2h_2)}\sixj{P_1}{P_2}{P}{P_2}{P_1}{P_0}\\
    &= \int_0^\infty \d P\, \rho_0(P) \e^{\pi i(2h_1+2h_2-2h_P)}\sixj{P_1}{P_2}{P}{P_2}{P_1}{P_0}\, .\label{eq:Shat and 6j}
\end{align}
In the second line above we have used the fact that the modular kernel satisfies the reality property
\begin{equation}\label{eq:Skernel reality}
    \widehat{\mathbb{S}}^*_{P_1P_2}[P_0] = \e^{-\pi i h_0}\,\widehat{\mathbb{S}}_{P_1P_2}[P_0].
\end{equation}
We can also go the opposite direction, and express the $6j$ symbol (with repeated external momenta) in terms of the modular kernel. This is the Virasoro-Verlinde formula \cite{Post:2024itb}:
\begin{equation}
    \sixj{P_3}{P_3}{P_0}{P_1}{P_1}{P_2} = \int_0^\infty \d P\, \skerhat{P_1}{P}{P_0} \,\sker{P}{P_2}{\id} \,\skerhat{P}{P_3}{P_0}^*.
\end{equation}
\paragraph{Consistency conditions of the Moore-Seiberg construction.}
The Virasoro fusion kernel/6j symbol and the modular crossing kernel describe the behaviour of Virasoro conformal blocks (or components of Virasoro blocks) under the action of elementary crossing moves. As such, they are subject to a number of consistency conditions \cite{Moore:1988qv}, which have been established in \cite{Teschner:2013tqy,Nidaiev:2013bda}. They are given as:
\begin{itemize}
    \item Idempotency:
    \begin{subequations}\label{eq:idempotency}
        \begin{align}
            \int_0^\infty \d P_b\, \rho_0(P_b)
        \e^{\pi i (h_0+h_{0'})}\widehat{\mathbb{S}}^*_{P_6P_2}[P_5]\ .
    \end{multline}
\end{itemize}

\paragraph{Verlinde lasso.}
The looping of a Wilson line by a ``Verlinde lasso'' as shown below produces a state that is proportional to the corresponding trivalent vertex by a factor of the modular crossing kernel \cite{Takahashi:2024ukk}:
\begin{equation}\label{eq:Verlinde lasso}
   \vcenter{\hbox{
   %
   }}
\end{equation}
This can be derived by inserting an identity line between the $1$ and $u$ Wilson lines, doing a fusion transformation, reducing the resulting Wilson triangle in terms of the $6j$ symbol using (\ref{eq:Wilson triangle}), and then recognizing the integral representation of the modular crossing kernel (\ref{eq:Shat and 6j}).
The opposite pattern of over- and under-crossings leads to the complex conjugate amplitude.

\paragraph{Verlinde loop.}
Taking the limit $P_3\to \id$ of the Verlinde lasso relation leads to the observation that Verlinde loops can be removed up to a factor of the modular crossing kernel \cite{Eberhardt:2023mrq}:
\begin{equation}\label{eq:verlindeloop}
   \vcenter{\hbox{

   }}
\end{equation}

\paragraph{Hooked lines.}
In studying knots and links in Virasoro TQFT we often encounter configurations involving Wilson lines linked as shown below, which can be simplified as follows: 
\begin{align}
   \vcenter{\hbox{
   %
   }} \label{eq:hookedlines}
\end{align}
These representations can be derived either by inserting a fictitious identity line between the two Wilson lines and then fusing the two lines, or by inserting an identity line between two ends of the same Wilson line, doing a fusion move, and then applying the Verlinde lasso rule (\ref{eq:Verlinde lasso}). The latter introduces a factor of the modular crossing kernel.
The linking with the opposite pattern of over- and under-crossings leads to the complex-conjugate amplitude.

\paragraph{Looped lines.}
With these basic rules amplitudes involving Verlinde loops surrounding more complicated configurations of Wilson lines can be evaluated by inserting various fictitious identity lines. For example, in the case of two looped lines as shown below, we have:
\begin{align}
        \vcenter{\hbox{
        \begin{tikzpicture}
            \draw[very thick, wilsonred] (0,0) [partial ellipse=150:-150:.2 and .5];
            \draw[very thick, wilsonred] (0,0) ellipse (.2 and .5);
            \draw[draw=none,fill=white] (-.15,.25) circle (.1);
            \draw[draw=none,fill=white] (-.15,-.25) circle (.1);
            \draw[very thick, wilsonred] (-1,.25) to (.1,.25);
            \draw[very thick, wilsonred] (-1,-.25) to (.1,-.25);
            \draw[very thick, wilsonred] (.3,.25) to (1,.25);
            \draw[very thick, wilsonred] (.3,-.25) to (1,-.25);
            \node[left] at (-1,.25) {$1$};
            \node[left] at (-1,-.25) {$2$};
            \node[above] at (0,.5) {$u$};
            \node[below,white] at (0,-.5) {$u$};
        \end{tikzpicture}
        }}
        &= \int_0^\infty \d P\,\rho_0(P)\mathsf{C}_{12P} \, \widehat{\mathbb{S}}_{uP}[\id]
        \vcenter{\hbox{
       \begin{tikzpicture}
          \draw[very thick, wilsonred] (-3/4,1/2) to (-1/4,0) to (1/4,0) to (3/4,1/2);
          \draw[very thick, wilsonred] (-1/4,0) to (-3/4,-1/2);
          \draw[very thick, wilsonred] (1/4,0) to (3/4,-1/2);
          \node at (-.95,.7) {$1$};
          \node at (-.95,-.7) {$2$};
          \node at (.95,.7) {$1$};
          \node at (.95,-.7) {$2$};
          \node [above] at (0,0) {$P$};
       \end{tikzpicture}
        }}\\
        &= \int_0^\infty \d P\, \rho_0(P)\sqrt{\mathsf{C}_{11P}\mathsf{C}_{22P}}\, \widehat{\mathbb{S}}^*_{uP_1}[P]\widehat{\mathbb{S}}_{uP_2}[P]
        \vcenter{\hbox{
       \begin{tikzpicture}
          \draw[very thick, wilsonred] (-1/2,3/4) to (0,1/4) to (0, -1/4) to (-1/2,-3/4);
          \draw[very thick, wilsonred] (0,1/4) to (1/2,3/4);
          \draw[very thick, wilsonred] (0,-1/4) to (1/2,-3/4);
          \node[left] at (0,0) {$P$};
          \node at (-.7,1.2-.25) {$1$};
          \node at (-.7,-1.2+.25) {$2$};
          \node at (.7,1.2-.25) {$1$};
          \node at (.7,-1.2+.25) {$2$};
       \end{tikzpicture}
       }} \label{eq:looped lines}
\end{align}
The first line is derived by inserting a fictitious identity line between the $1$ and $2$ Wilson lines, and the second by inserting an identity line between two ends of the $u$ Wilson line. As an aside, in a similar spirit to the comments made below (\ref{eq:Wilson square}), the above is proportional to the bulk two-point function in Liouville CFT on the disk with FZZT boundary conditions (where the Liouville momentum of the $u$ Wilson line corresponds to the FZZT parameter associated to the disk), with the expressions in the first and second lines corresponding to the closed- and open-string channels respectively.

\paragraph{Omega-loop.}
A special role is played by the so-called $\Omega$-loop, which is a closed Wilson line whose role is to project onto the identity Virasoro representation inside the loop. In VTQFT this is implemented by integrating the Liouville momentum of the privileged Wilson line weighted by the Cardy density of states $\rho_0(P) = \sker{\id}{P}{\id}$:
\begin{equation}
    \vcenter{\hbox{

    }}\, .
\end{equation}
We can obtain some nice identities by acting with the $\Omega$-loop on simple configurations of Wilson lines. For example, enclosing a single Wilson line it simply projects that Wilson line to the identity:
\begin{equation}\label{eq:omegaloop app}
   \vcenter{\hbox{
   %
   }}\,.
\end{equation}
To apply the $\Omega$-loop to more complicated configurations we use the rules of VTQFT to fuse Wilson lines such that a single representation runs through the loop and then project that representation to the identity. In this work a privileged role is played by the action of the $\Omega$-loop on a pair of Wilson lines, which can be computed by combining \eqref{eq:fictious identity} and \eqref{eq:omegaloop app}
\be \label{eq:omega loop on pair of lines}
        \vcenter{\hbox{
        %
        }}
\ee
This means that the $\Omega$-loop around two parallel Wilson lines exactly implements the cylinder surgery operation discussed in section~\ref{subsec:general gravitational machine}.

Let us also comment on the meaning of the delta-function $\delta(P-\id)$ appearing in \eqref{eq:omegaloop app}. It is meant to denote that the Wilson line originally labeled by the Liouville momentum $P$ is projected to the identity line, which is non-normalizable in VTQFT. To rigorously make sense of it, we would need to delve into the theory of distributions, see \cite{GelfandShilov2} or \cite{Maxfield:2019hdt} for an attempt within 2d CFT. For the purposes of this paper, in practice we only require the identity \eqref{eq:omega loop on pair of lines}. The presence of an $\Omega$-loop around a single line typically leads to non-hyperbolic manifolds. In fact, the condition \ref{cond:no Omega loops 2} for the cylinder surgery of the gravitational machine is precisely tailored to avoid this situation.

\paragraph{Action of the modular transform on knotted graphs.} Using the relationship between the modular and the fusion kernels \eqref{SwithF}, we can diagrammatically represent the action of a S-transform on a loop in a knotted trivalent graph: 
\begin{align}
    \int_0^\infty \d P_2 \,\sker{P_1}{P_2}{P_0} 
    \vcenter{
    \hbox{

    }
    }\ .\label{eq:modular_on_graphs}
\end{align}
Here the three lines are meant as a proxy for any more general configuration of Wilson lines.
This equation tells us that the S-transformed diagram now has an additional Dehn filling. On the right-hand side the Wilson line with Liouville momentum $P_2$ is integrated with the $\rho_0$ kernel and thus acts as an $\Omega$ loop.
One can derive this expression by utilizing \eqref{SwithF} and noting that:
\begin{equation}
    \vcenter{
    \hbox{
    %
    }
    }.
\end{equation}

\subsection{Kirby calculus} \label{subapp:Kirby calculus}
We can think of many of the above VTQFT identities as arising from a homeomorphism of 3-manifolds $M \cong M'$, since $Z_{\text{Vir}}$ is a topological invariant. The main tool that is used in topology to find homeomorphisms between 3-manifolds is known as `Kirby calculus'.
Here, we present some basic aspects of Kirby calculus relevant to the Virasoro TQFT. For a more detailed review of this topic, we recommend the following literature \cite{text1,text2,text3}. 

The Lickorish–Wallace theorem \cite{Lickorish1,Wallace1} states that any closed, connected, orientable 3-manifold can be obtained by integer Dehn surgery on the components of a link inside S$^3$.\footnote{There is also an extension to manifolds with boundary, see e.g.~appendix B.3 of \cite{text1}.} Integer Dehn surgery means that the coefficient $p/q$ associated to each link component can be chosen to be an integer, i.e.~with $q=1$. This result allows for the representation of any 3-manifold as a link with a framing. For example, from left to right, the following diagrams represent the manifolds $\mathrm{S}^2\times \mathrm{S}^1$, the 3-torus $T^3$, and the Lens space $L(7,5)$:
\begin{equation}
    \vcenter{
    \hbox{

    }
    }.
\end{equation}
The framing determines which Dehn filling to perform on the corresponding link component.

There are several standard conventions for indicating the framing. In the integer framing, each link component is assigned an integer. In the annular framing, the link is thickened into a ribbon which keeps track of these coefficients. The blackboard framing uses the diagram itself to record the framing. These three conventions (integer, annular, and blackboard) are illustrated below:
\begin{equation}
\vcenter{
\hbox{
%
}
}\ .
\end{equation}

The representation of closed 3-manifolds in terms of framed links is not unique, and the role of Kirby calculus is to determine when two framed links correspond to the same topology. For example, two links that are related via a sequence of \emph{framed} Reidemeister moves yield homeomorphic 3-manifolds.  There are three framed Reidemeister moves, with only the first differing from its classical version:
\begin{equation}
\begin{array}{c}
\text{R1:}\quad 
    \vcenter{\hbox{
\begin{tikzpicture}[scale=0.8]
    \draw[very thick, wilsonred] (0,0.75) arc (90:270:0.5 and 0.75);
    \draw[very thick,wilsonred] (0,0.75) to[out=0, in =-70, looseness=6] (-0.2,0.6);
    \draw[very thick, wilsonred] (-0.25,0.74) to[out = 110, in = 282] (-0.4,1.3);
    \draw[very thick,wilsonred] (0,-0.75) to[out=0, in =70, looseness=6] (-0.2,-0.6);
    \draw[very thick, wilsonred] (-0.25,-0.74) to[out = -110, in = -280] (-0.4,-1.3);
\end{tikzpicture}
    }}
    \quad \longleftrightarrow\quad 
    \vcenter{\hbox{
\begin{tikzpicture}[scale=0.8]
    \draw[very thick, wilsonred] (0,1.3) -- (0,-1.3); 
\end{tikzpicture}
    }}
    \quad \longleftrightarrow\quad 
 \vcenter{\hbox{
\begin{tikzpicture}[scale=0.8]
    \draw[very thick,wilsonred] (0,0.75) to[out=0, in =-70, looseness=6] (-0.2,0.6) to[out = 110, in = 285] (-0.4,1.3);
    \draw[very thick,wilsonred] (0,-0.75) to[out=0, in =70, looseness=6] (-0.2,-0.6) to[out = -110, in = -285] (-0.4,-1.3);
    \draw[very thick, wilsonred] (0,0.75) arc (90:110:0.5 and 0.75);
    \draw[very thick, wilsonred] ({0.5*cos(125)},{0.75*sin(125)}) arc (125:235:0.5 and 0.75);
    \draw[very thick, wilsonred] ({0.5*cos(250)},{0.75*sin(250)}) arc (250:270:0.5 and 0.75);
\end{tikzpicture}
    }},\\
\end{array}
\end{equation}
The need for framed moves instead of the standard Reidemeister moves is because the framed version preserves the blackboard framing.
Kirby \cite{Kirby1978} introduced two additional moves (K1 and K2) and proved that the surgery on a pair of framed links correspond to the same 3-manifold if and only if they are related by a finite sequence of R and K moves. The first Kirby move adds or deletes a disjoint unknot with framing $+1$ or $-1$ from the link $L$,
\begin{equation}
    \text{K1:} \quad L \quad \longleftrightarrow \quad L\; \sqcup\;
    \vcenter{
\hbox{
%
}
}.
\end{equation}
The second Kirby move, also known as a handle slide, is more involved and is better explained with an example:
\begin{equation}
\text{K2:}\quad
\vcenter{\hbox{
%
}}.
\end{equation}
This move involves two link components drawn in the blackboard frame. It slides one of the links along the framing of the other. 

From K1 and K2, one can build composite moves which make it easier to identify whenever two links correspond to the same manifold. Two of such operations are the blow up FR$_+$ and the blow down FR$_-$:  
\begin{equation}
    \vcenter{
    \hbox{
    %
    }}.
\end{equation}
These operation allow for the twisting (or untwisting) of $n$ strands via the deletion (or inclusion) of a $\pm 1$-framed unknot. The case where $n=0$ is equivalent to the Kirby move K1. These operations are also known in the literature as the Fenn-Rourke (FR) moves. Fenn and Rourke \cite{FENN19791} showed that two links in S$^3$ give homeomophic manifolds if and only if they are related by a sequence of blow ups and blow downs.  Below, we show that these moves are naturally realized within the Virasoro TQFT and follow directly from basic properties of the PSL$(2, \mathbb{Z})$ crossing kernels.

\paragraph{Blow ups and blow downs.} One can exploit the fact that framed links related by Kirby moves yield homeomorphic manifolds to derive useful VTQFT identities. However, care is required when manipulating VTQFT diagrams, as only those moves that do not involve cutting along non-hyperbolic manifolds are meaningful in VTQFT. In what follows, we show that blow-ups and blow-downs FR$_\pm$ are valid in VTQFT when the number of strands satisfies $n>0$. 

Using the PSL$(2,\mathbb{Z})$ relation
\begin{equation}
   \e^{2\pi i h_2} \sker{P_2}{P_1}{P_0} \e^{2\pi i h_1} = \int \text{d} P_3\,  \sker{P_2}{P_3}{P_0}\e^{-2\pi i h_3}
   \sker{P_3}{P_1}{P_0},
\end{equation} 
one can show that the FR$_+$ move is satisfied for a single strand:
\begin{equation}
\label{rrel1}
\int \text{d} P \rho_0(P)
    \vcenter{
    \hbox{

    }}.
\end{equation}
A similar result holds for FR$_-$.
Using the result for $n=1$, the result for any $n$ follows directly. For example, the case where $n=2$ is given by:
\begin{align}
\int \text{d} P \rho_0(P)
    \vcenter{
    \hbox{
    %
    }},
\end{align}
and so on for $n>2$.

\bibliography{bib}
\bibliographystyle{JHEP}

\end{document}